\documentclass[numberedappendix,twocolappendix,iop,apj]{emulateapj}
\usepackage{psfig,amsfonts,amsmath,graphicx,natbib,nicefrac,url,hyperref,breakurl,longtable,graphicx
}

\usepackage{booktabs}
\hypersetup{colorlinks=true, urlcolor=blue, citecolor=blue}

\newcommand{\Swift}{\textit{Swift}}
\newcommand{\Konus}{\textit{Konus-Wind}}

\newcommand{\EK}{\ensuremath{E_{\rm K}}}
\newcommand{\EKiso}{\ensuremath{E_{\rm K,iso}}}
\newcommand{\Egamma}{\ensuremath{E_{\gamma}}}
\newcommand{\Egammaiso}{\ensuremath{E_{\gamma,\rm iso}}}	     
\newcommand{\etarad}{\ensuremath{\eta_{\rm rad}}}
\newcommand{\epse}{\ensuremath{\epsilon_{\rm e}}}
\newcommand{\epsb}{\ensuremath{\epsilon_{\rm B}}}
\newcommand{\dens}{\ensuremath{n_{0}}}

\newcommand{\tjet}{\ensuremath{t_{\rm jet}}}
\newcommand{\thetajet}{\ensuremath{\theta_{\rm jet}}}

\newcommand{\AV}{\ensuremath{A_{\rm V}}}

\newcommand{\pcmsq}{\ensuremath{{\rm cm}^{-2}}}
\newcommand{\pcc}{\ensuremath{{\rm cm}^{-3}}}

\newcommand{\p}{\ensuremath{^{\prime}}}

\newcommand{\nua}{\ensuremath{\nu_{\rm a}}}

\newcommand{\numax}{\ensuremath{\nu_{\rm m}}}
\newcommand{\nuc}{\ensuremath{\nu_{\rm c}}}
\newcommand{\nuaf}{\ensuremath{\nu_{\rm a,f}}}
\newcommand{\numf}{\ensuremath{\nu_{\rm m,f}}}
\newcommand{\nucf}{\ensuremath{\nu_{\rm c,f}}}
\newcommand{\fnumf}{\ensuremath{F_{\nu,\rm m,f}}}
\newcommand{\nuar}{\ensuremath{\nu_{\rm a,r}}}
\newcommand{\numr}{\ensuremath{\nu_{\rm m,r}}}
\newcommand{\nucr}{\ensuremath{\nu_{\rm c,r}}}
\newcommand{\fnumr}{\ensuremath{F_{\nu,\rm m,r}}}
\newcommand{\gammae}{\ensuremath{\gamma_{\rm e}}}

\newcommand{\RB}{\ensuremath{R_{\rm B}}}
\newcommand{\RE}{\ensuremath{R_{\rm e}}}
\newcommand{\tE}{\ensuremath{t_{\rm E}}}

\newcommand{\nuopt}{\ensuremath{\nu_{\rm opt}}}
\newcommand{\nux}{\ensuremath{\nu_{\rm X}}}

\newcommand{\nucratio}{\ensuremath{\mathcal{Q}_{\rm c,fr}}}

\newcommand{\fnupk}{\ensuremath{F_{\nu,\rm pk}}}
\newcommand{\fnumax}{\ensuremath{F_{\nu,\rm m}}}
\newcommand{\td}{\ensuremath{t_{\rm d}}}

\shorttitle{A Refreshed RS in GRB\,161219B}
\shortauthors{Laskar et al.}

\def\nrao{1}
\def\ucb{2}
\def\janskyfellow{3}
\def\cfa{4}
\def\ferrara{5}
\def\northwestern{6}
\def\hubblefellow{7}
\def\ucsc{8}
\def\ariz{9}
\def\drouta{10}
\def\droutb{11}
\def\bath{12}
\def\lvjm{13}
\def\okc{14}
\def\cit{15}
\def\cssac{16}
\def\mpifr{17}
\def\yitf{18}

\begin{document} 
\submitted{Submitted 2017 October 17; revised 2018 May 8; accepted 2018 June 4}

\title{First ALMA Light Curve Constrains Refreshed Reverse Shocks \& Jet Magnetization in
GRB\,161219B}
\author{Tanmoy~Laskar\altaffilmark{\nrao,\ucb,\janskyfellow},
        Kate~D.~Alexander\altaffilmark{\cfa},
        Edo~Berger\altaffilmark{\cfa},
	Cristiano~Guidorzi\altaffilmark{\ferrara},
	Raffaella Margutti\altaffilmark{\northwestern},
	Wen-fai~Fong\altaffilmark{\northwestern,\hubblefellow},
	Charles~D.~Kilpatrick\altaffilmark{\ucsc},
	Peter~Milne\altaffilmark{\ariz},
	Maria~R.~Drout\altaffilmark{\drouta,\droutb},
	C.~G.~Mundell\altaffilmark{\bath},        
	Shiho~Kobayashi\altaffilmark{\lvjm},
        Ragnhild~Lunnan\altaffilmark{\okc,\cit},
	Rodolfo Barniol Duran\altaffilmark{\cssac},
        Karl~M.~Menten\altaffilmark{\mpifr},
        Kunihito~Ioka\altaffilmark{\yitf},        
        Peter~K.~G.~Williams\altaffilmark{\cfa}
}
\altaffiltext{\nrao}{National Radio Astronomy Observatory,
520 Edgemont Road, Charlottesville, VA 22903, USA} 
\altaffiltext{\ucb}{Department of Astronomy, University of California, 501 Campbell Hall, 
Berkeley, CA 94720-3411, USA} 
\altaffiltext{\janskyfellow}{Jansky Fellow}
\altaffiltext{\cfa}{Department of Astronomy, Harvard University, 60 Garden Street, Cambridge, MA 
02138, USA}
\altaffiltext{\ferrara}{Department of Physics and Earth Science, University of Ferrara, via Saragat 
1, I-44122, Ferrara, Italy}
\altaffiltext{\northwestern}{Center for Interdisciplinary Exploration and Research in Astrophysics 
(CIERA) and Department of Physics and Astrophysics, Northwestern University, Evanston, IL 60208, 
USA}
\altaffiltext{\hubblefellow}{Hubble Fellow}
\altaffiltext{\ucsc}{Department of Astronomy and Astrophysics, University of California, Santa 
Cruz, CA 95064, USA}
\altaffiltext{\ariz}{Steward Observatory, University of Arizona, 933 N. Cherry Ave, Tucson, AZ 
 85721, USA}
\altaffiltext{\drouta}{Hubble and Carnegie-Dunlap Fellow}
\altaffiltext{\droutb}{The Observatories of the Carnegie Institution for Science, 813 Santa Barbara 
Street, Pasadena, CA 91101, USA}
\altaffiltext{\bath}{Department of Physics, University of Bath, Claverton Down, Bath, BA2 7AY, 
United Kingdom}
\altaffiltext{\lvjm}{Astrophysics Research Institute, Liverpool John Moores University, 
IC2, Liverpool Science Park, 146 Brownlow Hill, Liverpool L3 5RF, United Kingdom}
\altaffiltext{\okc}{The Oskar Klein Centre \& Department of Astronomy, Stockholm University, 
AlbaNova, SE-106 91 Stockholm, Sweden}
\altaffiltext{\cit}{Department of Astronomy, California Institute of Technology, 1200 East 
California Boulevard, Pasadena, CA 91125, USA}
\altaffiltext{\cssac}{Department of Physics and Astronomy, California State University, Sacramento, 
6000 J Street, Sacramento, CA 95819, USA}
\altaffiltext{\mpifr}{Max-Planck-Institut f{\"u}r Radioastronomie, Auf dem Huegel 69, 53121 Bonn, 
Germany}
\altaffiltext{\yitf}{Center for Gravitational Physics, Yukawa Institute for Theoretical Physics, 
Kyoto University, Kyoto 606-8502, Japan}

\begin{abstract}
We present detailed multi-wavelength observations of GRB\,161219B at $z=0.1475$, spanning the radio 
to X-ray regimes, and the first ALMA light curve of a GRB afterglow. The cm- and mm-band 
observations before $8.5$\,d require emission in excess of that produced by the afterglow 
forward shock (FS). 
These data are consistent with radiation from a refreshed reverse shock (RS) produced by 
the injection of energy into the FS, signatures of which are also present in the X-ray and optical 
light curves. We infer a constant-density circumburst environment with an extremely low density, 
$\dens\approx 3\times10^{-4}$\,\pcc, and show that this is a characteristic of all strong RS 
detections to date. 
The VLA observations exhibit unexpected rapid variability on $\sim$ minute timescales, 
indicative of strong interstellar scintillation. 
The X-ray, ALMA, and VLA observations together constrain the 
jet break time, $\tjet\approx32$\,d, yielding a wide jet opening angle of 
$\thetajet\approx13^{\circ}$, implying beaming corrected $\gamma$-ray and kinetic energies of 
$\Egamma\approx4.9\times10^{48}$\,erg and $\EK\approx1.3\times10^{50}$\,erg, respectively. 
Comparing the RS and FS emission, we show that the ejecta are only 
weakly magnetized, with relative magnetization, $\RB\approx1$, compared to the FS. These 
direct, multi-frequency measurements of a refreshed RS spanning the optical to radio bands 
highlight the impact of radio and millimeter data in probing the production and nature of GRB 
jets.
\end{abstract}


\section{Introduction}
Long-duration $\gamma$-ray bursts (GRBs) have thus far been almost exclusively discovered through 
their prompt $\gamma$-ray emission, which unequivocally arises from relativistic outflows at high 
Lorentz factors, $\Gamma\gtrsim10^2$ \citep{kp91,feh93,wl95,bh95,bh97,ls01}. These outflows are 
understood to be produced by a nascent, compact central engine, such as a magnetar or accreting 
black hole, formed in the collapsing core of a dying massive star \citep{wb06,pir05,mgt+11}. The 
internal shock model proposed to explain the $\gamma$-ray emission invokes collisions between shells 
with a wide distribution of Lorentz factors ejected by the engine \citep{rm92,kps97,kp00a}. 
Understanding the distribution of ejecta energy as a function of their Lorentz factor is therefore a 
critical probe of the nature of the central engine, its energy source, and the energy extraction 
mechanism \citep{woo93,mw99,ami+00,npk01,zwm03,tmn08}. 

While monitoring the $\gamma$-ray sky remains an excellent means for detecting GRBs, a detailed 
description of the energetics of their jets and their progenitor environments is only possible 
through a study of the long-lasting X-ray to radio afterglow, generated when ejecta interact with 
their circumburst environment setting up the forward shock, and producing synchrotron radiation 
\citep[FS;][]{spn98}. Theoretical modeling of detailed multi-wavelength observations in the 
synchrotron framework yields the energy of the explosion, the degree of jet collimation, the density 
of the surrounding medium, and the mass loss history of the progenitor star, as well as information 
about the microphysical processes responsible for relativistic particle acceleration 
\cite{sph99,cl00,gs02}. 

Whereas GRB afterglows have traditionally been modeled as arising from jets with a uniform 
bulk Lorentz factor, radially structured ejecta profiles with energy spanning a range of Lorentz 
factors are gaining traction as viable models for the observed deviations of X-ray and optical 
light curves from the synchrotron model\footnote{Alternate explanations include circumburst density 
enhancements, structured jets, viewing angle effects, varying microphysical parameters, and 
gravitational microlensing 
\citep{zfd+06,nkg+06,pmg+06,tiyn06,eg06,gkp06,jyfw07,sd07,kwhc10,dm15,uz14}.
} 
\citep{np03,bhpf02,bgj04,hcg06,jbg06,mmk+08,mgk+09,tsg+12,vmp+13}. 
Ejecta released later, or at lower Lorentz factors than the initial impulsive shell responsible for 
the prompt emission, catch up with the contact discontinuity during the afterglow phase and inject 
energy into the FS \citep{rm98,sm00}. Energy injection through massive ejecta may explain late-time 
plateaus, re-brightening events, slow decays, and unexpected breaks observed in the X-ray 
and optical light curves of some afterglows 
\citep{kp00,zm02,gnp03,pmg+06,mhm+07,gvs+07,mgg+10,hdpm+12,llt+12,gkn+13,pvw13,nef+14,dpko+15,bm17},
 and forms a distinct class of models from late-time central engine activity, which has been 
invoked to explain some rapid X-ray and optical flares 
\citep{brf+05,ikz05,gngc09,nggc10,mgc+10,mbbd+11,llt+12}.

The process of energy transfer between the ejecta and the circumburst medium is expected to 
be mediated by a reverse shock (RS) propagating into the ejecta during the injection period. 
This RS is similar to the one expected from the deceleration of the ejecta by the circumburst 
environment as observed in exquisite detail in the afterglow of GRB\,130427A 
\citep{lbz+13,pcc+14,vdhpdb+14}; however, an RS supported by energy injection is expected to 
continue propagating into the ejecta during the entire injection period \citep{zm02}.
If injection takes place in the form of a violent shell collision, the resulting strong RS is 
expected to exhibit a detectable observational signature in the form of an optical flash or radio 
flare \citep{abb+99,sp99a,kfs+99,sr02,zm02,kz03,bsfk03,sr03,clf04}. In the case of gentle, or 
continuous energy injection, the RS is long-lasting, and its flux remains proportional to that of 
the FS during the entire injection period, $\fnumr\propto\Gamma\fnumf$ 
\citep{sm00,zm02,pk04,gdm07,lcj17}. Thus, it may be possible to detect reverse shocks arising from 
energy injection in cases both of violent collisions and of interactions at high enough ejecta 
Lorentz factor. Strong RS signatures are also excellent probes of the magnetization of the jets 
($\sigma_{\rm B}$), since high $\sigma_{\rm B}$ effectively increases the sound speed\footnote{In 
magnetized media, information travels at the speed of the fast magnetosonic wave.}, thereby 
suppressing shock formation \citep{gma08}. 

Our previous observations of GRB\,140304A at $z\approx5.3$ yielded the first multi-frequency, 
multi-epoch detection of a RS from a violent shell collision, lending credence to the multi-shell 
model \citep{lbm+17}. However, the high redshift of this event impacted the quality of data, 
limiting the strength of the inference feasible. In an analysis of four GRB afterglows exhibiting 
late-time optical and X-ray re-brightening events, we constrained the distribution of ejecta energy 
as a function of Lorentz factor \citep{lbm+15}. In one case, our observations were incompatible with 
RS radiation from the injection, suggesting collisions in at least some instances may 
be gentle processes; for the remaining three cases, the observations lacked the requisite 
temporal sampling and frequency coverage to conclusively rule out an injection RS. The reason may 
partly stem from the fact that the RS emission peaks in the mm-band for typical shock parameters, 
and no facilities in this observing window had the requisite sensitivity \citep{duplm+12}. However, 
the advent of the Atacama Large Millimeter/submillimeter Array (ALMA) now allows us to track the 
mm-band evolution of afterglows to a sensitivity $\sim 30$--$100\,\mu$Jy for the first time, 
re-energizing the search for refreshed reverse shocks.

Here we report detailed radio through X-ray observations of GRB\,161219B at $z=0.1475$, and present 
the first ALMA light curve of a GRB afterglow. The cm-band SEDs at $\lesssim 8.5$\,d exhibit 
unusual spectral features, which we discuss in detail in a separate work (Alexander et el., in 
prep; henceforth ALB18). Through multi-wavelength modeling of the X-ray, optical, and late 
radio data, we constrain the parameters of the FS powering the afterglow emission. The 
resulting model over-predicts the early X-ray emission, which can be explained by an episode of 
energy injection culminating at $\approx0.25$\,d. We interpret the early optical and radio 
observations as arising from a reverse shock launched by the same injection event. By tying the RS 
and FS parameters together, we show that the ejecta were not strongly magnetized. 
We employ standard cosmological parameters of $\Omega_m=0.31$, $\Omega_{\lambda}=0.69$, and 
$H_0=68$\,km\,s$^{-1}$\,Mpc$^{-1}$. All magnitudes are in the AB system and not corrected for 
Galactic extinction\footnote{Galactic extinction correction based on \cite{sf11} is built into our 
modeling software \citep{lbt+14}.}, all uncertainties are at 1$\sigma$, and all times are relative 
to the \Swift\ trigger time and in the observer frame, unless otherwise indicated.

\begin{deluxetable}{lc}
 \tabletypesize{\footnotesize}
 \tablecolumns{2}
 \tablecaption{XRT Spectral Analysis for GRB\,161219B\label{tab:xrtspect}}
 \tablehead{   
   \colhead{Parameter} &
   \colhead{Value} 
   }
 \startdata 
 $T_{\rm start}$ (s)                                    & $1.1\times10^{2}$ \\
 $T_{\rm end}$ (s)                                      & $1.1\times10^{7}$ \\
 $N_{\rm H, gal}$ ($10^{20}$ \pcmsq)                    & 3.06 \\
 $N_{\rm H, int}$ ($10^{21}$ \pcmsq)                    & $2.2\pm0.1$\\
 Photon index, $\Gamma_{\rm X}$                         & $1.86\pm0.03$\\
 Flux$^{\dag}$ (observed)   & $(1.86\pm0.05)\times10^{-12}$ \\
 Flux$^{\dag}$ (unabsorbed) & $(2.41\pm0.06)\times10^{-12\ddag}$ \\
 C statistic (dof) & 684 (699)
 \enddata
 \tablecomments{${}^\dag$ erg\,\pcmsq\,s$^{-1}$ (0.3--10\,keV); $^{\ddag}$ assuming the same 
fractional uncertainty as for the absorbed flux.}
\end{deluxetable}

\section{GRB Properties and Observations}
\label{text:GRB_Properties_and_Observations}
GRB\,161219B was discovered by the \Swift\ \citep{gcg+04} Burst Alert Telescope 
\citep[BAT,][]{bbc+05} on 2016 December 19 at 18:48:39\,UT \citep{gcn20296}. The burst duration  
is $T_{90} = 6.94\pm0.79$\,s, and the $\gamma$-ray spectrum is well fit with a power law plus 
exponential cut off model\footnote{Here, $dN_{\gamma}$ is the number of photons with energy in the 
range, $E$ to $E+dE$.}
\begin{align}
\frac{dN_{\gamma}}{dE_{\gamma}} = 
E_{\gamma}^{\alpha_{\gamma}}e^{-E_{\gamma}(2+\alpha_{\gamma})/E_{\gamma,\rm peak}}, 
\end{align}
with power law photon index, $\alpha_{\gamma} = -1.29\pm0.35$ and $E_{\gamma,\rm 
peak}=61.9\pm16.5$\,keV, yielding a fluence of $F_{\gamma} = (1.5 \pm 0.1) 
\times10^{-6}$\,erg\,\pcmsq\ \citep[15--150\,keV, $90\%$ confidence;][]{gcn20308}. The burst was 
also detected by \Konus\ with a duration of $T_{90}\approx10$\,s; the spectral fit to the \Konus\ 
light curve yields $\alpha_{\gamma} = -1.59\pm0.71$, $E_{\rm peak}=91\pm21$\,keV, and 
$F_{\gamma}=(3.1\pm0.8)\times10^{-6}$\,erg\,\pcmsq\ \citep[20-1000\,keV, $1\sigma$;][]{gcn20323}.
The optical afterglow was discovered by the \Swift\ UV/Optical Telescope \citep[UVOT;][]{rkm+05} in 
observations beginning 112\,s after the BAT trigger \citep{gcn20306}. Spectroscopic 
observations 36\,hr after the burst with the X-shooter instrument on the ESO VLT 8.2\,m telescope 
provided a redshift of $z = 0.1475$ \citep{gcn20321}.
At this redshift, the inferred isotropic equivalent $\gamma$-ray energy in the 1-$10^4$ keV rest 
frame energy band is $\Egammaiso=(1.8\pm0.4)\times10^{50}$\,erg from \Konus\ and 
$\Egammaiso=(1.1\pm0.1)\times10^{50}$\,erg from \Swift-BAT, respectively, based on a Monte Carlo 
analysis using the respective spectral parameters. Since the \Konus\ energy range is wider than the 
BAT band and therefore samples more of the $\gamma$-ray spectrum, we use the value of $\Egammaiso$ 
as determined from \Konus\ in this work. The corresponding isotropic-equivalent luminosity, 
$L_{\gamma,\rm iso}=E_{\gamma,\rm iso}(1+z)T_{90}^{-1}\approx10^{49}$\,erg\,s$^{-1}$, which makes 
this an intermediate-luminosity GRB \citep{bnp11}.

\subsection{X-ray: \Swift/XRT}\label{text:data_analysis:XRT}                            
The \Swift\ X-ray Telescope \citep[XRT,][]{bhn+05} began observing GRB\,161219B 108\,s after the 
BAT trigger. The X-ray afterglow was localized to RA = 06h\,06m\,51.37s, Dec = 
-26$^\circ$\,47\arcmin\,29.7\arcsec\ (J2000), with an uncertainty radius of 1.4\arcsec\ (90\% 
containment)\footnote{\url{http://www.swift.ac.uk/xrt_positions/727541/}}. XRT continued 
observing the afterglow for $123$\,d in photon counting mode. 

We extract XRT PC-mode spectra using the on-line tool on the \Swift\ website 
\citep{ebp+07,ebp+09}\footnote{\url{http://www.swift.ac.uk/xrt_spectra/727541/}}.
We downloaded the event and response files 
and fit them using the HEASOFT (v6.19) software package and corresponding calibration files. We 
used Xspec to fit the data, assuming a photoelectrically absorbed power law model (\texttt{tbabs 
$\times$ ztbabs $\times$ pow}), constraining the intrinsic absorption to remain constant across the 
epochs, and fixing the galactic absorption column to $N_{\rm H, Gal} = 
3.06\times10^{20}\,\pcmsq$ \citep{wsb+13}. We do not find strong evidence for evolution in 
the X-ray photon index.
Constraining the photon index to remain fixed, we find $\Gamma_{\rm X} = 1.86\pm0.03$ for a 
spectrum comprising all available PC-mode data (Table \ref{tab:xrtspect}). We use this value of the 
photon index and the unabsorbed counts-to-flux conversion rate from the \Swift\ 
website of $4.95\times10^{-11}$\,erg\,\pcmsq\,ct$^{-1}$ to convert the 0.3--10\,keV count rate 
light curve\footnote{Obtained from the \Swift\ website at 
\url{http://www.swift.ac.uk/xrt_curves/727541} and re-binned to a minimum signal-to-noise ratio per 
bin of 8.} to flux density at 1\,keV for subsequent analysis. We combine the uncertainty in flux 
calibration based on our spectral analysis (2.4\%) in quadrature with the statistical uncertainty 
from the on-line light curve. 

\begin{deluxetable}{clccc}
\tabletypesize{\scriptsize}
\tablecaption{\Swift\ UVOT Observations of GRB\,161219B \label{tab:data:UVOT}}
\tablehead{
\colhead{Mid-time,} & \colhead{UVOT} & \colhead{Flux density} & \colhead{Uncertainty}
& \colhead{Detection?}\\ 
\colhead{$\Delta t$ (d)} & \colhead{band} & \colhead{(mJy)} & \colhead{(mJy)} & \colhead{(1=Yes)}}

\startdata
$2.16\times10^{-3}$ &  \textit{uwh} & $6.19\times10^{-1}$ & $1.74\times10^{-2}$ & 1 \\
$5.19\times10^{-3}$ &  \textit{uvu} & $5.97\times10^{-1}$ & $1.67\times10^{-2}$ & 1 \\
$6.81\times10^{-3}$ &  \textit{uvb} & $6.92\times10^{-1}$ & $7.38\times10^{-2}$ & 1 \\
$7.10\times10^{-3}$ &  \textit{uwh} & $5.30\times10^{-1}$ & $2.50\times10^{-2}$ & 1 \\
$7.39\times10^{-3}$ &  \textit{uw2} & $2.99\times10^{-1}$ & $4.12\times10^{-2}$ & 1 \\
\ldots & \ldots & \ldots
\enddata
\tablecomments{This is a sample of the full table available on-line.}
\end{deluxetable}

\begin{deluxetable*}{ccccccccc}
\tabletypesize{\scriptsize}
\tablecaption{Optical and Near-IR Observations of GRB\,161219B \label{tab:data:opt}}
\tablehead{
\colhead{$\Delta t$} & \colhead{Observatory} & \colhead{Instrument} &
\colhead{Filter} & \colhead{Frequency} & \colhead{Flux density} & 
\colhead{Uncertainty\tablenotemark{\dag}} & \colhead{Detection?} & \colhead{Reference}\\ 
\colhead{(d)} & & & & \colhead{(Hz)} & \colhead{(mJy)} & \colhead{(mJy)} & \colhead{1=Yes} }

\startdata
$5.44\times10^{-4}$ & SAAO & MASTER & \textit{CR} & $4.67\times10^{14}$ & $1.34\times10^{0}$ & 
$4.26\times10^{-1}$ & 1 & \cite{gcn20330} \\
$7.33\times10^{-3}$ & Terksol & K-800 & \textit{CR} & $4.67\times10^{14}$ & $6.52\times10^{-1}$ & 
$2.35\times10^{-2}$ & 1 & \cite{gcn20309} \\
$8.82\times10^{-3}$ & Terksol & K-800 & \textit{CR} & $4.67\times10^{14}$ & $7.29\times10^{-1}$ & 
$2.81\times10^{-2}$ & 1 & \cite{gcn20309} \\
\ldots & \ldots & \ldots & \ldots & \ldots & \ldots & \ldots & \ldots & \ldots
\enddata
\tablecomments{\dag An uncertainty of 0.2~AB mag is assumed where not provided. The data have 
not been corrected for Galactic extinction. This is a sample of the full table available on-line.}
\end{deluxetable*}

\subsection{UV, optical, and near-IR}
\label{text:data_analysis:optical}
We analyzed the UVOT data using HEASOFT (v. 6.19) and corresponding calibration files. 
The afterglow was detected in all seven optical and UV filters. The background near the source was 
dominated by diffracted lighted from a nearby $R\sim13$\,mag USNO-B1 star (RA = 06h\,06m\,50.65s, 
Dec = -26$^\circ$\,47\arcmin\,53.3\arcsec; J2000) 21\arcsec\ SE of the afterglow.
We performed photometry using the recommended $5\arcsec$ aperture centered on the source, 
but estimated the background contribution using an annulus with inner radius 21\arcsec\ and outer 
radius 31\arcsec centered on the nearby star, masking out one other contaminating source from the 
background region. 
The uncertainty in the background measurement contributes an additional, unknown source of 
systematic uncertainty in the target flux density near the end of the UVOT light curve (Table 
\ref{tab:data:UVOT}).

We began observing GRB\,161219B with two 1-m telescopes in Sutherland (South Africa), which are 
operated by Las Cumbres Observatory Global Network (LCOGT; \citealt{bbb+13}) on 2016 December 19, 
20:43 UT, at $1.9$~hours since the GRB, in SDSS $r\p$ and $i\p$ filters. Observations 
with 1-m and 2-m LCOGT telescopes (formerly Faulkes Telescopes North and South) both in Hawaii and 
in Siding Springs (Australia) proceeded on a daily basis for four days, followed by a regularly 
increasing spacing until 2017 January 14 (25 days post GRB). Additional optical observations with 
the 2-m Liverpool Telescope (LT; \citealt{ssr+04}) in the same filters culminated on January 23 (35 
days post GRB). Bias and flat-field corrections were applied using the specific pipelines of the 
LCOGT and of the LT. The optical afterglow magnitudes were obtained by PSF-fitting photometry, after 
calibrating the zero-points with four nearby stars with SDSS $r\p$ and $i\p$ magnitudes from the 
AAVSO Photometric All-Sky Survey (APASS) catalog \citep{htt+16}. A systematic error of $0.02$~mag 
due to the zero-point scatter of the calibrating stars was incorporated as an additional source of 
uncertainty in the magnitudes.

We obtained \textit{uBVgri} imaging of GRB\,161219B from 2016 December 22 to 2017 March 21 using 
the Direct CCD Camera on the Swope 1.0 m telescope at Las Campanas Observatory in Chile.  We 
reduced the data using the {\tt photpipe} imaging and photometry package \citep{rsb+05} following 
the methods described in \citet{kfd+17}. We performed aperture photometry using a $4\arcsec$ 
circular aperture on the position of GRB\,161219B. We calibrated the photometry 
in $u\p$-band using Tycho2 standards, and the other filters using PS1 standard-star 
fields observed in the same instrumental configuration and at a similar airmass, after 
transforming the $gri$ magnitudes to the Swope natural system using the corresponding filter 
functions
\citep{scr+15}. 

We observed GRB\,161219B with the Low Resolution Imaging Spectrometer \citep[LRIS;][]{occ+95} on 
the 10-m Keck I telescope on 2017 March 29 in $UBgRIz$ bands. The images were 
bias-subtracted, flat-fielded and cleaned of cosmic rays using 
LPipe\footnote{\url{http://www.astro.caltech.edu/~dperley/programs/lpipe.html}}. The host galaxy is 
well detected in all filters. We performed photometry relative to the PS1, Tycho2, and APASS 
standards using a $4\arcsec$ aperture. 

We obtained 7 epochs of near-IR observations in the $JHK$-bands with the Wide-field Camera (WFCAM; 
\citealt{caado+07}) mounted on the United Kingdom Infrared Telescope (UKIRT) spanning $\approx 2.5$ 
to $\approx270$\,d. We obtained pre-processed images from the WFCAM Science Archive \citep{hcc+08} 
which are corrected for bias, flat-field, and dark current by the Cambridge Astronomical Survey 
Unit\footnote{\url{http://casu.ast.cam.ac.uk/}}. For each epoch and filter, we co-add the images 
and perform astrometry relative to 2MASS using a combination of tasks in 
Starlink\footnote{\url{http://starlink.eao.hawaii.edu/starlink}} and IRAF. We perform aperture 
photometry using standard tasks in IRAF using an aperture of 4.5 times the full-width at 
half-maximum of the seeing measured from stars in the field, in order to capture the combined light 
of the afterglow, supernova, and host galaxy.

We present the results of our optical and NIR photometry, together with a compilation of all other 
optical observations reported in GCN circulars in Table \ref{tab:data:opt}. We include the GROND, 
Gran Telescopio Canarias (GTC), Nordic Optical Telescope (NOT), and Pan-STARRS1 (PS1) light curves 
presented by \cite{cidup+17}, together with the European Southern Observatory (ESO) Very Large 
Telescope (VLT), the 3.6 m Telescopio Nazionale Galileo (TNG), LT, and Keck observations presented 
by \cite{apm+17} in our modeling.  

\begin{deluxetable}{clcc}
\tabletypesize{\scriptsize}
\tablecaption{GRB\,161219B: Log of ALMA observations \label{tab:data:alma}}
\tablehead{
\colhead{$\Delta t$} & \colhead{Frequency} & \colhead{Flux density} & 
\colhead{Uncertainty}\\
\colhead{(d)} & \colhead{(GHz)} & \colhead{($\mu$Jy)} & \colhead{($\mu$Jy)}
}
\startdata
1.30  & 91.5  & 1332 & 32\\
1.30  & 103.5 & 1244 & 31 \\
3.30  & 91.5  & 853  & 34 \\
3.30  & 103.5 & 897  & 33 \\
8.31  & 91.5  & 505  & 15 \\
8.31  & 103.5 & 500  & 19 \\
24.45 & 91.5  & 314  & 41 \\
24.45 & 103.5 & 285  & 43 \\
78.18 & 91.5  & 64   & 14 \\
78.18 & 103.5 & 51   & 20
\enddata
\end{deluxetable}

\subsection{Millimeter: ALMA}
\label{text:data_analysis:millimeter}
We obtained ALMA observations of the afterglow at 1.3\,d after the burst through program 
2016.1.00819.T (PI: Laskar) in Band 3, with two 4\,GHz-wide base-bands centered at 91.5 and 103.5 
GHz, respectively. Prompt data reduction, facilitated through rapid data release by the Joint ALMA 
Observatory (JAO), yielded a strong ($\gtrsim 50\sigma$) detection in our 31~min scheduling block, 
with 16~min on source. We acquired two additional epochs with an identical setup at $\approx3.3$ and 
$\approx8.3$\,d, respectively. Given the brightness of the afterglow and the unusual nature of the 
radio SEDs, we requested and were granted Director's Discretionary Time through program 
2016.A.00015.S for two additional epochs. All observations utilized J0522-3627 as bandpass and flux 
density calibrator, and J0614-2536 as phase calibrator. We imaged the pipeline products using the 
Common Astronomy Software Application \citep[CASA;][]{mws+07}. The afterglow was detected in all 5 
epochs, and the superb sensitivity of ALMA allowed us to measure the flux density in the two 
side-bands separately, yielding the first ALMA light curve of a GRB afterglow. We report the results 
of our ALMA observations in Table 
\ref{tab:data:alma}. 

\begin{deluxetable}{clccc}
\tabletypesize{\scriptsize}
\tablecaption{GRB\,161219B: Log of VLA observations \label{tab:data:vla}}
\tablehead{
\colhead{$\Delta t$} & \colhead{Frequency} & \colhead{Flux density} & 
\colhead{Uncertainty} & \colhead{Det.?}\\
\colhead{(d)} & \colhead{(GHz)} & \colhead{($\mu$Jy)} & \colhead{($\mu$Jy)}
}
\startdata
0.51  &  19.0   & 278.1  &  28.6   &  1\\
0.51  &  21.0   & 156.2  &  41.7   &  1\\
0.51  &  23.0   & 184.7  &  36.1   &  1\\
0.51  &  25.0   & 242.4  &  47.5   &  1\\
\ldots & \ldots & \ldots & \ldots & \ldots
\enddata
\tablecomments{The last column indicates a detection (1) or non-detection (0); in the latter case, 
the flux density is a 3$\sigma$ upper limit and the uncertainty refers to the mean map noise. 
We include the GMRT detection reported by \cite{gcn20344}. 
This is a sample of the full table available on-line.}
\end{deluxetable}

\subsection{Centimeter: VLA}\label{text:data_analysis:radio}
We observed the afterglow using the Karl G. Jansky Very Large Array (VLA) starting 0.5\,d after 
the burst through program 15A-235 (PI: Berger). We detected and tracked the flux density of the 
afterglow from 1.2\,GHz to 37\,GHz over 9 epochs until $\approx159$\,d after the burst. We used 
3C48 as the flux and bandpass calibrator and J0608-2220 as gain calibrator. Some of the 
high-frequency observations ($\gtrsim15$\,GHz) suffered from residual phase errors, which we 
remedied using phase-only self-calibration in epochs with sufficient signal-to-noise. We 
carried out data reduction using CASA, and list the results of our VLA observations in Table 
\ref{tab:data:vla}.

\section{Basic Considerations}
\label{text:basic_considerations}
We now interpret the X-ray, optical, and radio observations in the standard synchrotron framework 
\citep{spn98,gs02}, in which the observed spectra are characterized by power law segments 
connected at characteristic break frequencies: the self-absorption frequency (\nua), the 
characteristic synchrotron frequency (\numax), and the cooling frequency (\nuc). The electrons 
responsible for the observed radiation are assumed to form a power law distribution in energy with 
index, $p$.

\begin{deluxetable}{lr}
 \tabletypesize{\footnotesize}
 \tablecolumns{2}
 \tablecaption{XRT Light Curve Fit\label{tab:xrtlcfit}}
 \tablehead{   
   \colhead{Parameter} &
   \colhead{Value} 
   }
 \startdata 
 $t_{\rm b,1}$ (d) & $(6.0\pm2.3)\times10^{-2}$ \\
 $t_{\rm b,2}$ (d) & $(1.3\pm0.4)\times10^1$    \\
 $F_{\rm b}$   (mJy) & $(9.7\pm2.6)\times10^{-3}$ \\
 $\alpha_{\rm X,1}$    & $-0.37\pm0.09$  \\
 $\alpha_{\rm X,2}$    & $-0.82\pm0.02$  \\
 $\alpha_{\rm X,3}$    & $-1.32\pm0.08$  \\
 $\chi^2$/dof          & 126/75
 \enddata 
\end{deluxetable}

\begin{figure}
 \includegraphics[width=\columnwidth]{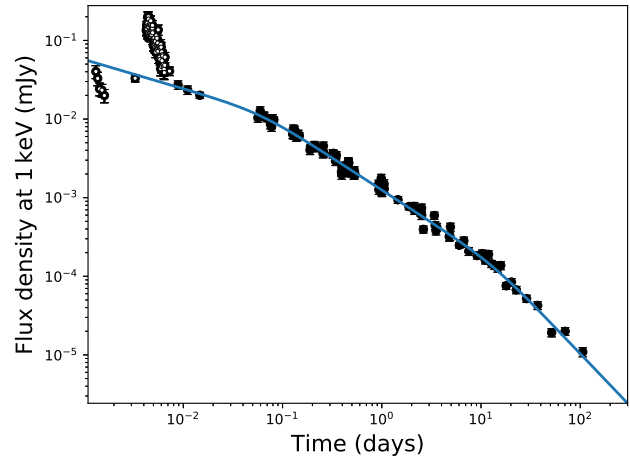}
 \caption{\Swift\ XRT light curve of GRB\,161219B at 1\,keV (black points), together with a 
twice-broken power law fit (blue; equation \ref{eq:dbpl}). Data before $\approx8.8\times10^{-3}$~d 
are dominated by flaring activity and are not included in the fit.}
\label{fig:XRT-lc}
\end{figure}

\begin{deluxetable*}{lccccc}
 \tabletypesize{\footnotesize}
 \tablecolumns{6}
 \tablecaption{UV/Optical Light Curve Fit\label{tab:uvlcfit}}
 \tablehead{   
   \colhead{Band} & \colhead{$t_{\rm b}$}  & \colhead{$F_{\rm b}$} &
                \colhead{$\alpha_1$}   & \colhead{$\alpha_2$} & \colhead{$\chi^2$/dof}\\
   \colhead{} & \colhead{(d)}          & \colhead{(mJy)} &
                \colhead{}             & \colhead{}      & \colhead{}
   }
 \startdata 
 $r\p$    &  $0.15^{\dag}$   & $0.31^{\ddag}$ & $-0.29\pm0.03$ & $-0.68\pm0.01$       &$6.3/14$ \\
 \textit{uvw1}   &  $0.21\pm0.04$   & $0.13\pm0.02$  & $-0.33\pm0.04$ & $-0.95\pm0.05$&$7.9/10$ \\
 \textit{uvw2}   &  $0.12\pm0.03$   & $0.14\pm0.02$  & $-0.20\pm0.05$ & $-0.79\pm0.05$&$4.7/11$ \\
 \textit{white}  &  $0.07\pm0.02$   & $0.29\pm0.03$  & $-0.17\pm0.02$ & $-0.70\pm0.03$&$5.6/11$ \\
 Avg$^{\S}$      &  $0.094\pm0.013$ &    \ldots      & $-0.22\pm0.02$ & $-0.76\pm0.02$&\ldots
 \enddata 
 \tablecomments{${}^{\dag}$ Fixed. ${}^{\ddag}$ This parameter is strongly correlated with $t_{\rm 
b}$. ${}^{\S}$ Weighted average of the UV fits.}
\end{deluxetable*}

\subsection{X-rays and optical -- and location of $\nuc$}  
\label{text:basic_x}
\begin{figure}
 \includegraphics[width=\columnwidth]{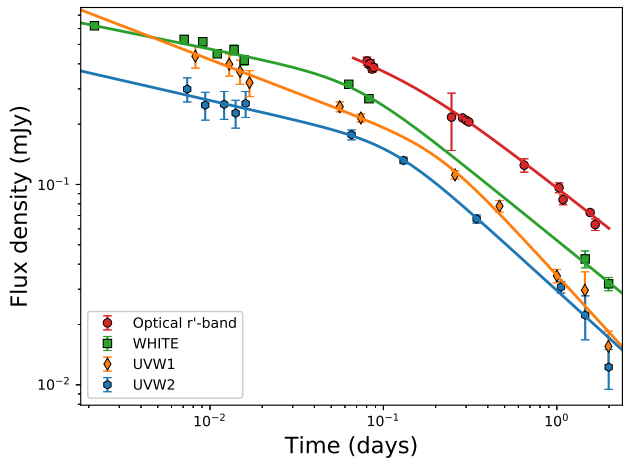}
 \caption{Optical $r\p$-band (red circles), and \Swift/UVOT \textit{uwh}- 
\textit{uw1}- and \textit{uw2}-band light curves of GRB\,161219B (green squares, orange diamonds, 
and blue hexagons respectively) before 2.4\,d, together with broken power law fits (lines; equation 
\ref{eq:bpl}).}
\label{fig:uvotlc}
\end{figure}

The XRT light curve exhibits a flare at $4.2\times10^{-3}$ to $6.2\times10^{-3}$\,d  (Figure 
\ref{fig:XRT-lc}). Such flares in early X-ray light curves are relatively common and may arise from 
residual central engine activity or the collisions of relativistic shells 
\citep{bfc+07,mgc+10,lbm+17}, and we do not include this portion of the light curve in our 
multi-wavelength analysis. 
The PC-mode light curve after $8.8\times10^{-3}$\,d can be fit with a power law with two breaks, 
described by
\begin{align}
\label{eq:dbpl}
 F_{\nu}(t) &= F_{\rm b} \left[
 \frac{1}{2} \left(\frac{t}{t_{\rm b,1}}\right)^{-y_1\alpha_{\rm X,1}} + 
 \frac{1}{2} \left(\frac{t}{t_{\rm b,1}}\right)^{-y_1\alpha_{\rm X,2}}
 \right]^{-1/y_1}\nonumber \\
 &\times \left[1+\left(\frac{t}{t_{\rm b,2}}\right)^{y_2 (\alpha_{\rm 
X,2}-\alpha_{\rm X,3})}\right]^{-1/y_2},
\end{align}
breaking\footnote{We fix the smoothness parameters here and in equation \ref{eq:bpl} at 
$y_1=y_2=y=5$, and use the convention $F_{\nu}\propto t^{\alpha}\nu^{\beta}$.} first from 
$\alpha_{\rm X,1} = -0.37\pm0.09$ to $\alpha_{\rm X,2} = -0.82\pm0.02$ at $\approx0.06$\,d and then 
to $\alpha_{\rm X,3} = -1.32\pm0.08$ at $\approx13$\,d 
(Table \ref{tab:xrtlcfit}).
We also fit the \Swift/UVOT light curves in three well-sampled bands 
at $\lesssim2.4$\,d 
(\textit{uwh, uvw1, uvw2}) with a broken power law model,
\begin{align}
\label{eq:bpl}
 F_{\nu}(t) &= F_{\rm b} \left[
 \frac{1}{2} \left(\frac{t}{t_{\rm b}}\right)^{-y\alpha_1} + 
 \frac{1}{2} \left(\frac{t}{t_{\rm b}}\right)^{-y\alpha_2} 
 \right]^{-1/y}, 
\end{align}
and provide a fit to the $r\p$ in this same period for reference in Table 
\ref{tab:uvlcfit}. The \Swift/UVOT light curves in these three bands exhibit a shallow decline 
with $\alpha_{\rm UV,avg,1}=-0.22\pm0.02$, followed by a steepening with $\alpha_{\rm 
UV,avg,2}=-0.76\pm0.02$ at $t_{\rm b,UV,avg}=(9.4\pm1.3)\times10^{-2}$\,d (weighted 
averages; Figure \ref{fig:uvotlc}).
The prominent re-brightening in the optical light curves after 
$\approx2.4$\,d is associated with an emerging supernova 
(SN2016jca\footnote{\url{https://wis-tns.weizmann.ac.il/object/2016jca/}}), and the subsequent 
flattening at $\gtrsim 50$\,d can be attributed to contamination from an underlying host galaxy, 
the latter also detectable as an extension in the optical images \citep{cidup+17}.

The break time of $t_{\rm b,avg}\approx 9\times10^{-2}$\,d in the UV light curves is consistent 
with the time of the first break in the X-ray light curve at $t_{\rm 
b,X,1}\approx6\times10^{-2}$\,d. Such an achromatic break is unusual in GRB afterglows 
and in the standard synchrotron model can only be explained as (i) onset of the afterglow 
\citep{sp99,kz07}, (ii) viewing angle effects \citep{gpkw02} and (iii) jet breaks 
\citep{rho99,sph99}. Of these, the first two are preceded by rising light curves, and the third 
results in a steeply decaying light curve ($\alpha \approx-p$), neither of which is the case here.
We investigate the origin of this feature in Section \ref{text:model}.

We now interpret the observed light curves at $\gtrsim0.1$\,d in the synchrotron 
framework, beginning with the location of the cooling frequency, $\nuc$. We investigate four 
possibilities, $\nuc>\nux$, $\nuc<\nuopt$, $\nuopt<\nuc<\nux$, and $\nuc\approx\nux$. In the first 
scenario ($\nuc>\nux$), we note that the observed X-ray spectral index $\beta_{\rm 
X}=-0.86\pm0.03=(1-p)/2$ implies $p=2.72\pm0.06$, which yields $\alpha_{\rm X}=-1.29\pm0.04$ (ISM) 
or $\alpha_{\rm X}=-1.79\pm0.04$ (wind). However, the measured decline rate is $\alpha_{\rm 
X}=-0.82\pm0.02$. Thus $\nuc>\nux$ between $0.1$\,d and $\approx13$\,d is ruled out. 

If $\nuc<\nuopt$, then $\beta_{\rm X}=-0.86\pm0.03$ implies $p=1.76\pm0.06$. In this case, we 
expect the X-ray and optical to lie on the same spectral slope, with $\beta_{\rm opt}\approx-0.86$. 
We find that the host-subtracted GROND $grizJHK$ photometry at $\approx1.5$\,d can be fit with a 
single power law, $\beta_{\rm opt} = -0.5\pm0.1$. This is shallower than expected, and cannot be 
explained if $\nuc<\nuopt$ (extinction in the host galaxy would further steepen the optical 
spectral index). Thus $\nuc<\nuopt$ is ruled out.

If $\nuopt<\nuc<\nux$, then the observed X-ray spectral index once again implies $p=1.76\pm0.06$. 
The expected optical spectral index is $(1-p)/2=-0.38\pm0.03$, and the steeper  could be explained 
as arising from extinction. For $p<2$, the expected light curves depend upon assumptions regarding 
the normalization of the total 
energy of accelerated electrons relative to the energy of the forward shock \citep{bha01,dc01}. If 
we assume the electron spectrum cuts off above a maximal electron Lorentz 
factor\footnote{One possibility for a high energy cutoff in the electron spectrum ($\gamma_{\rm M}$) 
is afforded by balancing the electron acceleration timescale and the dynamical timescale, 
$\gamma_{\rm M}\sim\frac{\Gamma t q_{\rm e} B}{m_{\rm p}c}$, where $\Gamma$ is the shock Lorentz 
factor, $q_{\rm e}$ is the fundamental electron charge, $B$ is the post-shock magnetic field, 
$m_{\rm p}$ is the proton mass, and $c$ is the speed of light.}, that a constant 
fraction of the shock energy is given to the electrons, and that the total electron energy must be 
finite \citep{glz+13}, then we would have 
$\alpha_{\rm X}=-(3p+10)/16=-0.96\pm 0.01$, 
inconsistent with the observed value of $\alpha_{\rm X}=-0.82\pm0.02$, as well as  
$\alpha_{\rm opt}=-(3p+2)/16=-0.46\pm0.01$, inconsistent with the observed value of $\alpha_{\rm 
opt}=-0.76\pm0.02$. On the other hand, if we assume the closure relations of \cite{gs02} apply 
for $p<2$, then we expect $\alpha_{\rm X} = (2-3p)/4=-0.82\pm0.04$, and $\alpha_{\rm 
opt}=3(1-p)/4=-0.57\pm0.04$ (ISM) or $\alpha_{\rm opt}=(1-3p)/4=-1.07\pm0.04$ (wind). Whereas the 
observed X-ray decline rate matches this prediction, the optical decline rate does not. Thus, the 
$p=1.76\pm0.06$ model is disfavored. 

We therefore investigate the last possibility, $\nuc\approx\nux$ at $\gtrsim0.1$\,d. Anchoring 
this model to the optical spectral index, $\beta_{\rm opt}=-0.5\pm0.1$, we infer 
$p=1-2\beta=2.0\pm0.2$ for the spectral ordering, $\numax<\nuopt<\nuc\approx\nux$. 
The observed UV spectral index, $\beta_{\rm UV} =-1.2\pm0.2$, is steeper than the 
optical, and indicates extinction in the host galaxy.
The observed optical decline rate of $\alpha_{\rm opt}=-0.76\pm0.02$ 
is not consistent with the predicted value of $\alpha_{\rm opt}=(1-3p)/4=-1.25\pm0.15$ for the wind 
case, but agrees with the expected value of $\alpha_{\rm opt}=3(1-p)/4=-0.75\pm0.15$ for the ISM 
case. 

The spectral index between the NIR $K$-band and the X-rays is $\beta_{\rm ox}=-0.68\pm0.02$; this 
is steeper than $\beta_{\rm opt}$ and consistent with $\nuopt<\nuc$ at this time.
If $\nuc\approx\nux$, we expect the X-ray spectral index to be intermediate between 
$(1-p)/2\approx-0.5$ and $-p/2\approx-1$, which is satisfied by the measured index, $\beta_{\rm 
X}=-0.86\pm0.03$. If we place $\nuc\approx1$\,keV and use the rounded shape of the cooling break as 
derived by \cite{gs02}, we heuristically calculate a spectral index across the ends of the \Swift\ 
X-ray band between 0.3\,keV and 10\,keV of $\beta\approx-0.82$, consistent with the observed index. 
Finally, the observed decline rate, $\alpha_{\rm X} = -0.82\pm0.02$ is also intermediate between 
$3(1-p)/4\approx-0.75$ and $(2-3p)/4\approx-1$, further indicating $\nuc\approx\nux$.

Thus the optical and X-ray observations indicate an ISM environment with 
$\numax<\nuopt<\nuc\approx\nux$ at $\gtrsim0.1$\,d and moderate extinction in the host galaxy. 
Whereas the final X-ray decline rate of $\alpha_{\rm X,3}\approx-1.3$ appears too shallow for a jet 
break, where we expect $\alpha_{\rm X}\approx-p\approx-2$ \citep{rho99,sph99}, we note that the 
break time and post-break decay rate are degenerate with the break smoothness. A later break time 
in a physically-motivated model may be consistent with the steeper post-break decay expected. Our 
subsequent multi-wavelength analysis, described in Section \ref{text:model}, confirms this 
interpretation.

%
%
%
 
\begin{figure*}
\begin{tabular}{ccc}
 \centering
 \includegraphics[width=0.31\textwidth]{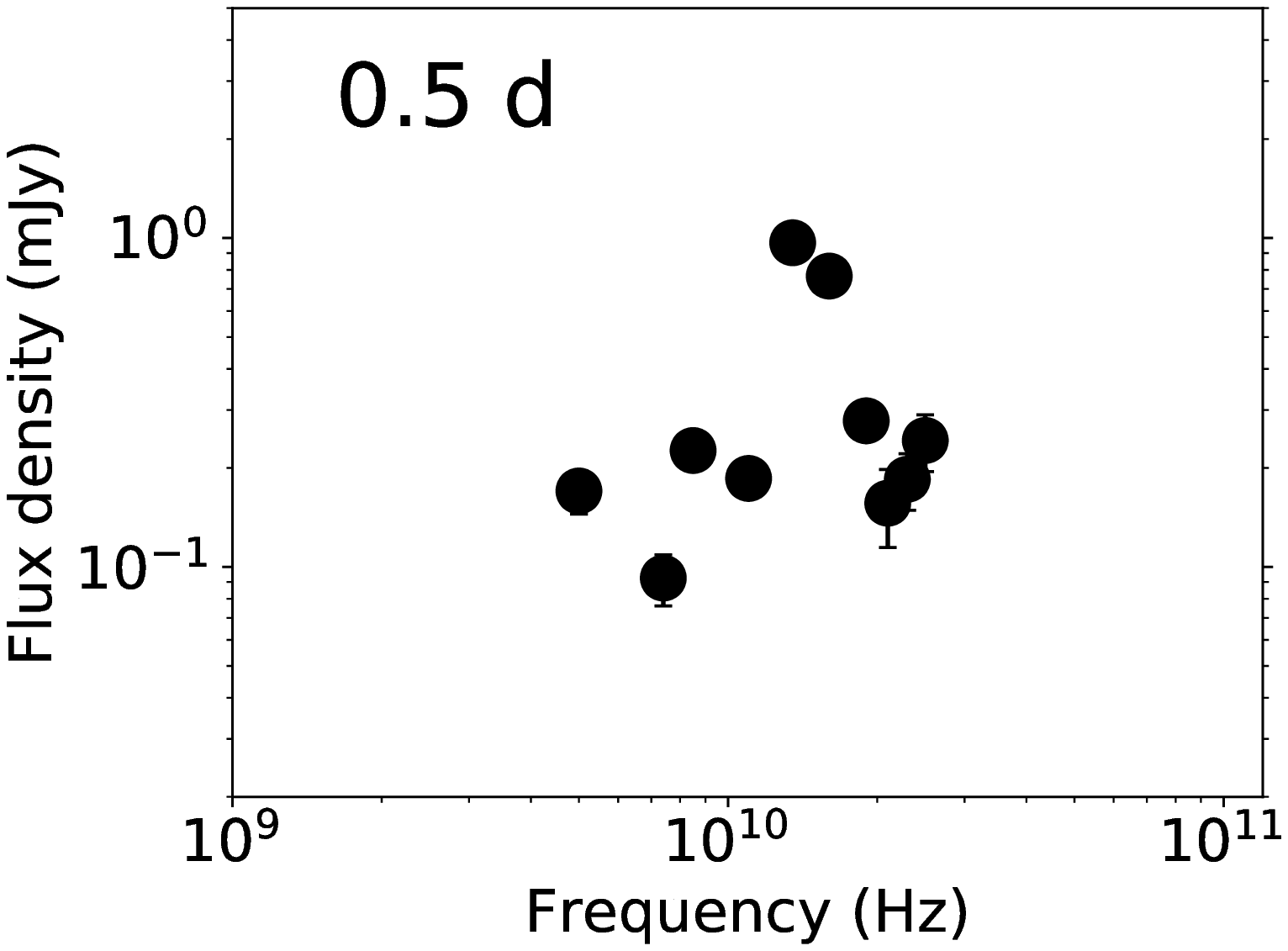} &
 \includegraphics[width=0.31\textwidth]{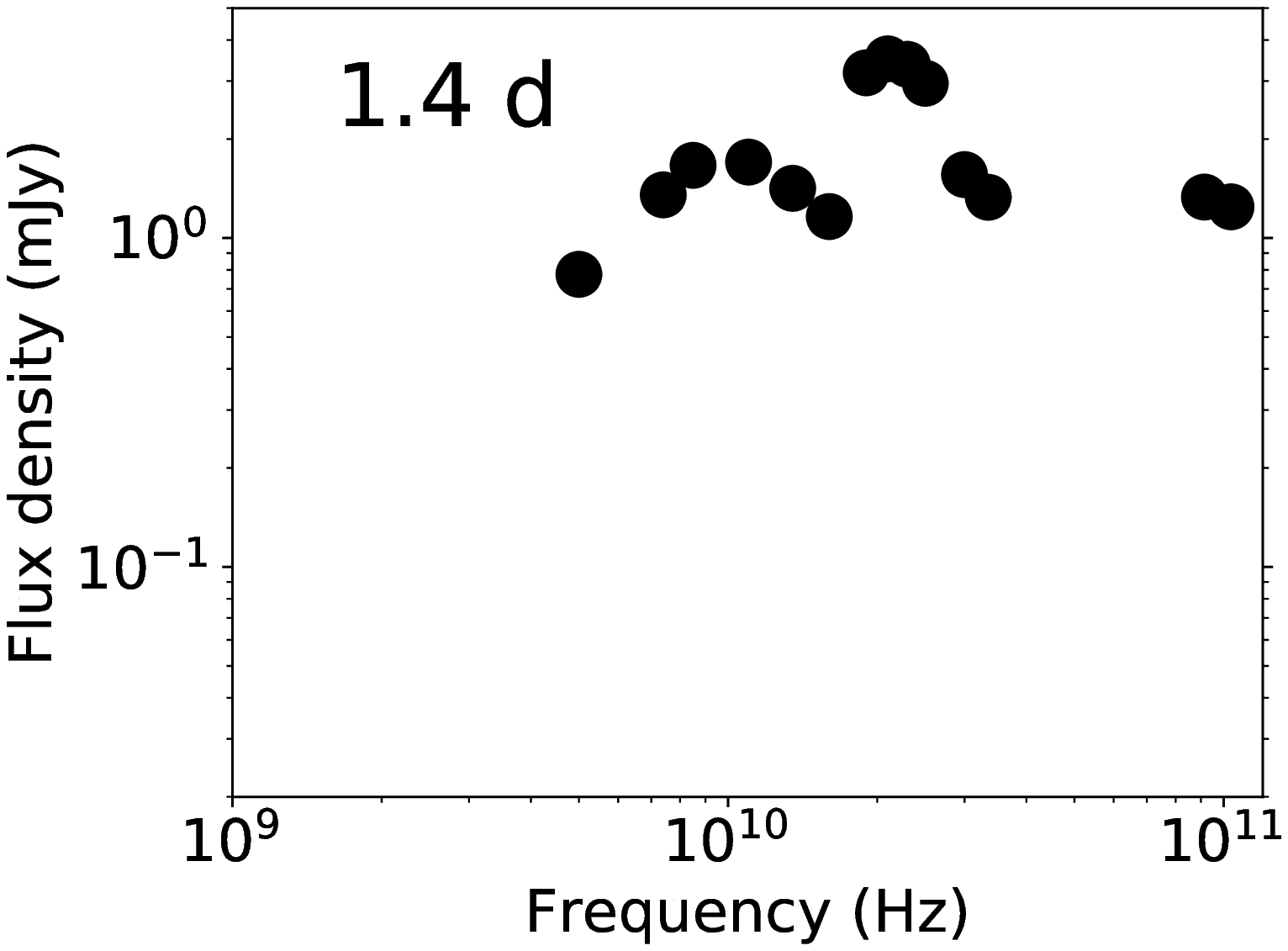} &
 \includegraphics[width=0.31\textwidth]{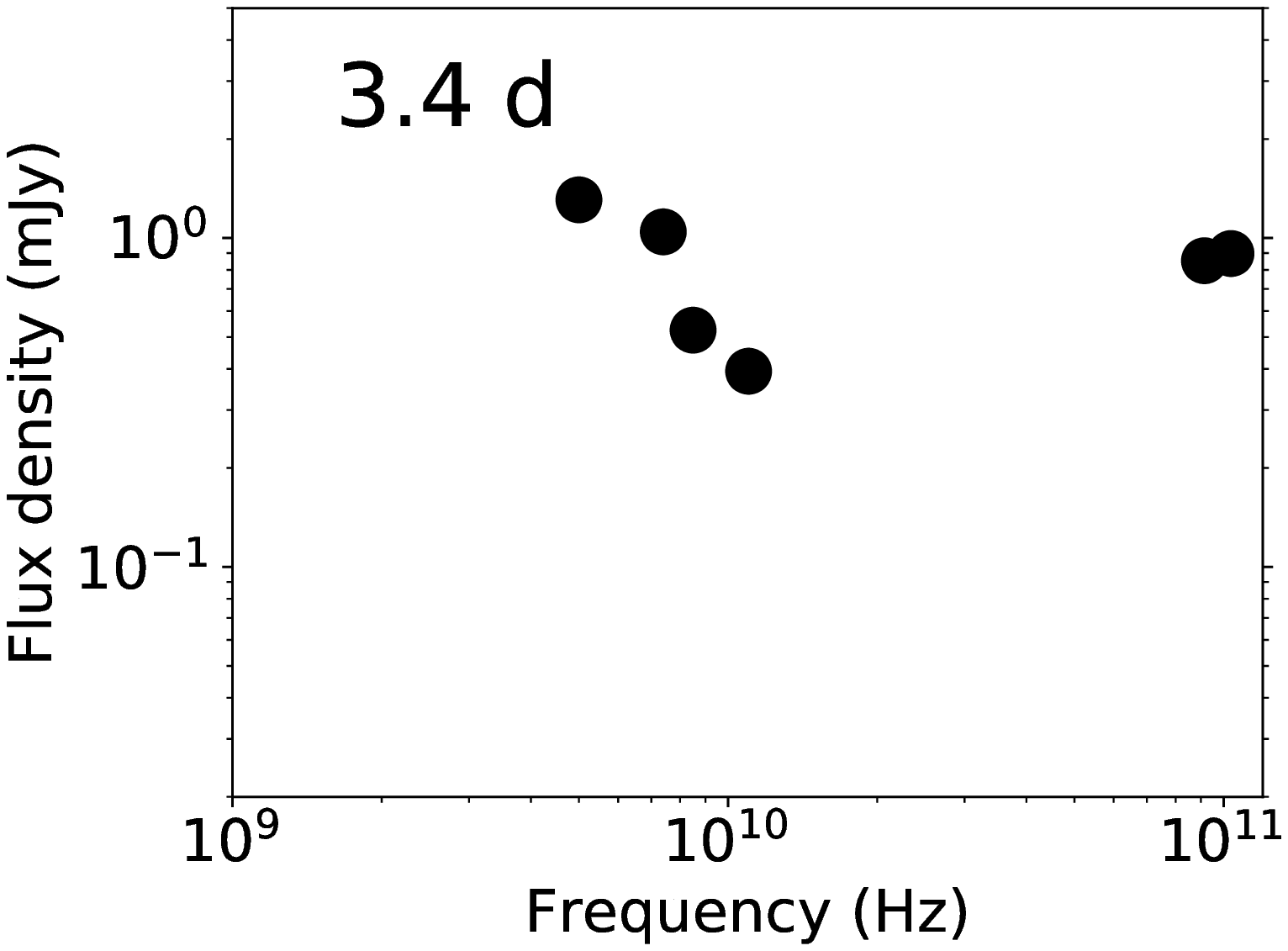} \\
 \includegraphics[width=0.31\textwidth]{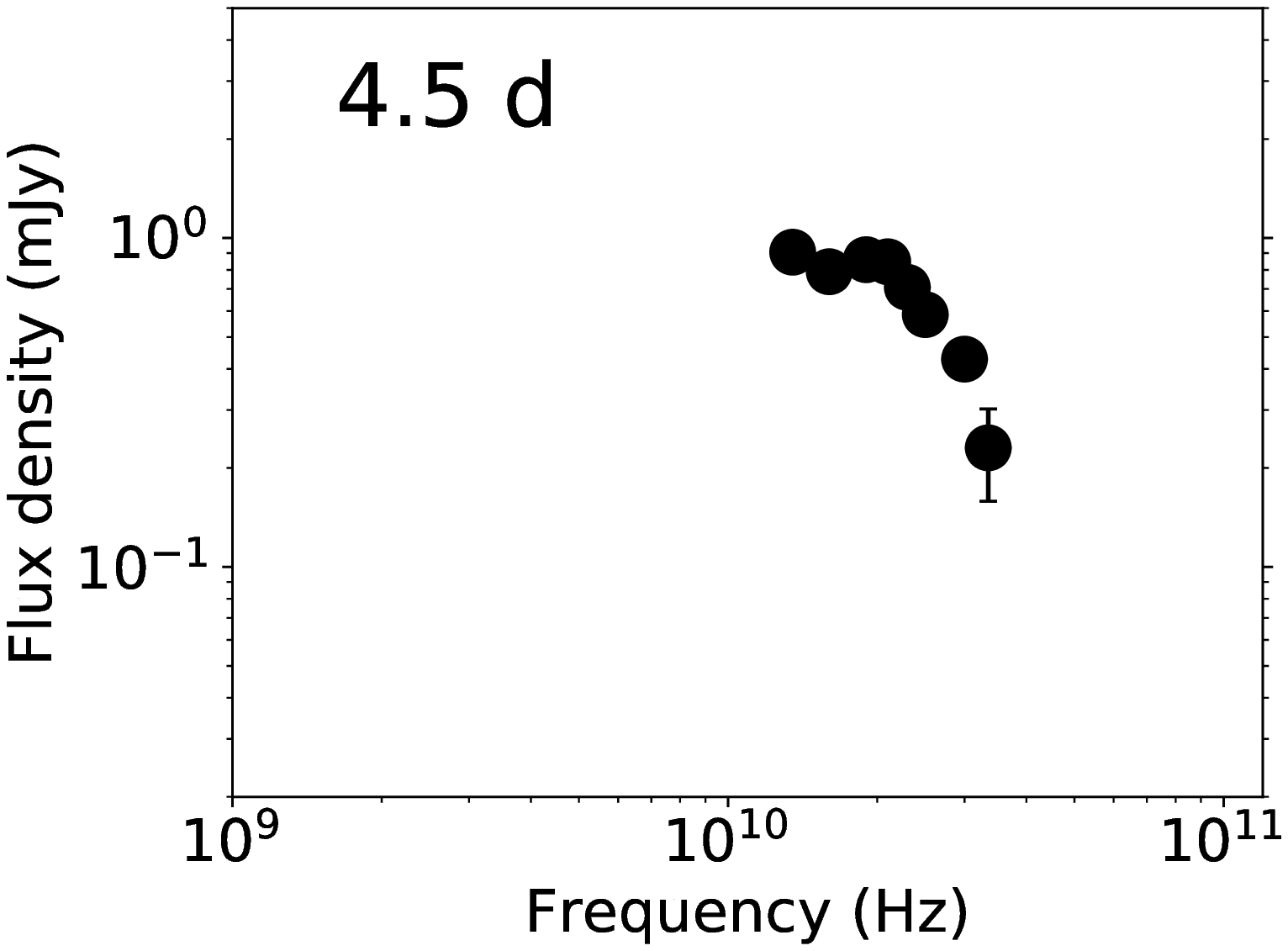} &
 \includegraphics[width=0.31\textwidth]{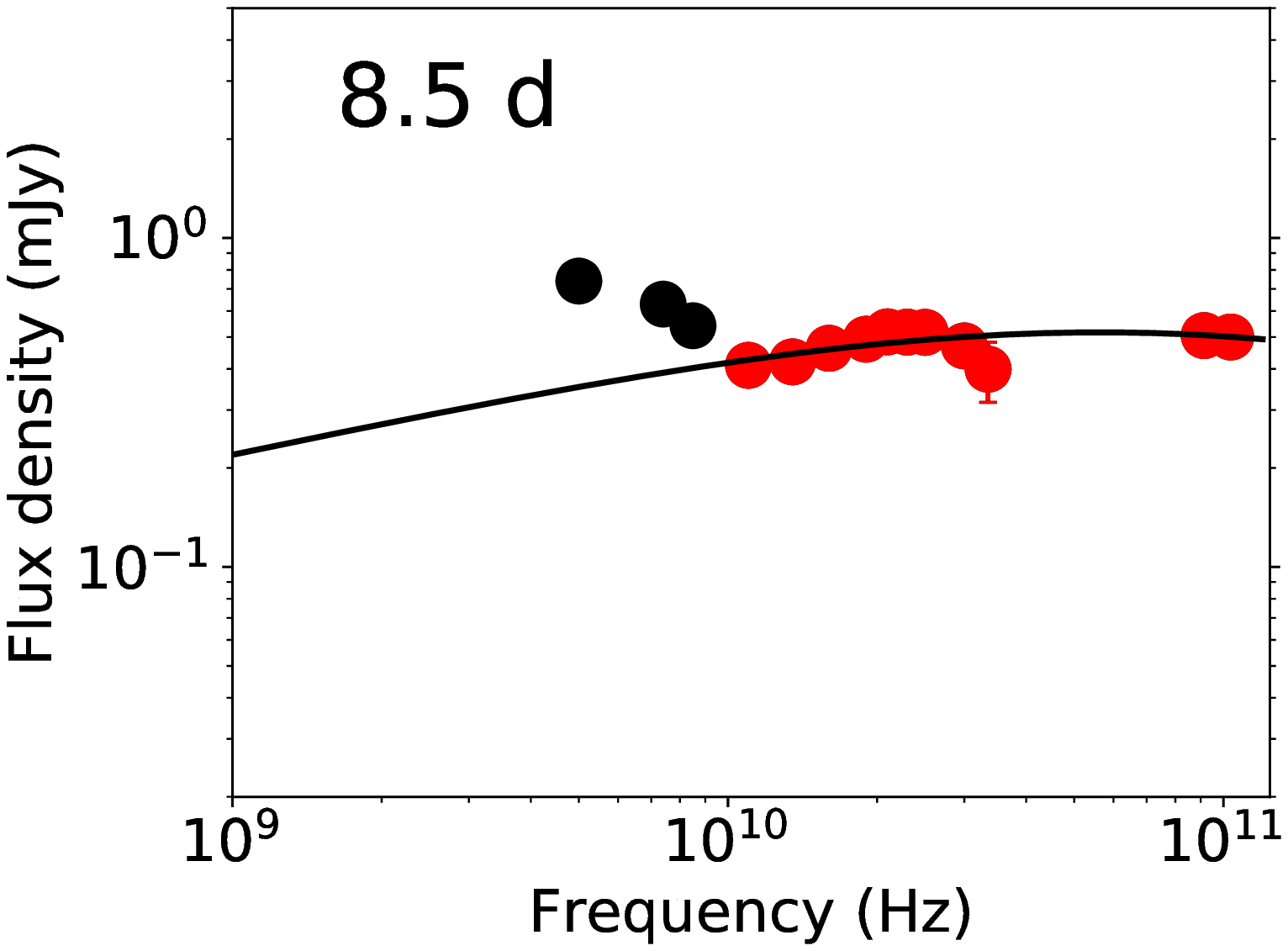} &
 \includegraphics[width=0.31\textwidth]{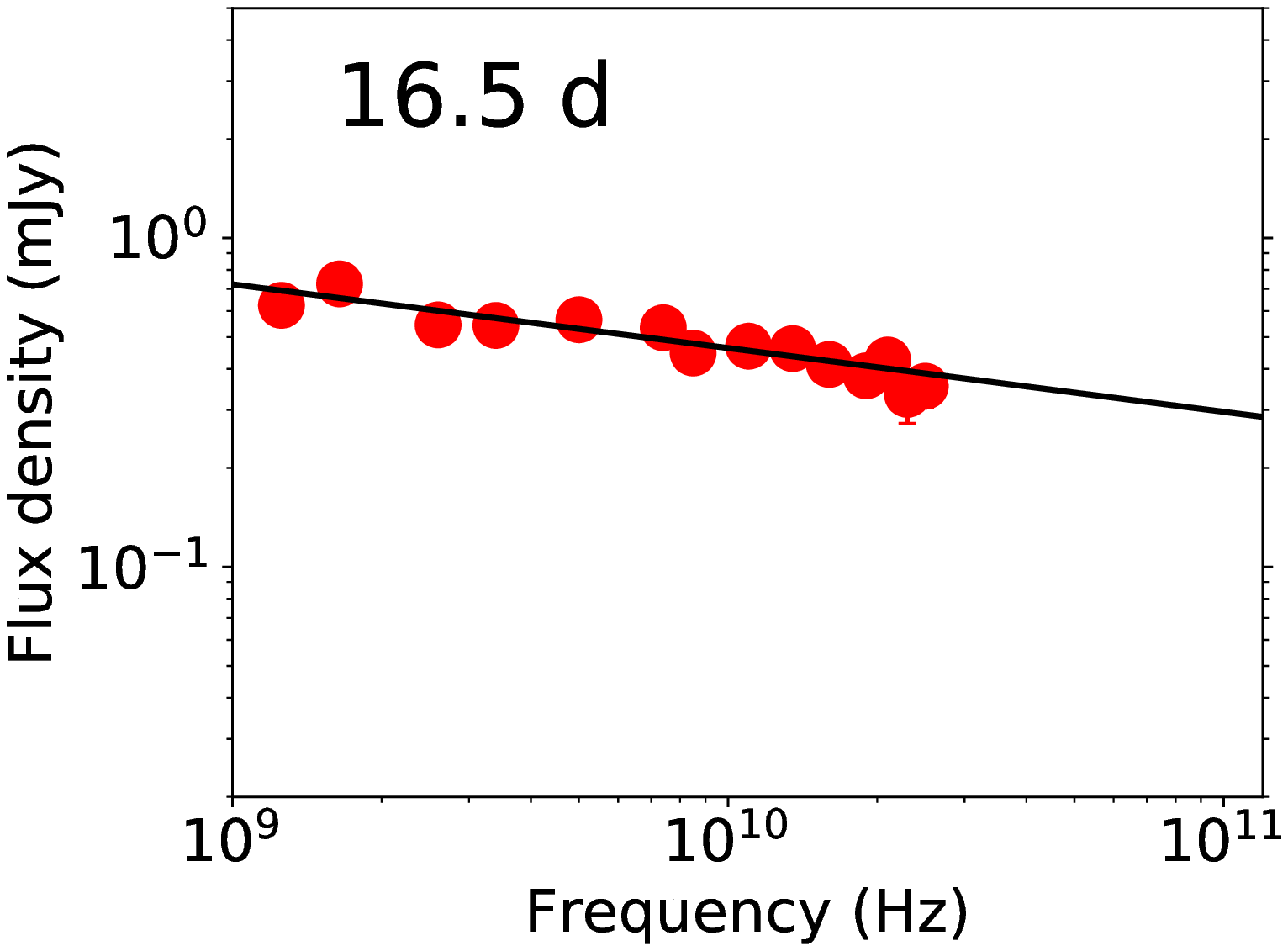} \\
 \includegraphics[width=0.31\textwidth]{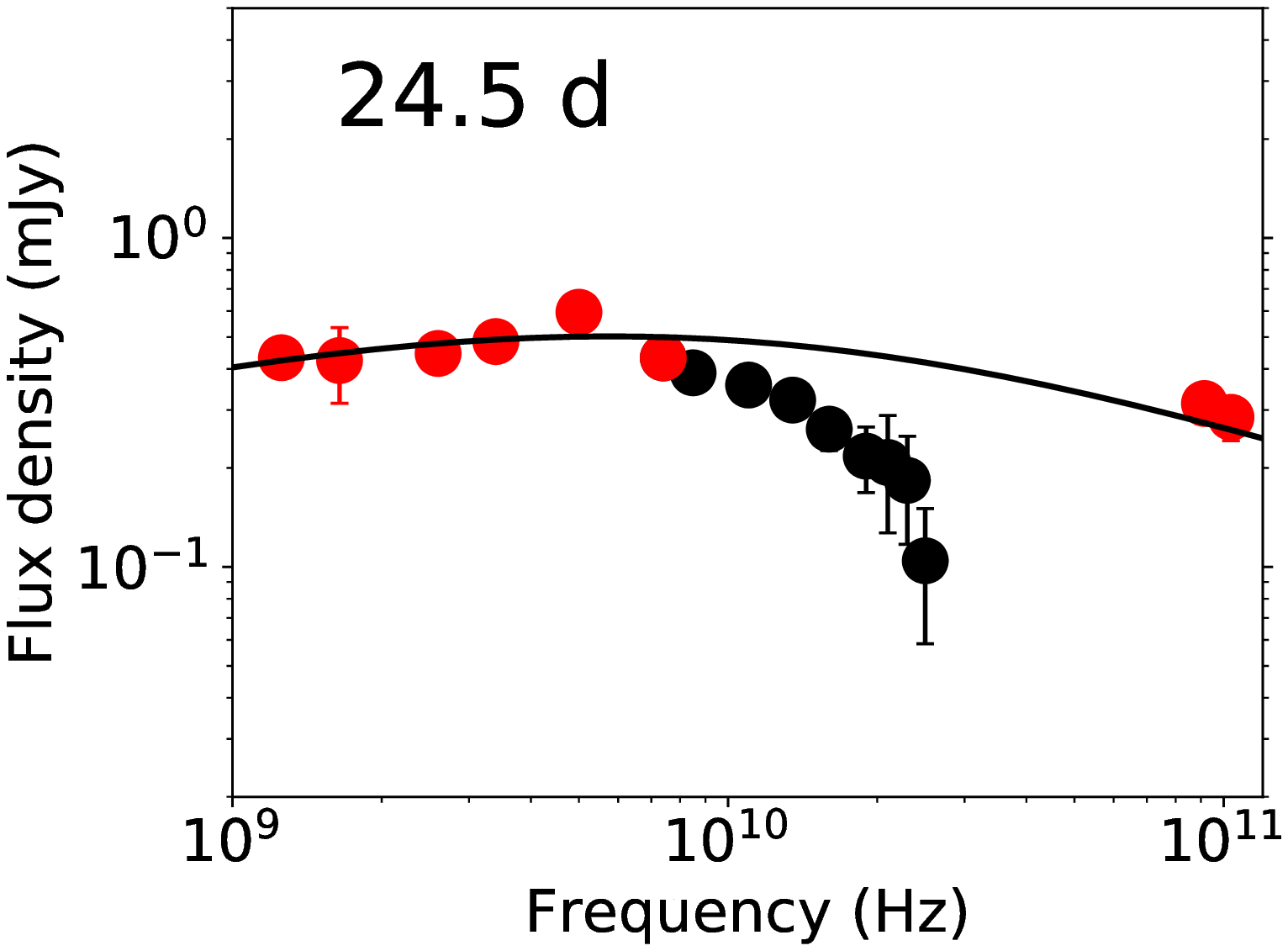} &
 \includegraphics[width=0.31\textwidth]{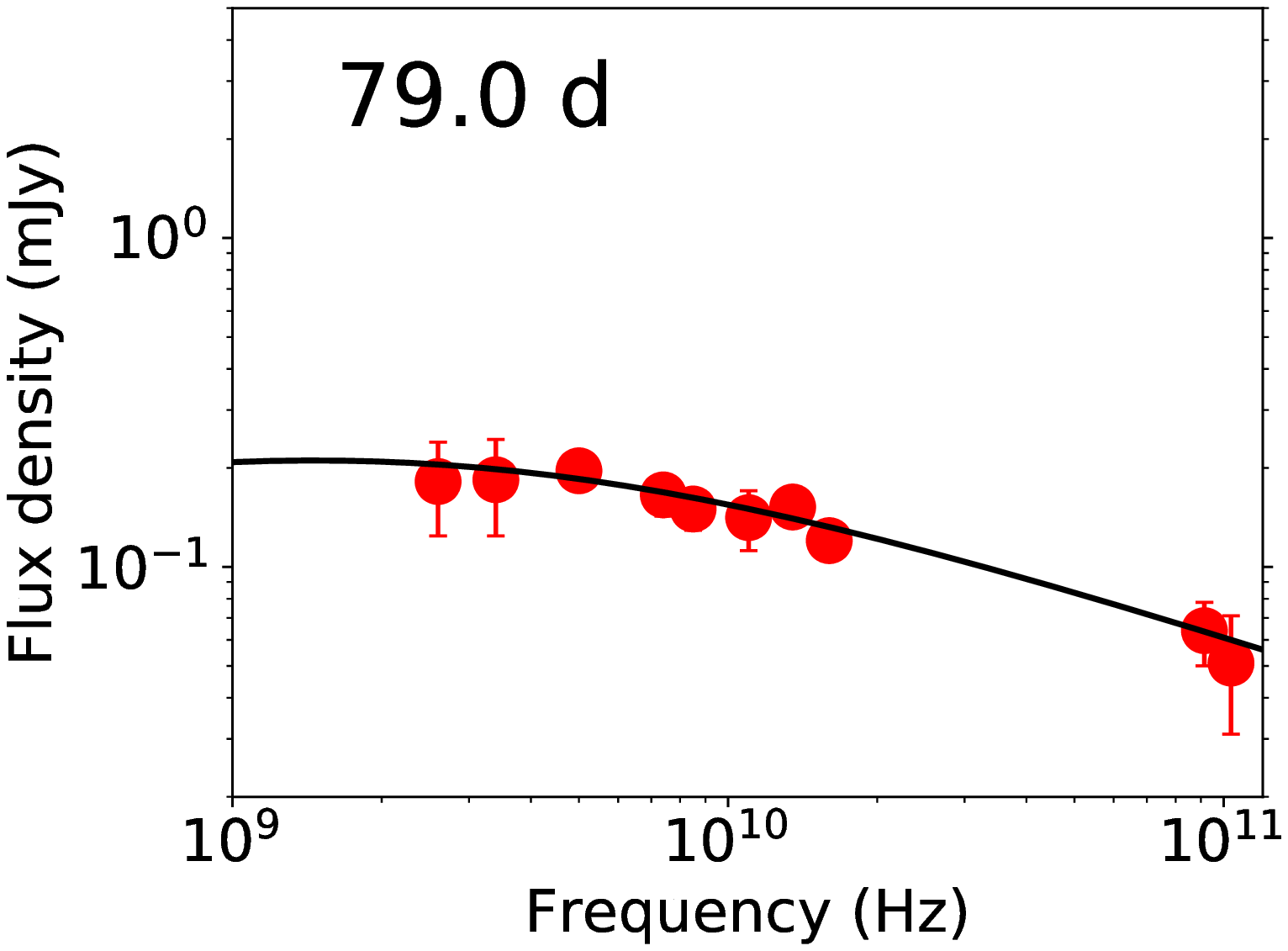} &
 \includegraphics[width=0.31\textwidth]{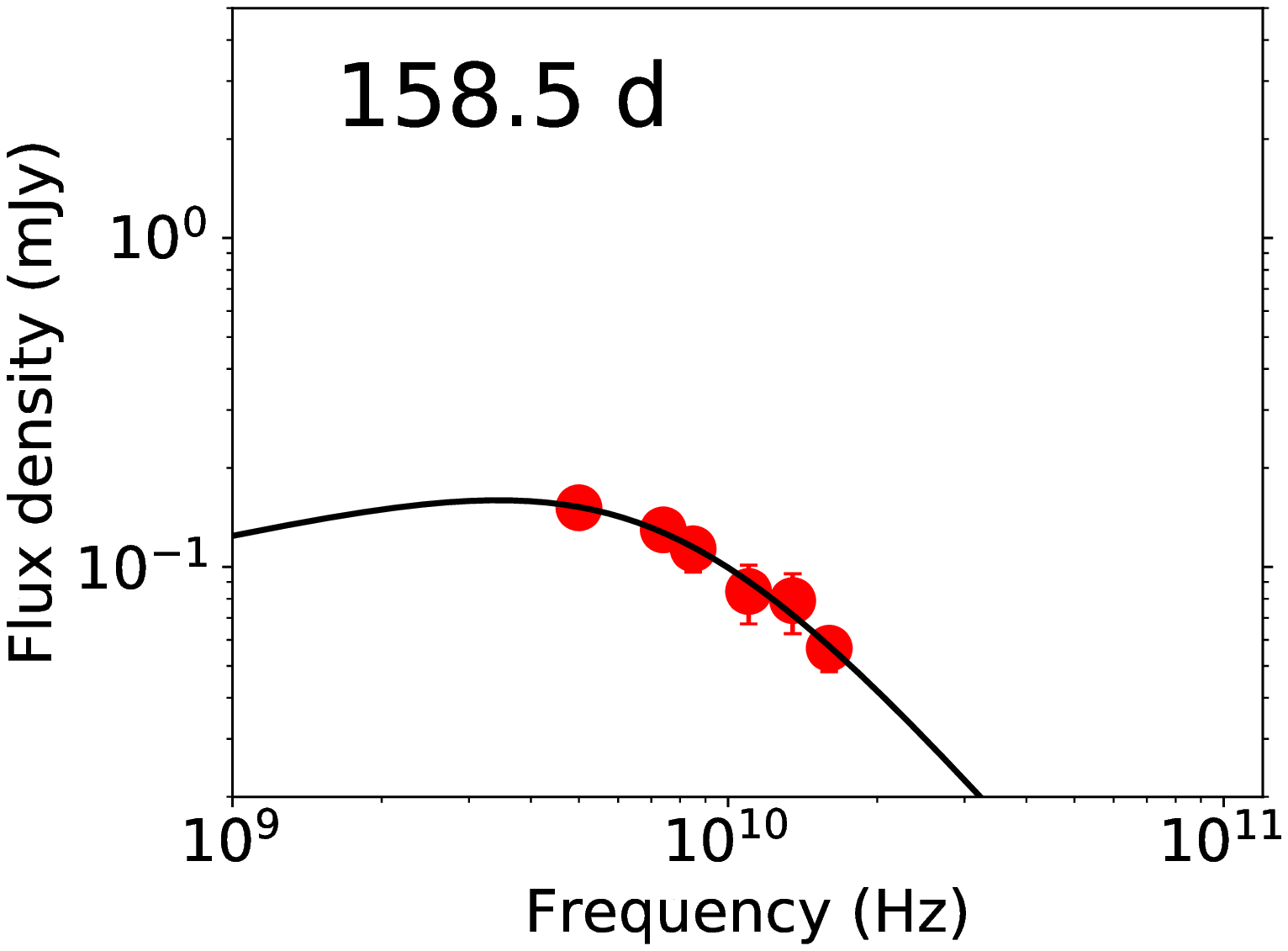} 
\end{tabular}
\caption{
Multi-frequency cm-band (VLA) and mm-band (ALMA) spectral energy distributions of the afterglow of 
161219B from 0.5\,d to $\approx159$\,d, together with power law (16.5\,d) and broken power 
law (8.5\,d, 24.5\,d, 79\,d, and 158.5\,d) fits (solid) to some of the observations (red points; 
see Section \ref{text:basic_radio} for details). The radio SEDs exhibit unexpected variability in 
the cm-band (see also Figure~\ref{fig:var}). }
\label{fig:radiodata}
\end{figure*}

\subsection{Radio -- unexpected variability and location of $\numax$}
\label{text:basic_radio}
The cm-band data of GRB\,161219B are truly remarkable, exhibiting spectro-temporal variability on 
time and frequency scales shorter than ever observed for a GRB radio afterglow. The SEDs at 
0.5, 1.4, 3.4, 4.5, and 8.5\,d exhibit spectral features with $\delta \nu/\nu\lesssim1$, too narrow 
for production via standard synchrotron emission (Figure \ref{fig:radiodata}). The observations at 
24.5\,d appear to exhibit a deficit at $\approx 5$--$30$\,GHz. Only the epochs at 
$\approx16.5$\,d, $\approx79$\,d, and $\approx159$\,d exhibit simple SEDs that can be understood as 
power laws, or combinations 
thereof. 

These unexpected spectral features appear to be due to a combination of factors. At frequencies 
above $\approx 10$\,GHz, atmospheric phase decoherence can reduce the measured flux density. 
Whereas most of the data were obtained in A-configuration and a fast cycle time of 2--4
minutes was employed, the phase referencing is not perfect and there are residual errors on 
all baselines. To check this, we performed phase-only self-calibration on the afterglow itself in 
epochs where the target was detected at a signal-to-noise of $\gtrsim3$ per solution interval, and 
found the process to yield an increased flux density by 10--30\%, and a reduced map noise in the 
vicinity of the afterglow. However, self-calibration is not possible when the target is fainter 
than $\approx$ 0.5--1\,mJy, and even where feasible, this process does not completely explain the 
observed spectral features. We also find the observed rapid variability at cm-bands to be robust to 
self-calibration.

\begin{figure*}
\begin{tabular}{cc}
 \centering
 \includegraphics[width=0.49\textwidth]{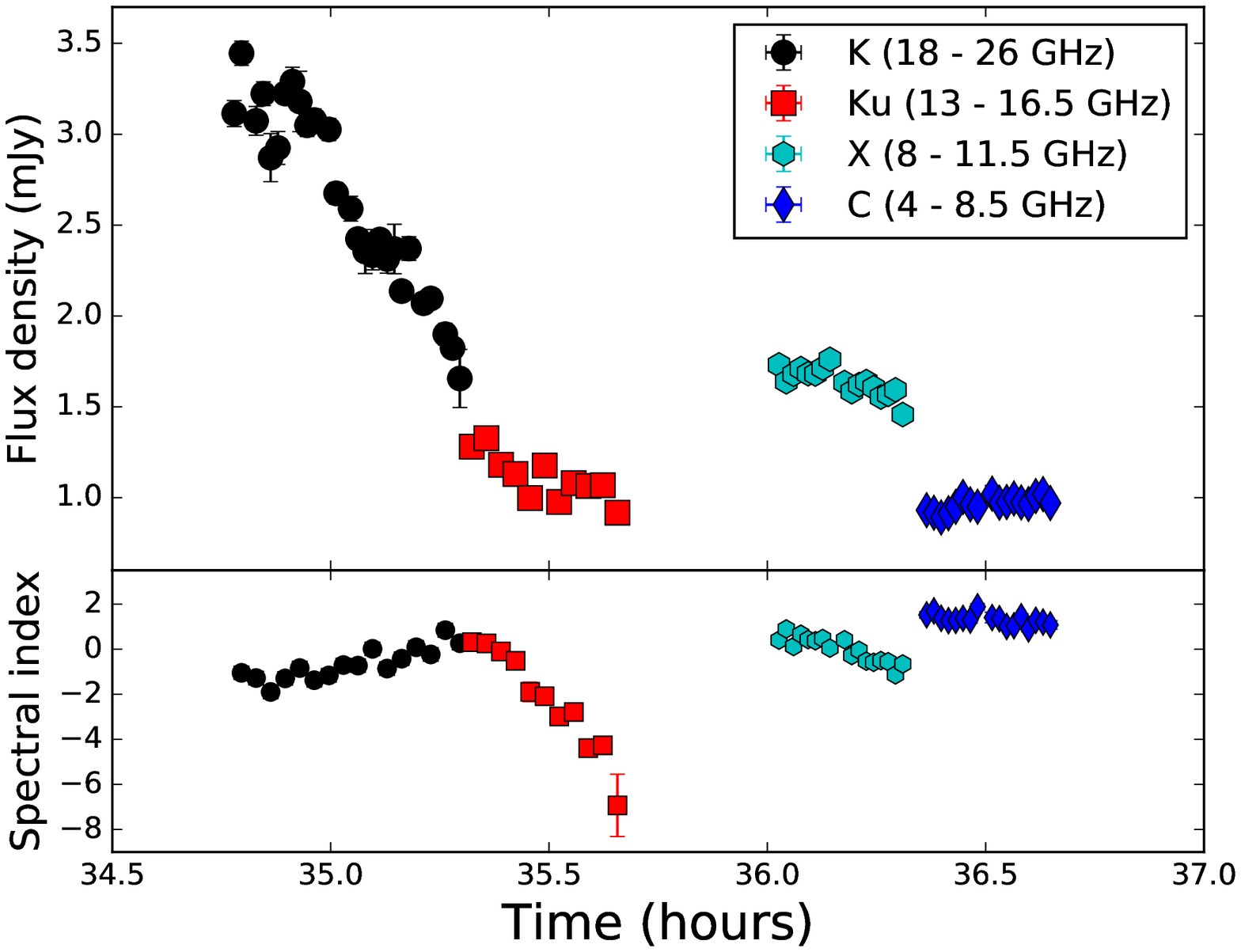} &
 \includegraphics[width=0.49\textwidth]{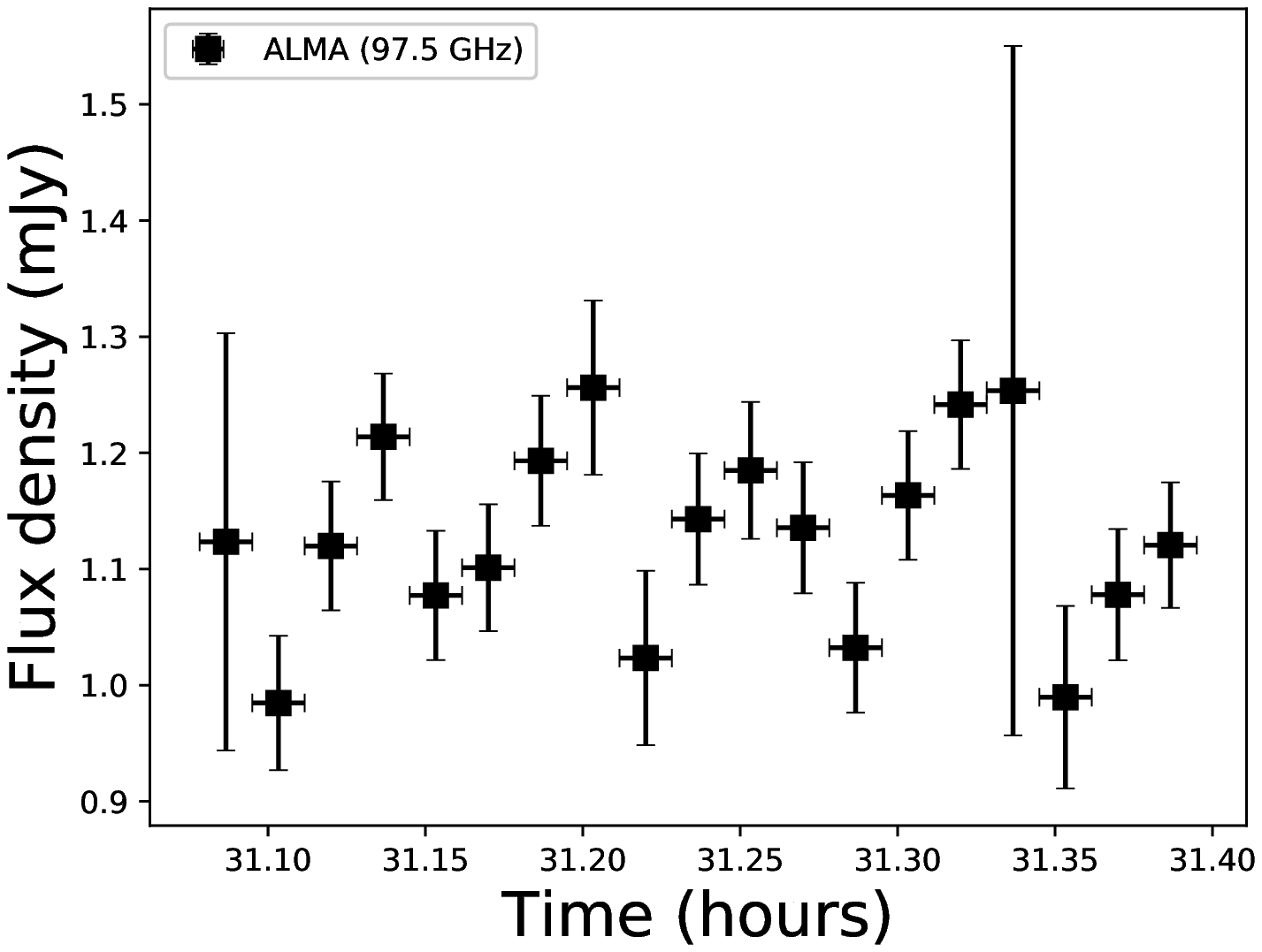}
\end{tabular}
\caption{Variability of the radio flux density and spectral index over $\approx2$\,hr at 
$\approx1.49$\,d after the burst in the cm-band (left), obtained by imaging the observations at 
1\,min intervals. The mm-band observations a few hours prior do not exhibit significant variability 
(right), suggesting an effect localized in the frequency domain. }
\label{fig:var}
\end{figure*}

High-cadence $uv$-domain fitting of the visibilities at time resolution of minutes reveals another 
unexpected variability: the afterglow light curve exhibits rapid brightening and fading within a 
single receiver base-band (2 GHz at K-band, 1 GHz otherwise) in the first four epochs on time scales 
of minutes, while the spectral index across base-bands within the same receiver tuning changes 
rapidly (Figure \ref{fig:var}). The mm-data do not exhibit comparable levels of variability, with 
the scatter in the time series being consistent with the mean uncertainty of the measurements. 
Whereas variability on short time scales is a known characteristic of diffractive interstellar 
scintillation, such effects have not been observed at frequencies $\gtrsim10$\,GHz, as apparent for 
this event \citep{ric90}. A detailed discussion of the cm-band variability is presented in ALB18; 
for the purposes of our broad-band analysis, we use the time-averaged data for each epoch, together 
with the ALMA light curve to study the behavior of the afterglow in the cm and mm bands.
The cm-band data in the first three epochs exhibit the greatest degree of variability, and we do 
not include them while computing the goodness of fit; however, they are important components for our 
final model, and we return to discussing the full cm-band data set in Section \ref{text:RS}. 

As the observed variability appears to decrease at $\gtrsim8.5$\,d, we attempt to derive 
the properties of the intrinsic emission by fitting the radio SEDs after this time. As the precise 
fits depend on the data selected for fitting, the true uncertainty on the measured numbers below are 
likely larger than those quoted, which are purely statistical.

The radio SED at 
$\approx 8.5$\,d exhibits a rising spectrum at $\gtrsim10$\,GHz. Fitting 
the data above 10\,GHz with a broken power law, 
\begin{align}
\label{eq:bpl2}
 F_{\nu}(\nu) &= F_{\rm b} \left[
 \frac{1}{2} \left(\frac{\nu}{\nu_{\rm b}}\right)^{-y\beta_1} + 
 \frac{1}{2} \left(\frac{\nu}{\nu_{\rm b}}\right)^{-y\beta_2} 
 \right]^{-1/y},
\end{align}
fixing $\beta_1=1/3$, $\beta_2 = (1-p)/2 \approx-0.5$, and $y = 1.84-0.40p\approx1.0$ 
\citep[appropriate for \numax;][]{gs02}, yields $\nu_{\rm b}=(9.2\pm1.0)\times10^{10}$\,Hz with 
flux density, $\fnupk = 0.508\pm0.007$\,mJy. 
The data at $\lesssim 10$\,GHz are in excess of the $\nu^{1/3}$ power law, while the spectrum at 
16.5\,d is relatively flat and can be fit as a single power law with $\beta=-0.19\pm0.03$. We 
address both points together in Section \ref{text:RS}.

The SED at $\approx 24.5$\,d exhibits a steep spectrum, $\beta=-1.1\pm0.2$ at $\approx 10$--30\,GHz, 
which underpredicts the ALMA observations at this time by a factor of $\approx 10$. It is 
possible that the decrement in the VLA observations at $\nu\gtrsim10$\,GHz relative to lower 
frequencies is due to phase decoherence, which systematically reduces the observed 
flux\footnote{If the phase fluctuations induced by the atmosphere on a given baseline can be 
approximated as a Gaussian random process with zero mean and standard deviation, $\sigma$, then the 
expectation of the interferometric visibility is $\langle V\rangle=Ve^{-\sigma^2/2}$, where 
$V$ is the true visibility. See Chapter 13 of \cite{tms01} for a derivation.}, as the data were 
acquired under marginal weather conditions; we therefore remove these data also from our model fit. 
Fitting the cm-band data at 
$\lesssim 10$\,GHz together with the ALMA observations, we find $\nu_{\rm b} = 9.4\pm4.4$\,GHz and 
$\fnupk=0.49\pm0.03$\,mJy at $\approx 24.5$\,d. The constancy of the peak flux density from 8.5\,d 
to 24.5\,d identifies this break as $\numax$ and confirms the circumburst medium as an ISM 
environment, for which we expect $\fnumax\propto t^0$; the observed decline rate of this 
frequency, $\alpha_{\nu,\rm peak}=-2.2\pm0.5$ is also consistent at 1.4$\sigma$ with the 
expectation of $\alpha_{\nu_{\rm m}}=-1.5$. 

Projecting this frequency back to the optical bands at earlier times with $\alpha_{\nu}=-1.5$, we 
expect the break to have crossed $R$-band at $\approx3\times10^{-2}$\,d. Clear filter observations 
calibrated to $R$-band from Terksol at 0.29\,d yield $f_{\nu,\rm R}=0.56\pm0.01$\,mJy 
\citep{gcn20309}, in excellent agreement with $\numax\approx\nu_{\rm opt}$ at this time. This is 
further consistent with the subsequent decline rate and spectral index in the optical 
bands (Section \ref{text:basic_x}), confirming the optical emission at 
$\gtrsim3\times10^{-2}$\,d and radio observations at $\gtrsim8.5$\,d as synchrotron emission from 
the FS. 
In the slow cooling regime, the afterglow peak flux density is given by,
\begin{align}
 F_{\nu} &=9.93(p+0.14)(1+z)\epsb^{1/2}\dens^{1/2}E_{\rm K,iso,52} d_{\rm L,28}^{-2}\,{\rm mJy}
 \nonumber \\
         &\sim 50 {\rm mJy} \left(\epsilon_{{\rm B},-2}\dens\right)^{1/2}
                  \left(\frac{1-\etarad}{\etarad}\right)E_{\rm \gamma,iso},
\end{align}
for 161219B and $p\approx2$, where $\etarad$ is the radiative efficiency.
Taking $\fnumax\approx0.5$\,mJy and assuming $\etarad\approx10\%$, we find 
$\dens\approx6\times10^{-4}\epsilon_{\rm B, -2}^{-1}\pcc$, indicating a low density environment. 

The ALMA light curve can be fit with a single power law with $\alpha_1 = -0.52\pm0.02$ from the 
first observation at 1.3\,d to the fourth epoch at $\approx24.5$\,d. This is shallower than the 
expected decline rate of FS emission, and is best described as a combination of two emitting 
components declining at different rates (Section \ref{text:RS}). This best-fit power law 
over-predicts the flux density at the fifth epoch at 78.2\,d by a factor of $\approx3$, which 
suggests a jet break has occurred between $24.5$ and $78.2$\,d, as indicated by the X-ray 
observations (Section \ref{text:basic_x}).

The radio SED fades at all frequencies between 24.5\,d and 79.0\,d, and the best fit broken 
power law model at 79.0\,d yields $\nu_{\rm b}=2.4\pm0.8$\,GHz with flux density 
$\fnupk=0.20\pm0.02$\,mJy (fixing the same parameters as at 24.5\,d). 
The drop in peak flux further indicates a jet break has taken place between 24.5\,d and 79.0\,d.
The SED in the last epoch at 159.5\,d can be fit either with a broken power law with 
$\beta_1=1/3$ (fixed), $\beta_2\approx-0.5$ (fixed), $\nu_{\rm b}=7.9\pm1.5$\,GHz and 
$\fnupk=0.12\pm0.01$\,mJy, or as a single power law with $\beta=-0.8\pm0.1$. An increase in the 
break frequency $\nu_{\rm b}$ with time is physically implausible, and it is possible that the 
lowest frequency observations in this epoch have contribution from the host galaxy. We discuss this 
epoch further in Section \ref{text:radioexcess}.

To summarize, the optical and X-ray light curves require a constant density environment with 
$p\approx2$. The multi-band X-ray through radio observations of the afterglow are consistent with 
a slow cooling FS ($\numax < \nuc$) in an ISM environment with $\numax\approx\nu_{\rm opt}$ at 
$\approx3\times10^{-3}$\,d, while the NIR to X-ray SED indicates $\nuc\approx\nux$ for the 
duration of the X-ray observations. The UV spectral slope is marginally steeper than that in the 
optical bands, indicating possible extinction in the host galaxy. 
The peak flux density of the FS is $\fnumax\approx0.5$\,mJy, implying that 
the optical and X-ray light curves prior to $\approx 3\times10^{-2}$\,d and radio SEDs prior to 
$8.5$\,d are dominated by emission from a separate mechanism.


\begin{figure*} 
 \begin{tabular}{cc}
  \includegraphics[width=0.47\textwidth]{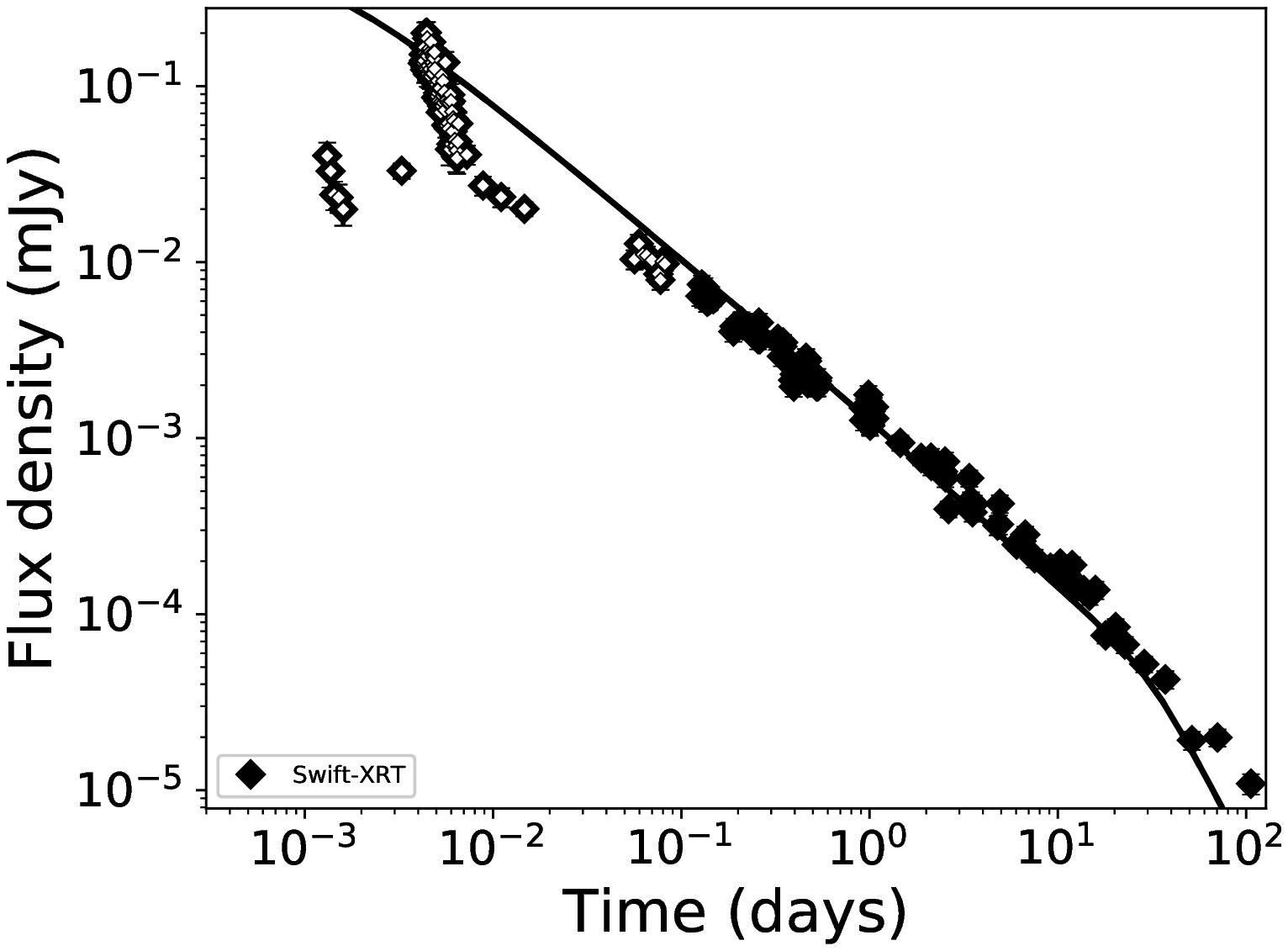} &
  \includegraphics[width=0.47\textwidth]{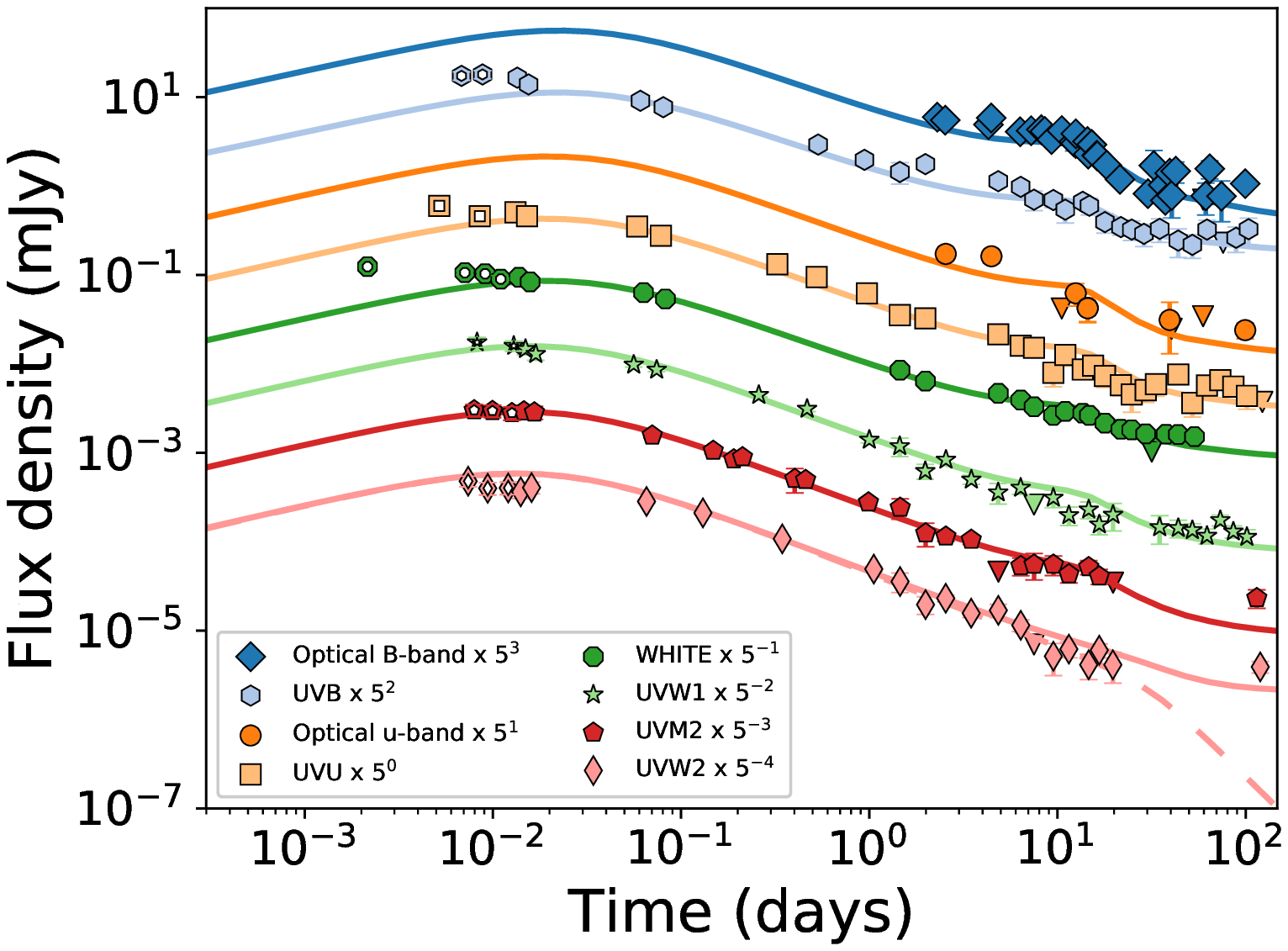} \\ [-3pt]
  \includegraphics[width=0.47\textwidth]{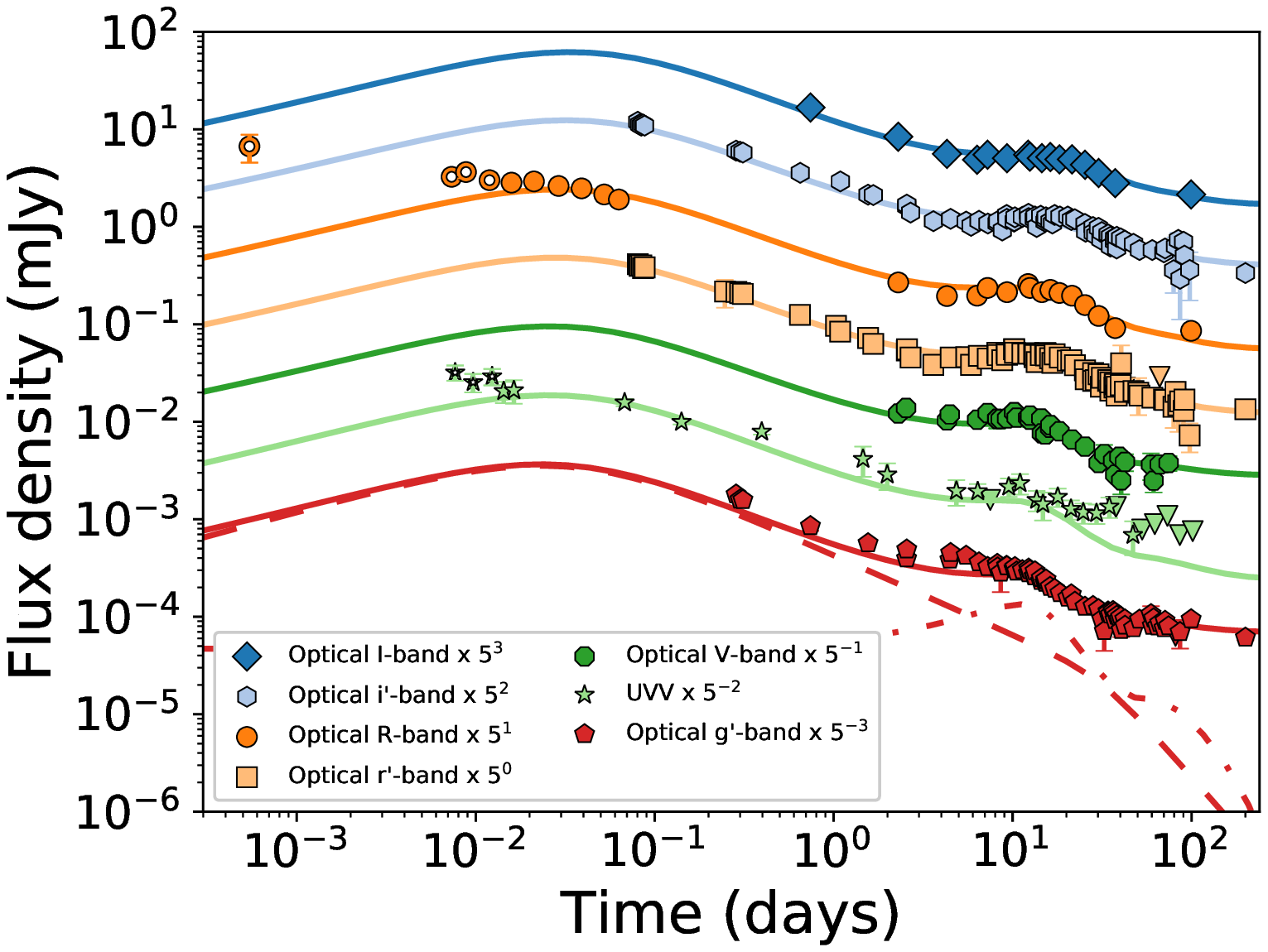} &
  \includegraphics[width=0.47\textwidth]{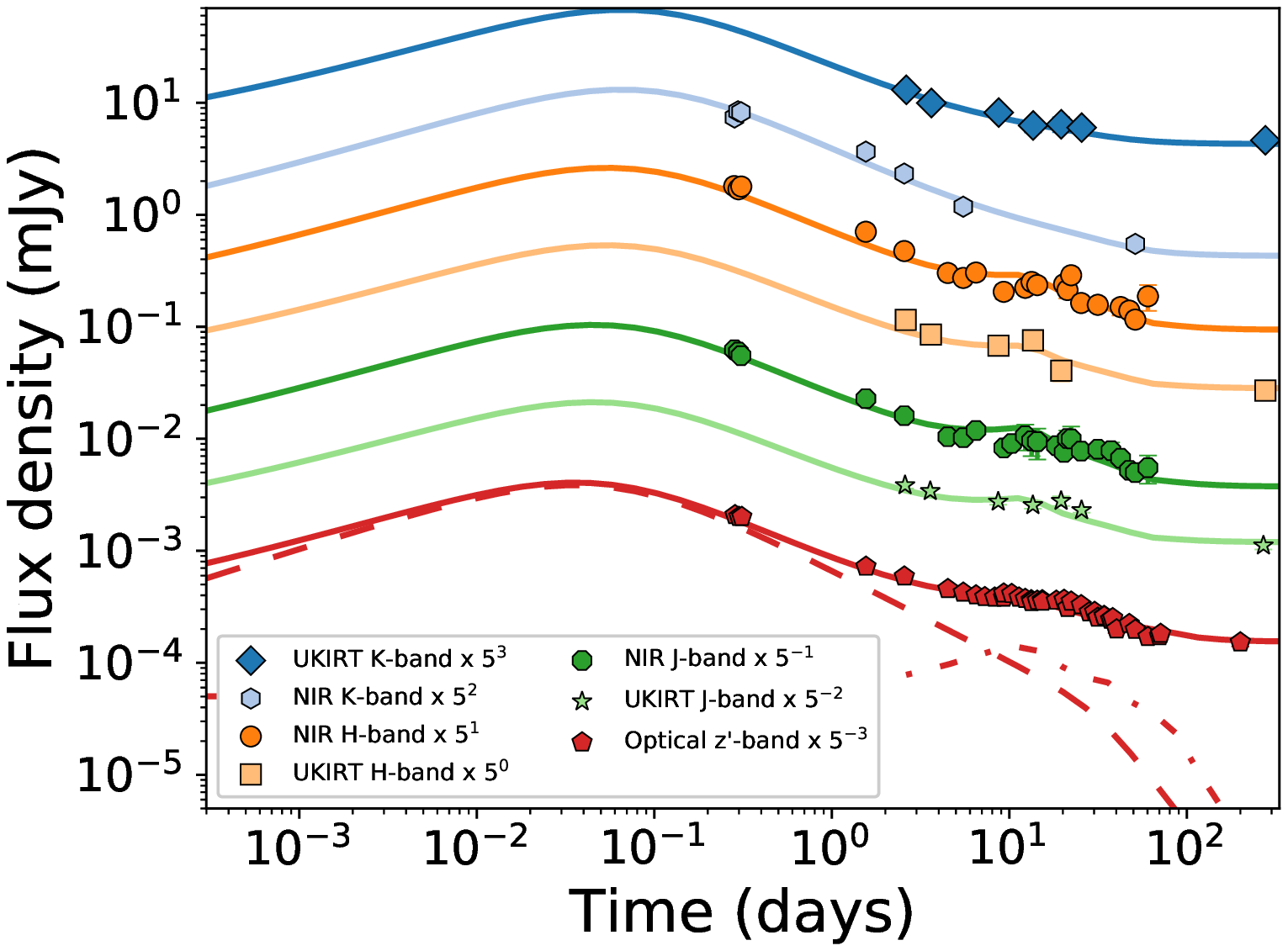} \\ [-3pt]
  \includegraphics[width=0.47\textwidth]{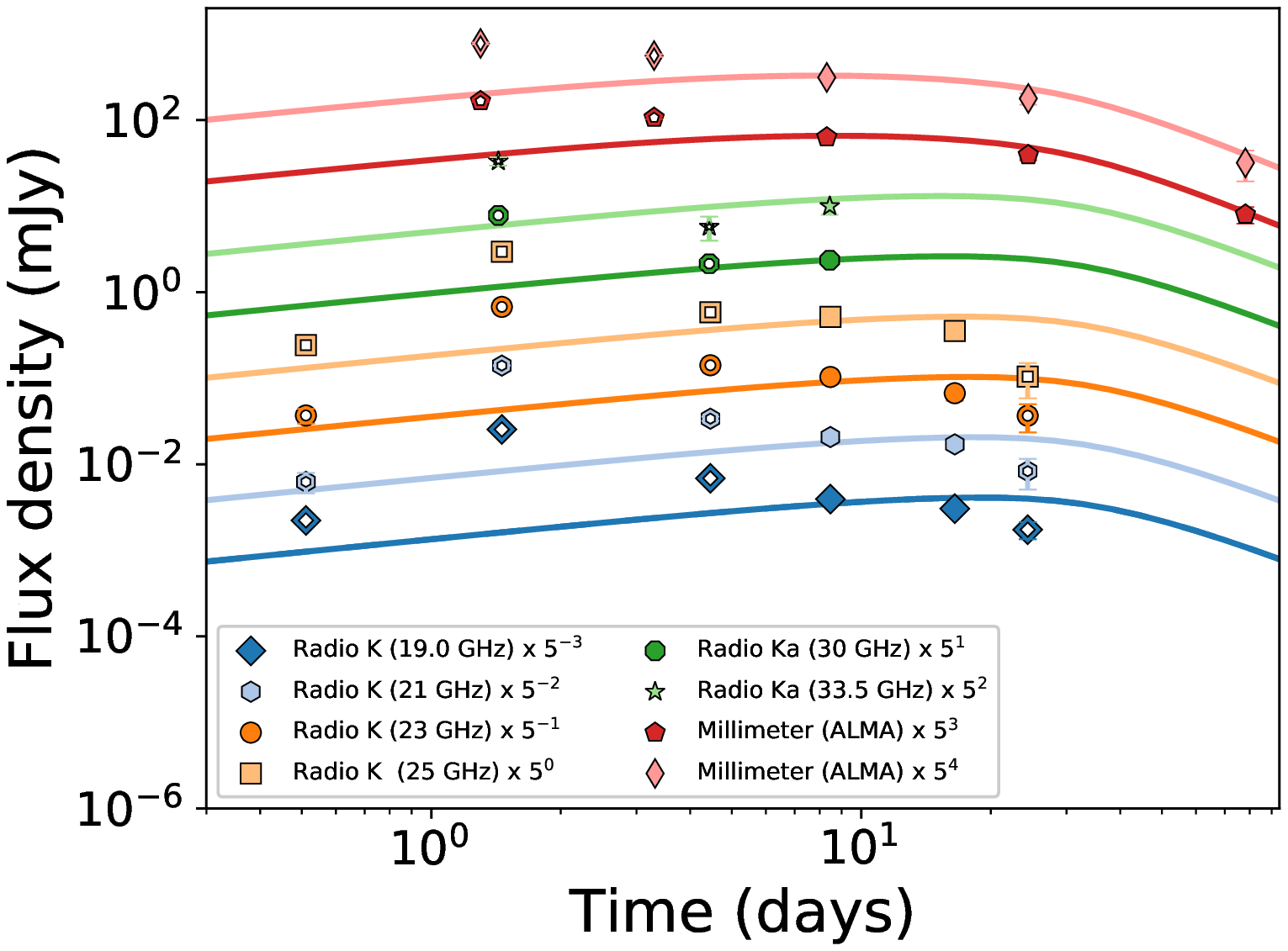} &
  \includegraphics[width=0.47\textwidth]{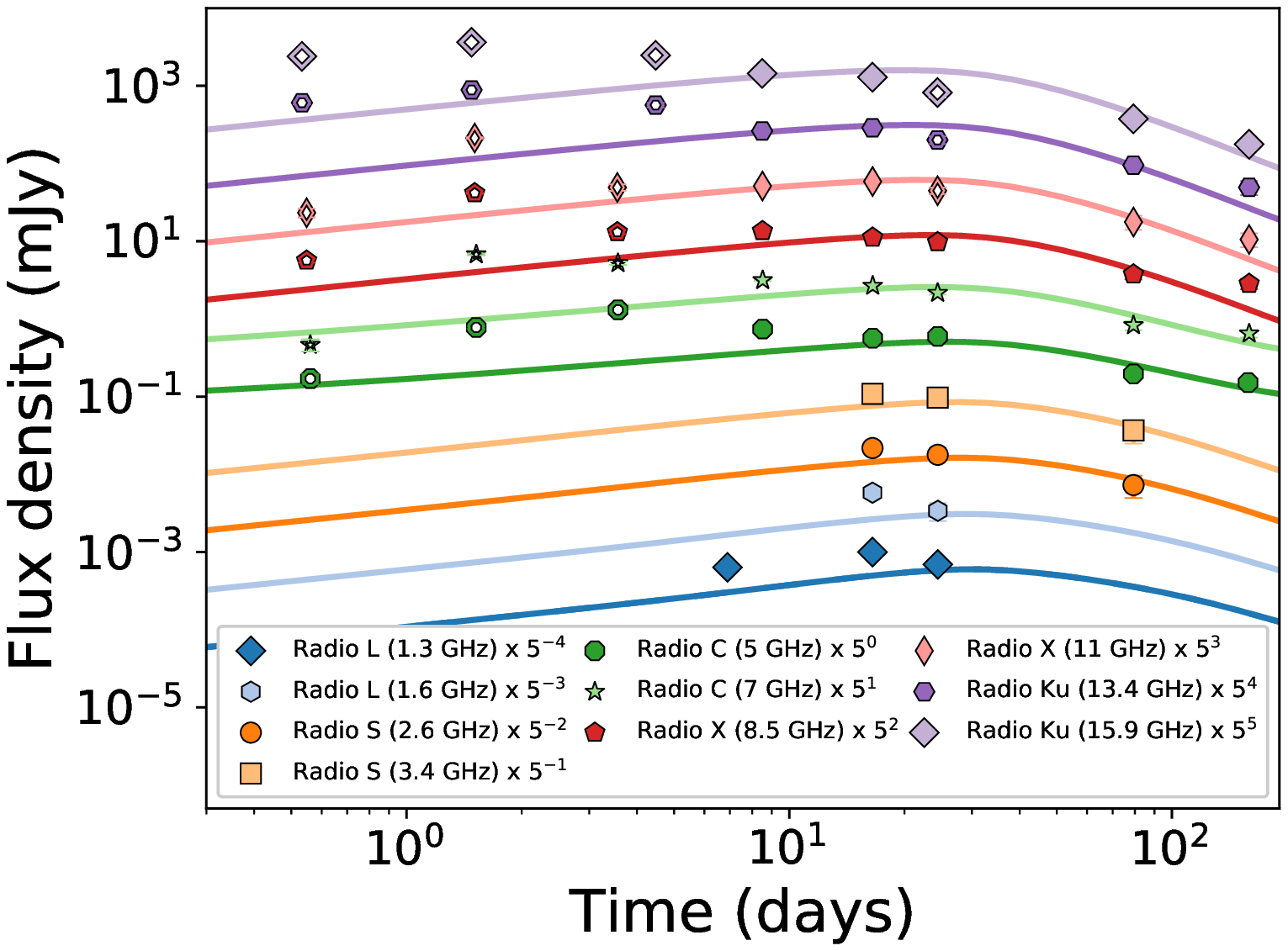} \\ [-3pt]
 \end{tabular}
 \caption{X-ray (top left), UV (top right), optical (center left), NIR (center right), and radio 
(bottom) light curves of GRB\,161219B/SN2016jca, together with an FS ISM model including 
contributions from the supernova light (solid lines). We show a decomposition of the \Swift/{\textit 
w2}-band, optical $g^{\prime}$-band, and optical $z^{\prime}$-band light curves into FS 
(dashed) and supernova (dash-dotted) components.
Data represented by open symbols are not included in the model fit. The $JHK$ photometry from GROND, 
NOT, and GTC was reported in a 2.2\arcsec\, aperture and does not include the full light of the 
host; these bands are therefore treated separately from the UKIRT photometry, which does include all 
contributions from the host (Section \ref{text:model}). This FS-only model over-predicts the X-ray 
data at $\lesssim0.1$\,d, and under-predicts the optical observations at $\lesssim3\times10^{-3}$\,d 
as well as the radio observations at $\lesssim8.5$\,d; both deficiencies are overcome in the 
refreshed RS model presented 
in Figure \ref{fig:modellc_RS_splits}.}
\label{fig:modellc_FS_splits}
\end{figure*}

\section{Multi-wavelength modeling}
\label{text:model}
Our preliminary considerations described above indicate that the X-ray light curve after 
$\approx0.1$\,d, the UV, optical, and NIR data at $1.3\times10^{-2}$ -- 2\,d and at $>50$\,d, as 
well as the radio SEDs at $\gtrsim8.5$\,d can be understood in the context of synchrotron radiation 
from an FS propagating into a constant density ISM environment. We derive the parameters of this 
shock using the smoothly connected power-law synchrotron spectra described by \cite{gs02} modified 
using the weighting schemes presented in \cite{lbt+14}, including the effects of inverse Compton 
cooling \citep{se01, lbm+15}. The free parameters in this model are the total isotropic-equivalent 
kinetic energy ($\EKiso$), the circumburst density ($\dens$), the fraction of shock energy imparted 
to relativistic electrons (\epse), the fraction imparted to magnetic fields (\epsb), and the 
index of the electron energy spectrum ($p$). We incorporate the effects of collimation using the 
jet break time ($\tjet$) as an additional free parameter \citep{rho99,sph99,cl00,lbt+14}. 

We include the contribution of the supernova using a template constructed from the 
extinction-corrected spectra of SN\,1998bw, performing K-corrections for the UV, optical and 
NIR bands \citep{pcd+01,lnf+05}. The template is scaled by three additional parameters: the 
relative peak time ($\delta t_{\rm sn}$), a temporal stretching factor ($\Upsilon_{\rm sn}$), 
and a flux density scaling ($\Xi_{\rm sn}$) using
\begin{align}
t_{\rm obs} &= \Upsilon_{\rm sn} t_0 + \delta t_{\rm sn}, \nonumber \\
f_{\nu,\rm obs} (t_{\rm obs}) &= \Xi_{\rm sn} f_{\nu,\rm 0},
\end{align}
where $f_0(t_0)$ is the SN\,1998bw template scaled to the redshift of GRB\,161219B. Thus $\Xi_{\rm 
sn}$ measures the intrinsic luminosity ratio of SN\,2016jca to SN\,1998bw, while $\delta t_{\rm 
sn}/(1+z)$ represents the rest-frame delay of the SN\,2016jca peak relative to that of 
SN~1998bw.

We model the extinction in the host galaxy using the SMC extinction curve \citep{pei92,lbt+14}, and 
include contributions from the host galaxy at all bands that exhibit a flattening at $\gtrsim 
50$\,d. Since the observations presented in \cite{cidup+17} are measured using a $2.2\arcsec$ 
aperture, the contribution of the host galaxy to their photometry is different from that in our 
observations. The largest differences appear in the $JHK$ bands, and we therefore keep the $JHK$ 
bands from the two sets of observations separate in our analysis. We include the pre-explosion 
$griz$ Pan-STARRS1 photometry of the host (in the same $2.2\arcsec$ aperture) as additional data 
points at the fiducial time of 200\,d, at which time the contribution of the supernova and 
afterglow is minimal, allowing us to constrain the contribution of the host to the light curve in 
all UV, optical, and NIR bands.

To efficiently sample parameter space and fully characterize the joint posterior density of 
the free parameters in our model, we carry out a Markov Chain Monte Carlo (MCMC) analysis using the 
Python-based code \textsc{emcee} \citep{fhlg13} following the procedure described in \cite{lbt+14} 
and \cite{lbm+15}. We assume uninformative, uniform priors on all free parameters.
Priors for all scale parameters (\epse, \epsb, \dens, \EKiso, \tjet, and $\Upsilon_{\rm SN}$) 
are uniform in logarithmic space \citep{jef46}. 
%
We initialized 128 Markov chains with parameters clustered around their best fit values (with a 3\% 
dispersion). After discarding samples prior to the stabilization of the average likelihood across 
chains as `burn-in'. we obtained $10^4$ samples from the posterior. Further details about our 
MCMC analysis method, convergence tests, and quantile and summary statistic calculation are 
available in \cite{lbt+14}. To account for calibration offsets in UV/optical/NIR data from 
different observatories as well as potential systematic intrinsic flux calibration uncertainties, we 
impose an uncertainty floor of $10\%$ prior to fitting with our modeling software.

\subsection{Afterglow}
In confirmation of the basic analysis presented in Section \ref{text:basic_considerations}, we find 
that an ISM model with $p\approx2.08$ describes the data well. Our highest-likelihood (best-fit) 
model has $\numax\approx4\times10^{12}\,{\rm Hz}<\nu_{\rm opt}$ and 
$\nuc\approx2\times10^{17}$\,Hz $\approx\nux$ at 1 day, 
as inferred from the optical and X-ray light curves. The extrapolated peak flux density is 
$\fnumax\approx1$\,mJy. The peak of the rounded spectrum\footnote{Here $y=1.84-0.40p\approx1.0$ is 
the smoothness of the $\numax$ break.} is then $\fnupk = 2^{-1/y}\fnumax\approx0.5$\,mJy, consistent 
with the radio SEDs and the optical observations at $\approx3\times10^{-2}$\,d (Section 
\ref{text:basic_radio}). The afterglow remains in the slow cooling regime for the duration of the 
observations.

\begin{figure*}
\begin{tabular}{ccc}
 \centering
 \includegraphics[width=0.31\textwidth]{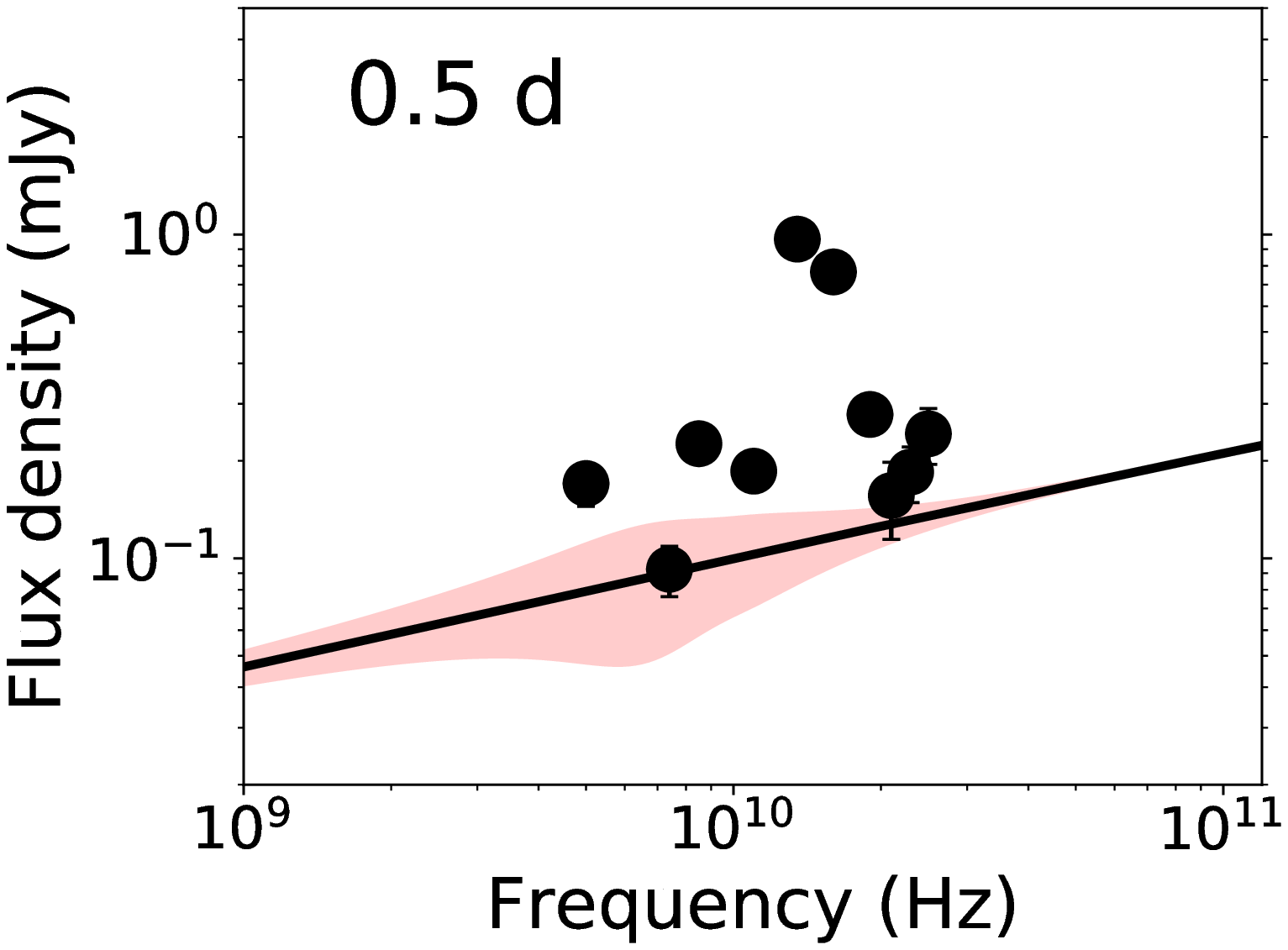} &
 \includegraphics[width=0.31\textwidth]{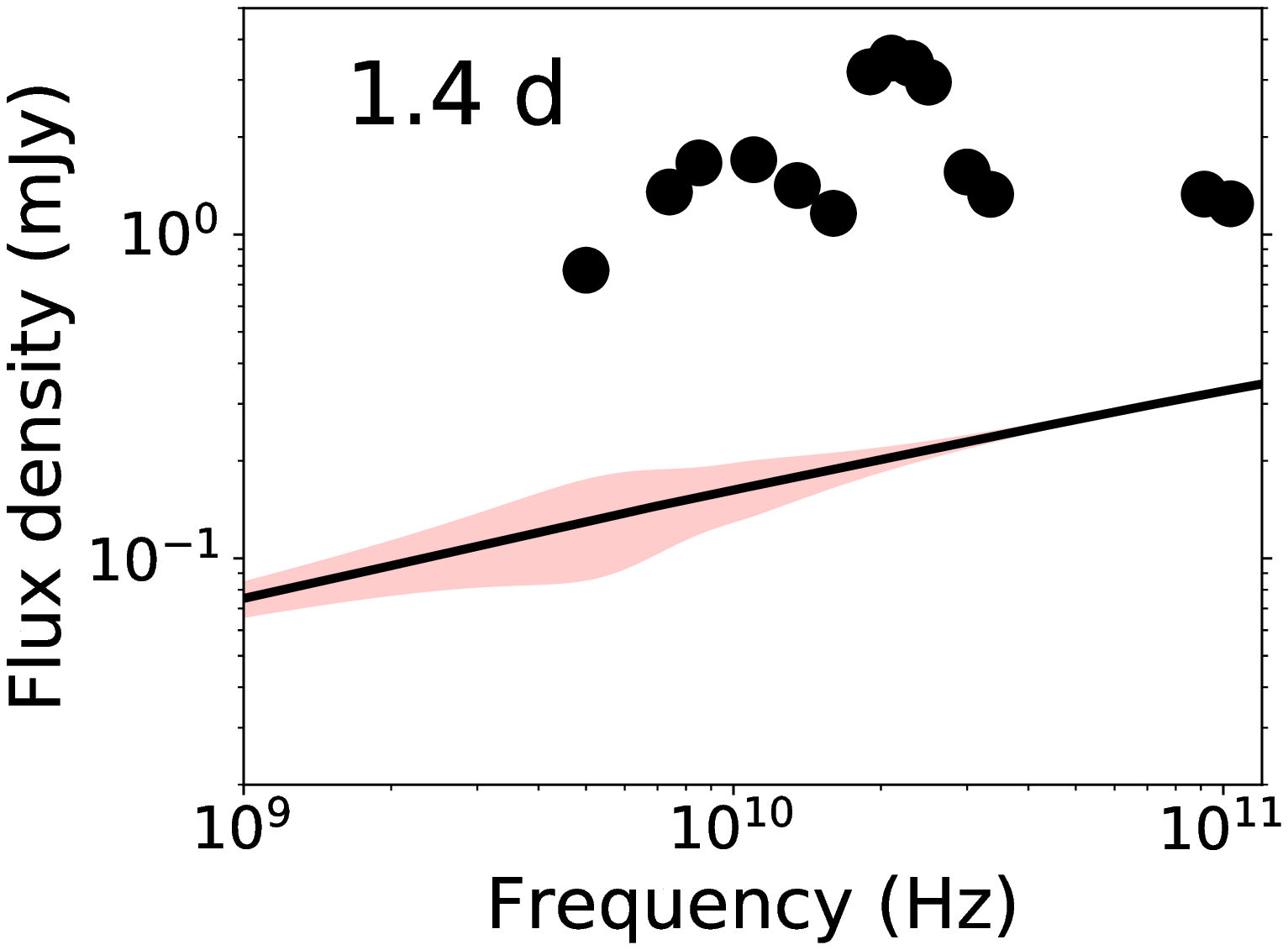} &
 \includegraphics[width=0.31\textwidth]{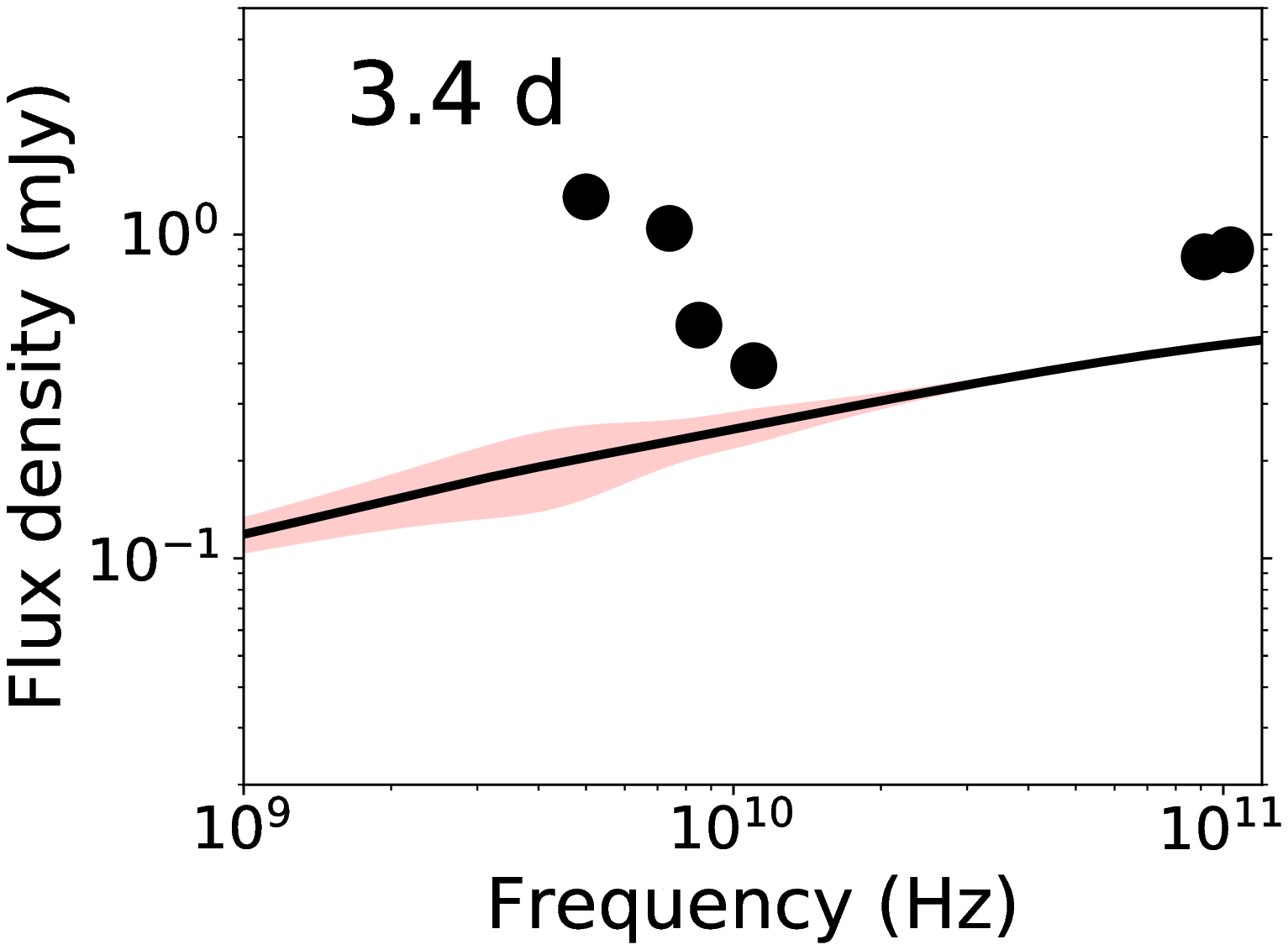} \\
 \includegraphics[width=0.31\textwidth]{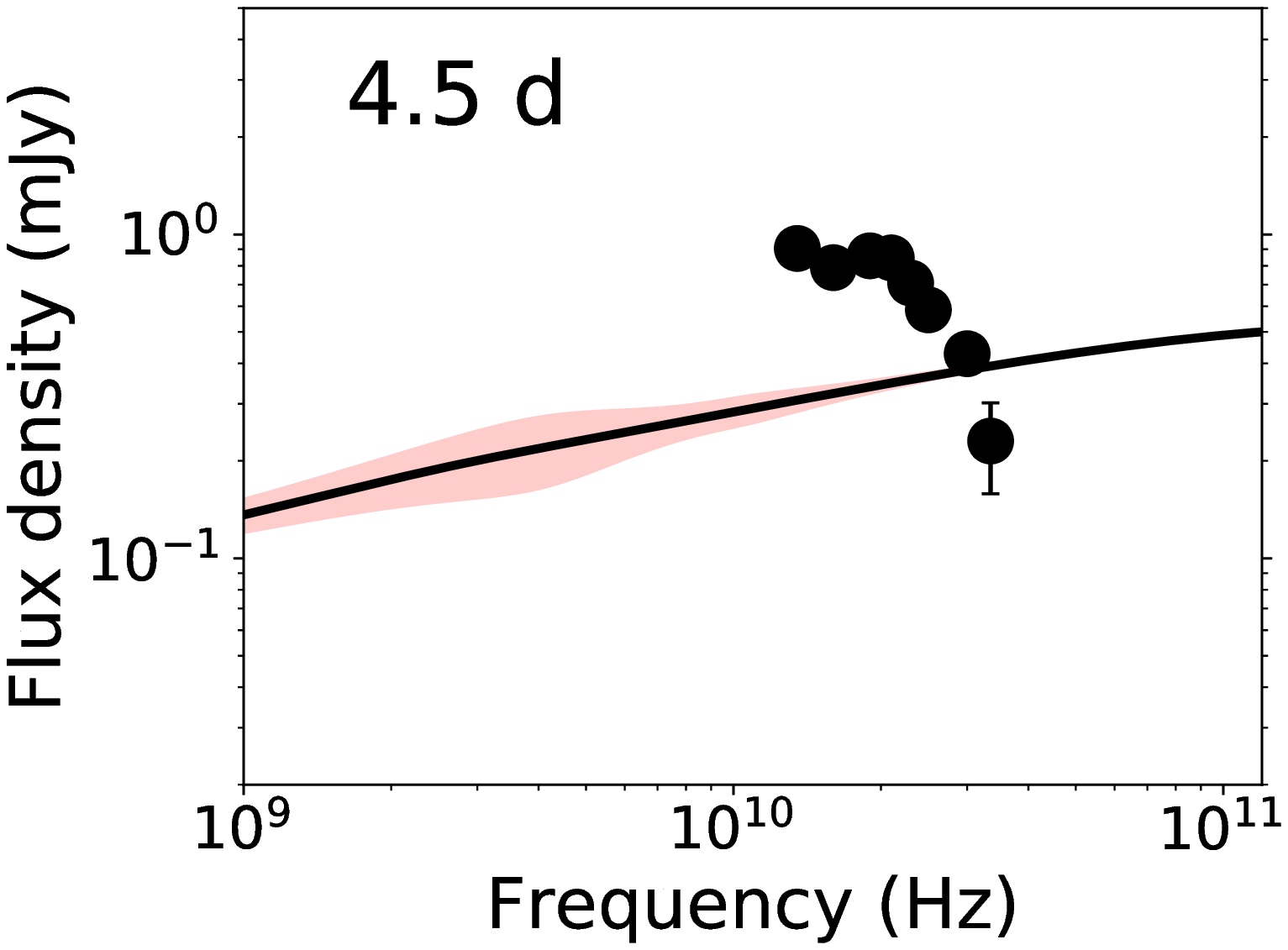} &
 \includegraphics[width=0.31\textwidth]{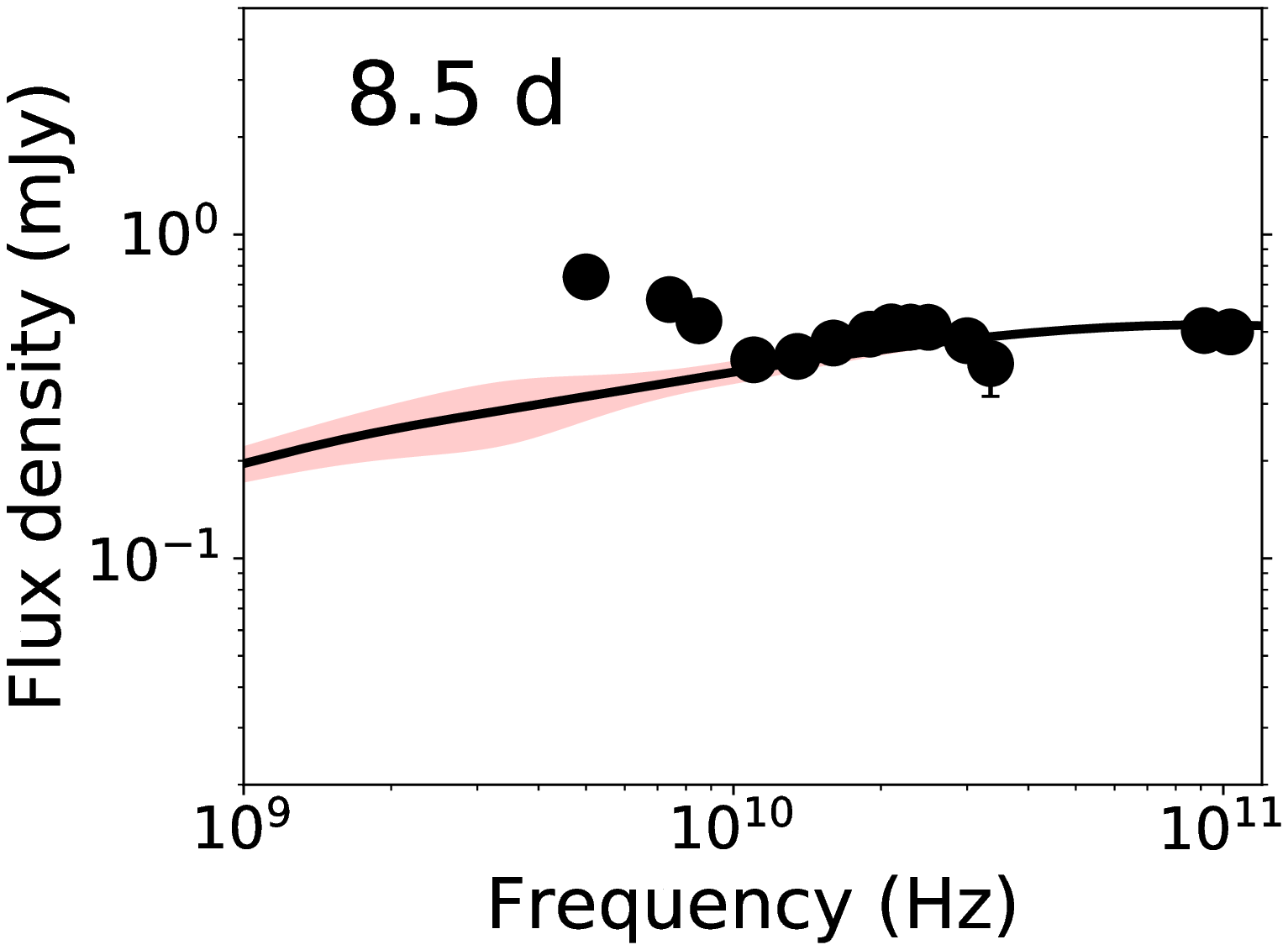} &
 \includegraphics[width=0.31\textwidth]{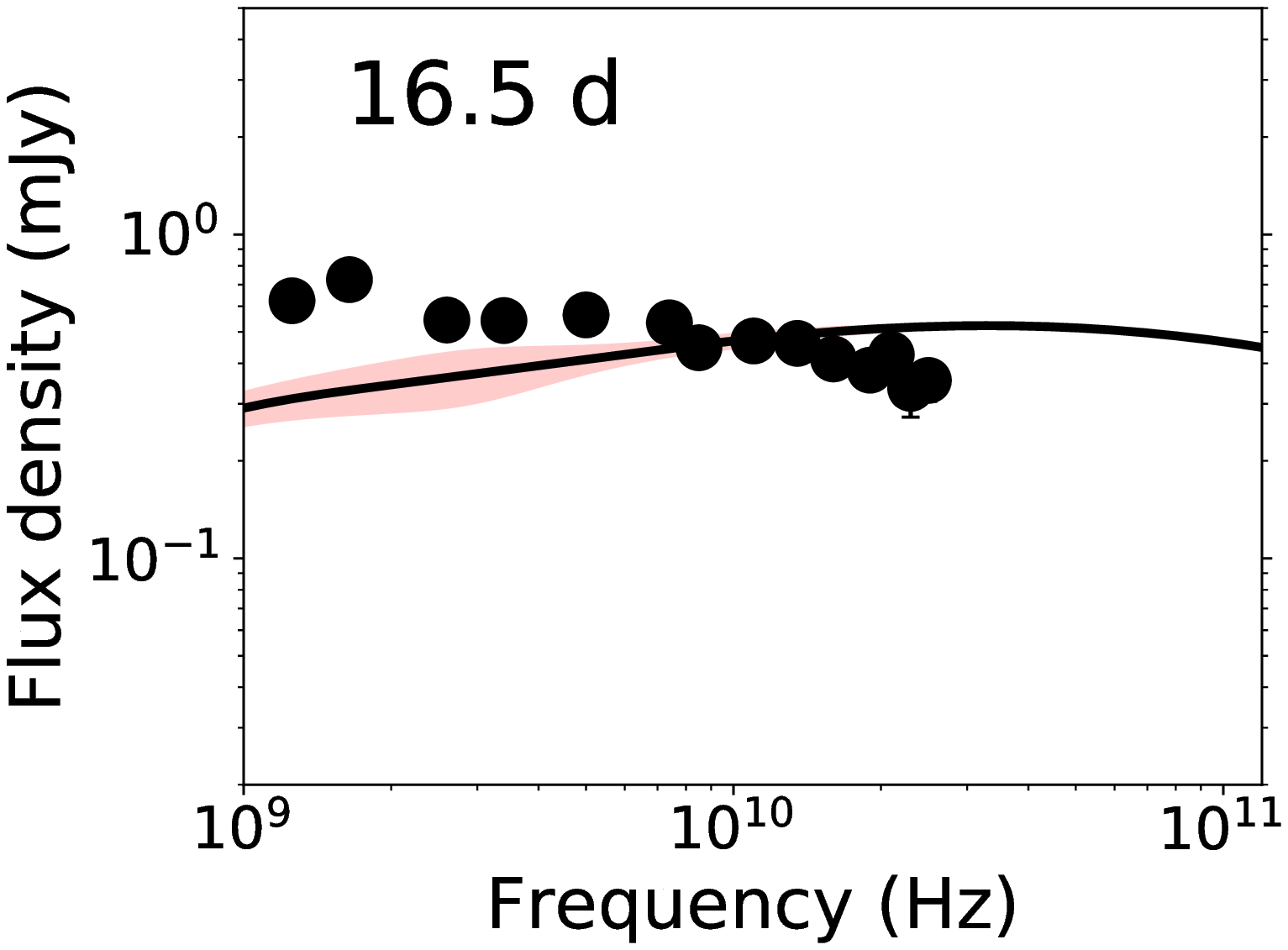} \\
 \includegraphics[width=0.31\textwidth]{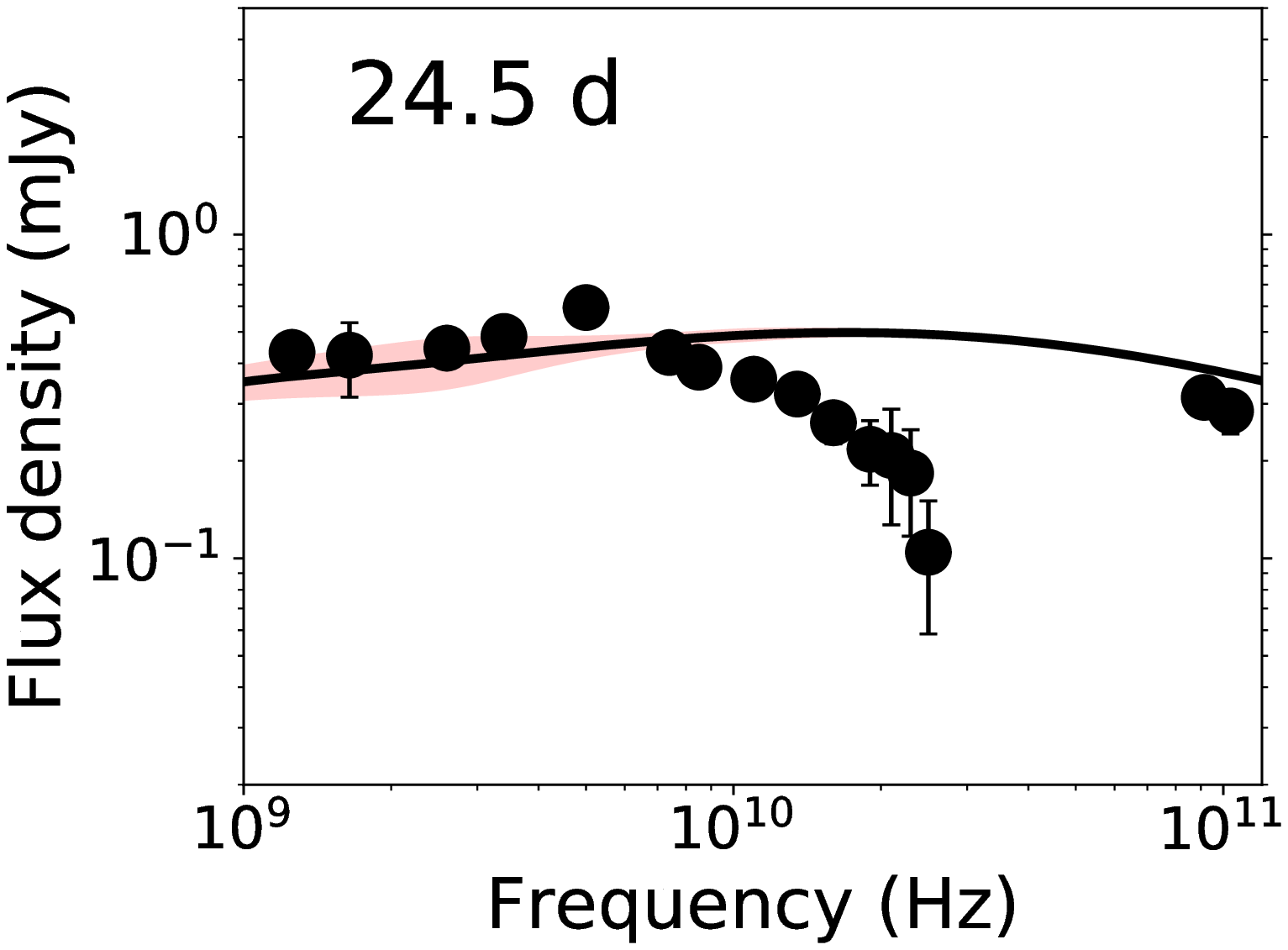} &
 \includegraphics[width=0.31\textwidth]{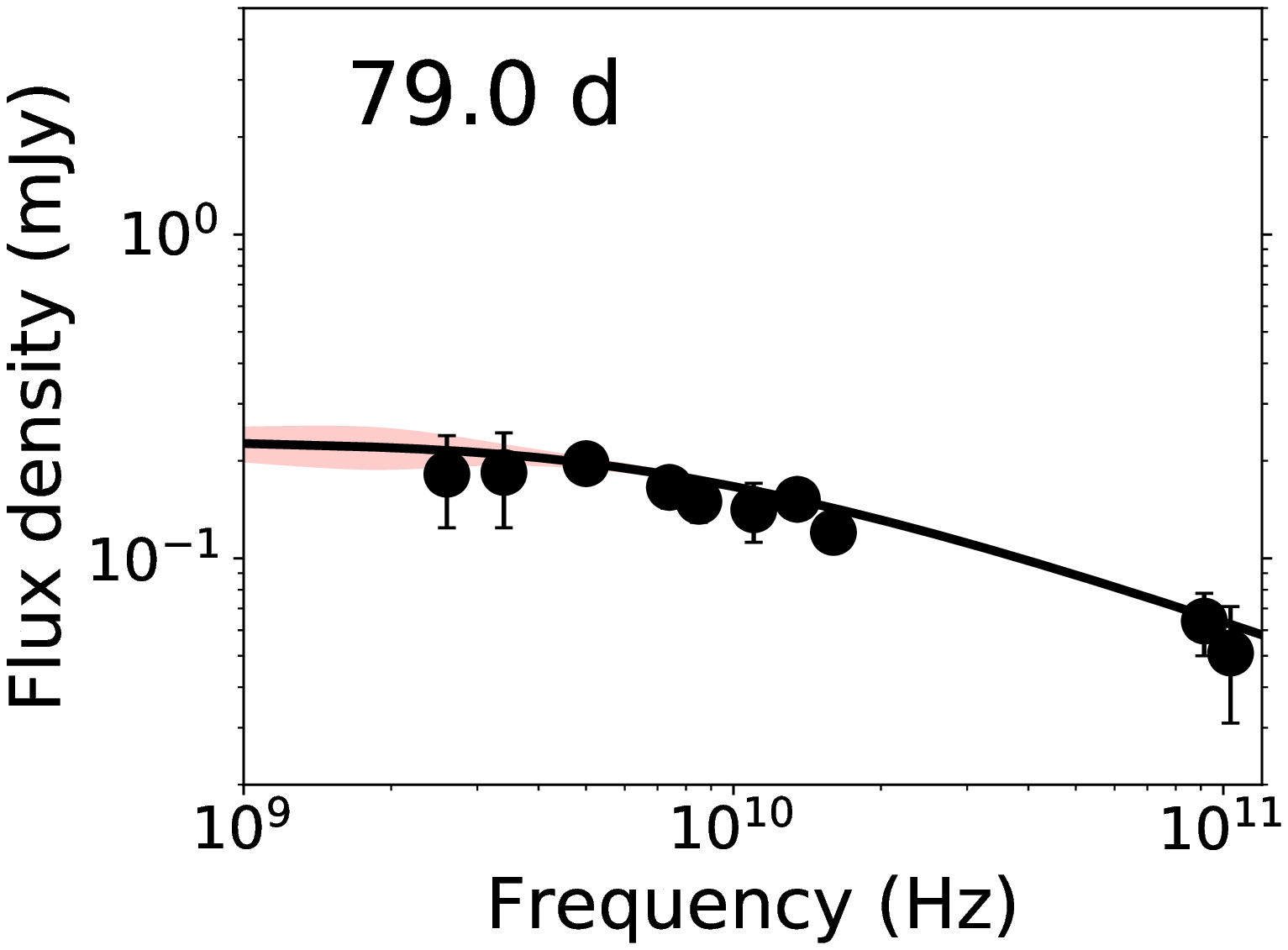} &
 \includegraphics[width=0.31\textwidth]{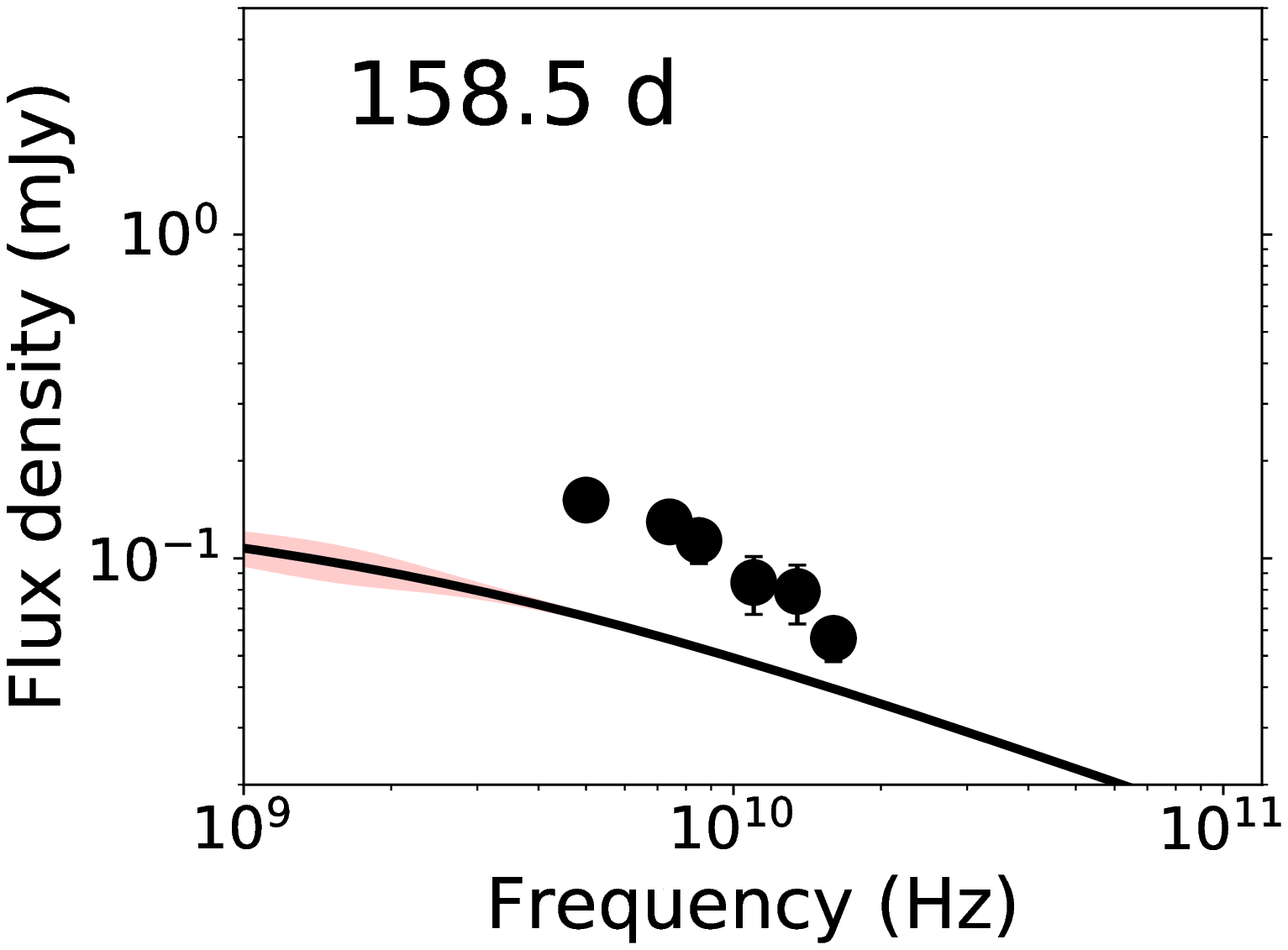} 
\end{tabular}
\caption{
Multi-frequency cm-band (VLA) and mm-band (ALMA) spectral energy distributions of the afterglow of 
161219B, together with a forward shock ISM model (solid lines; Section \ref{text:model}). The red 
shaded regions represent the expected variability due to scintillation. This FS-only model 
under-predicts the radio observations at $\lesssim8.5$\,d, and requires an additional component 
(Section \ref{text:RS} and Figure \ref{fig:modelsed_RS}).}
\label{fig:modelsed_FS}
\end{figure*}

\begin{figure}
\begin{tabular}{cc}
 \centering 
 \includegraphics[width=0.23\textwidth]{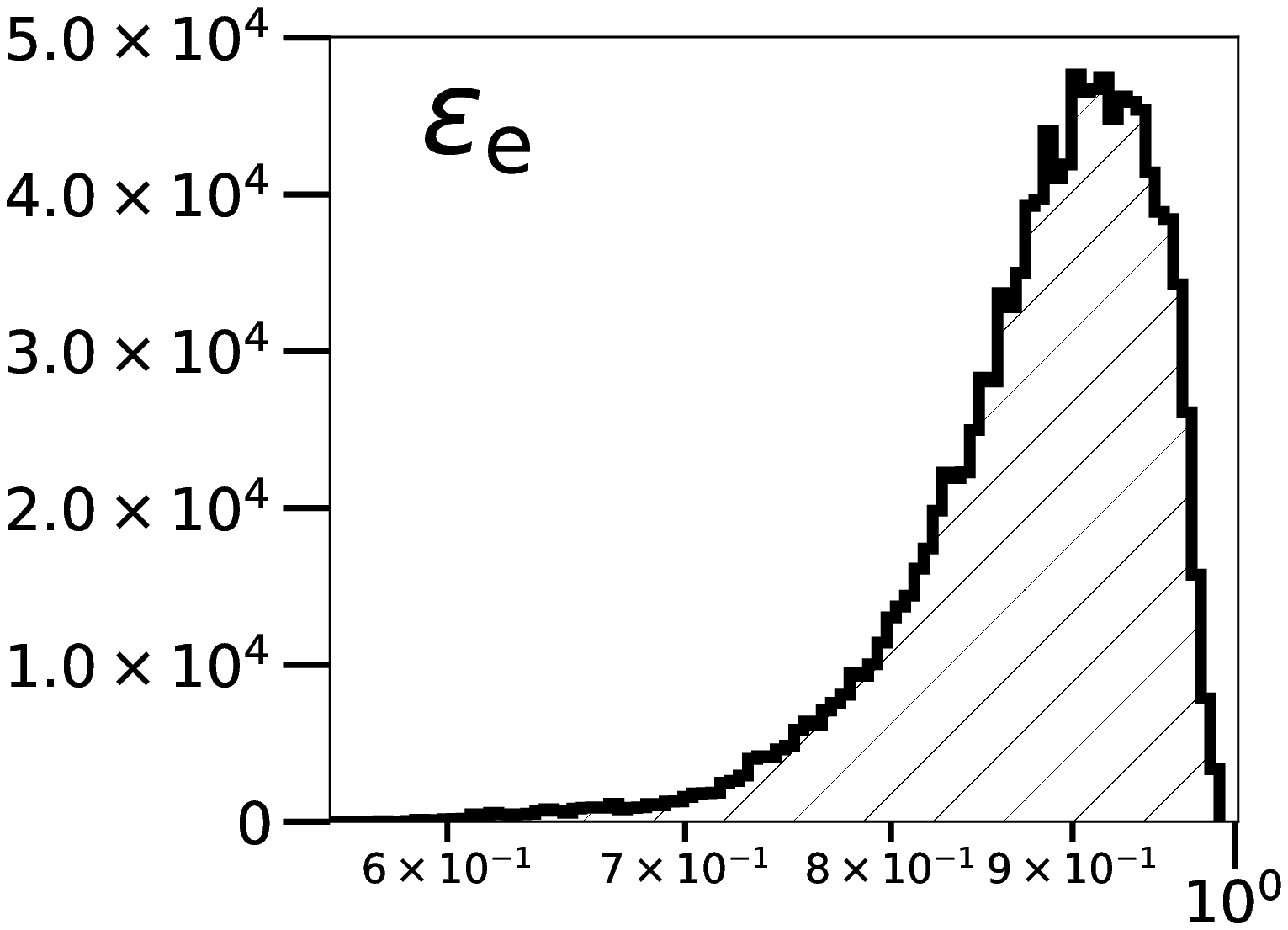} &
 \includegraphics[width=0.23\textwidth]{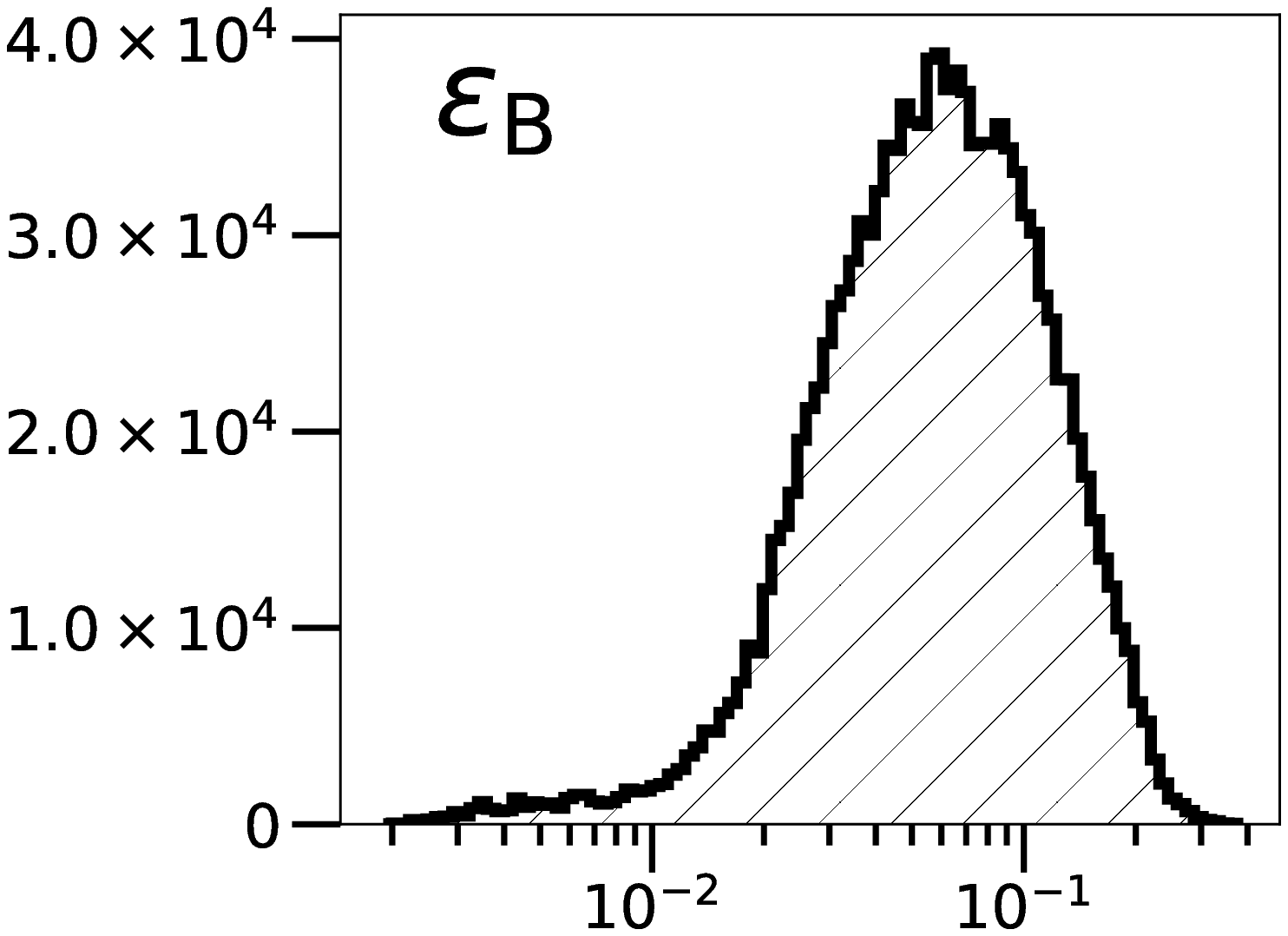} \\
 \includegraphics[width=0.23\textwidth]{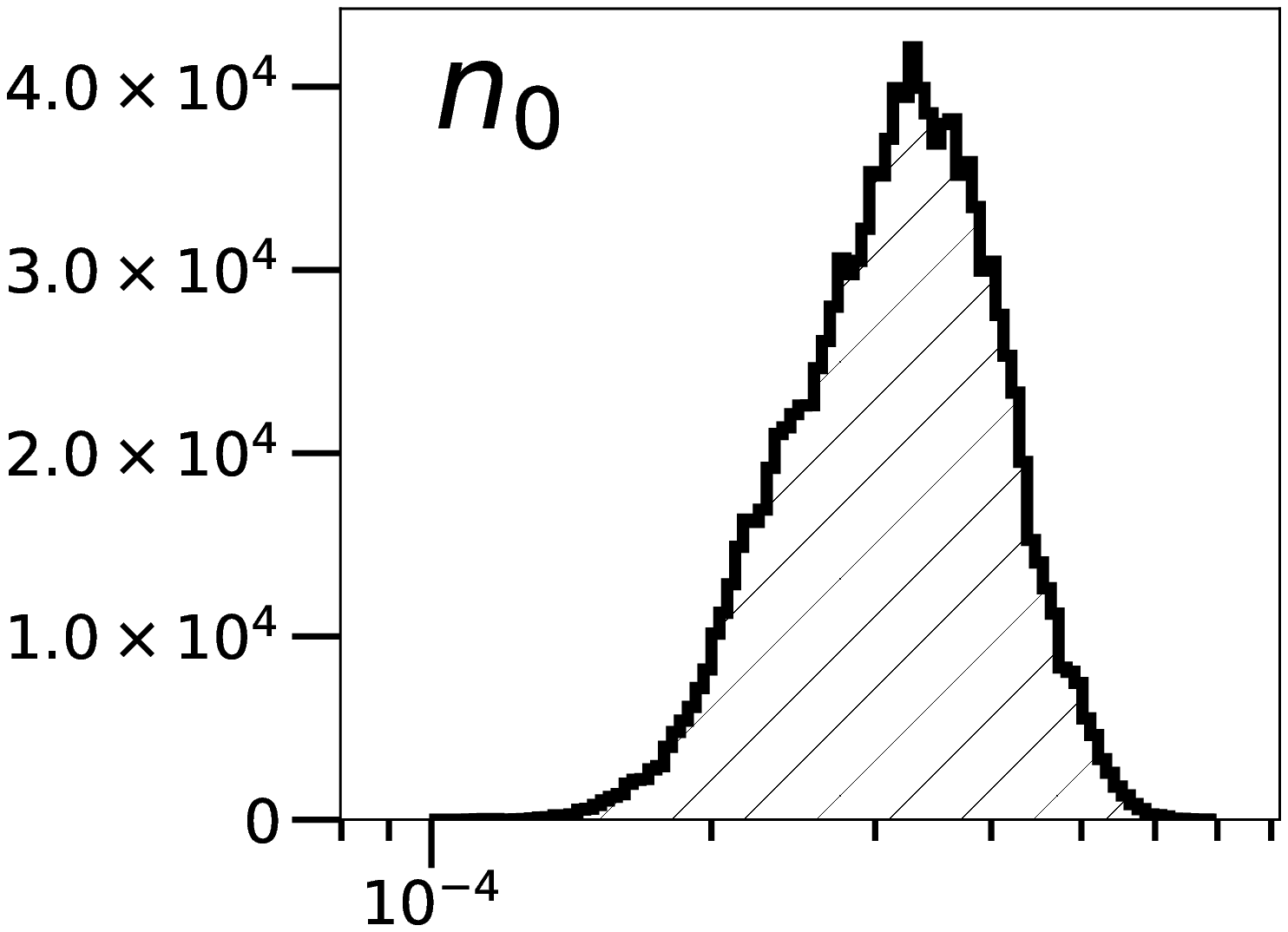} &
 \includegraphics[width=0.23\textwidth]{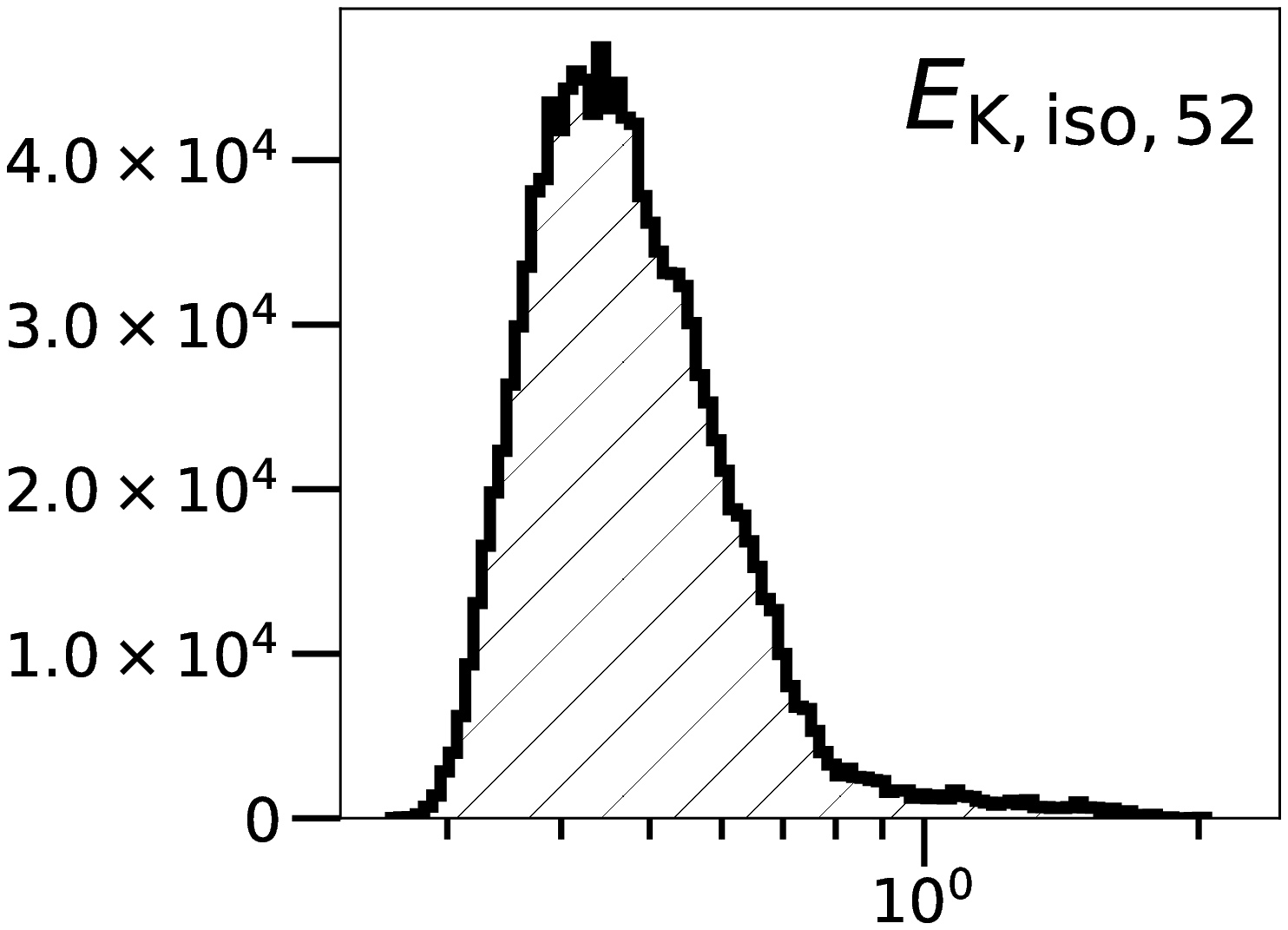} \\
 \includegraphics[width=0.23\textwidth]{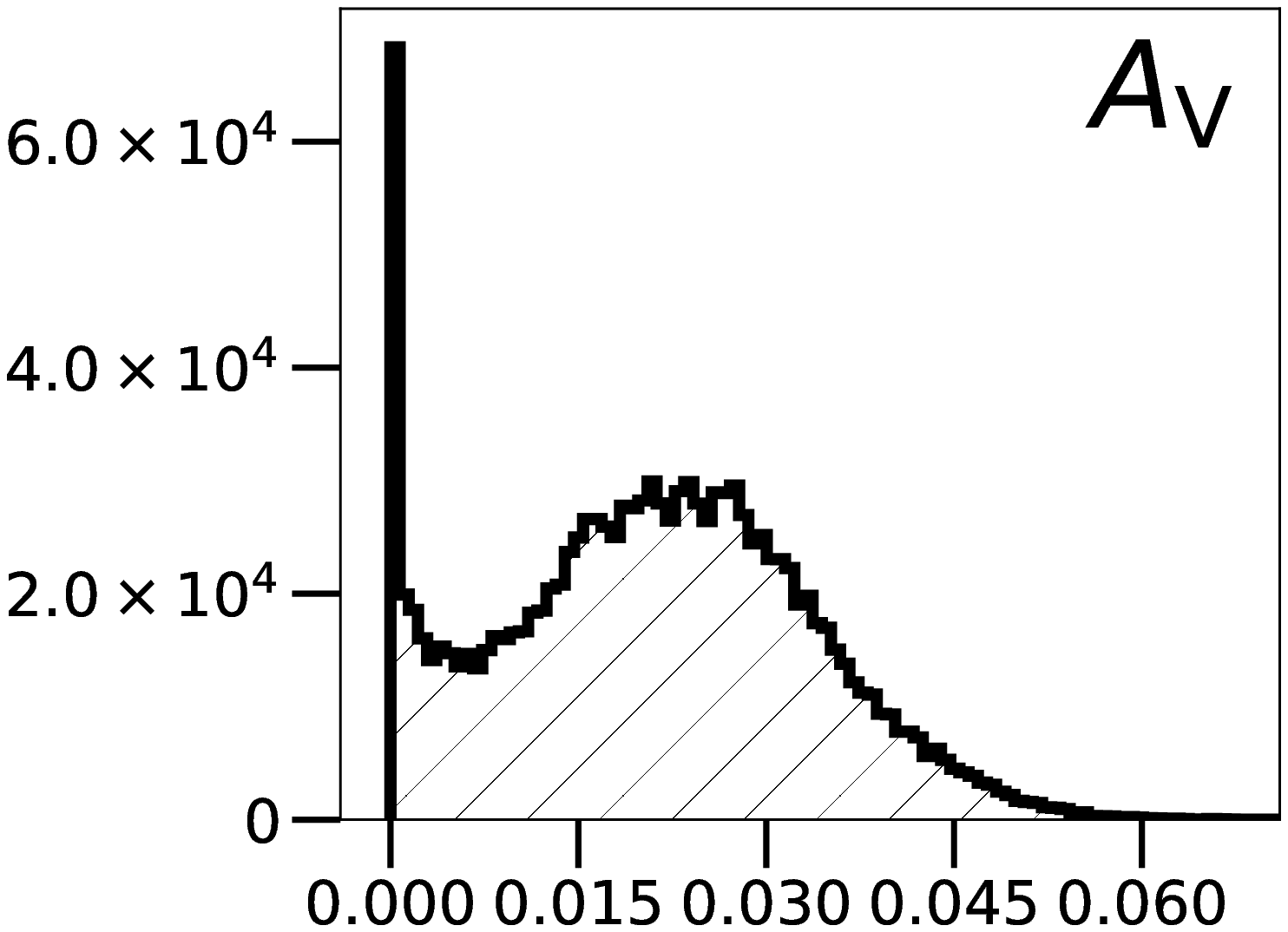} &
 \includegraphics[width=0.23\textwidth]{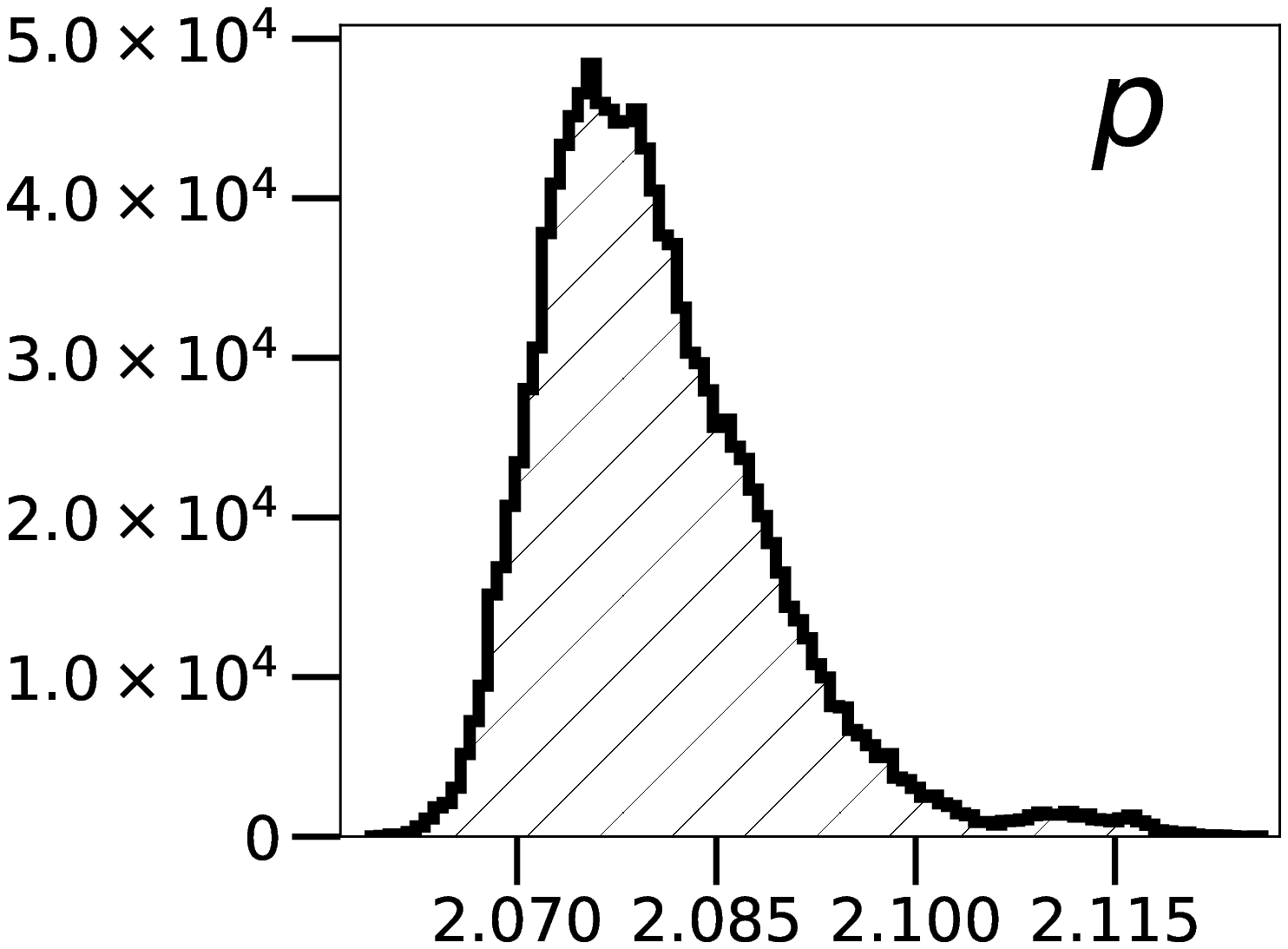} \\
 \includegraphics[width=0.23\textwidth]{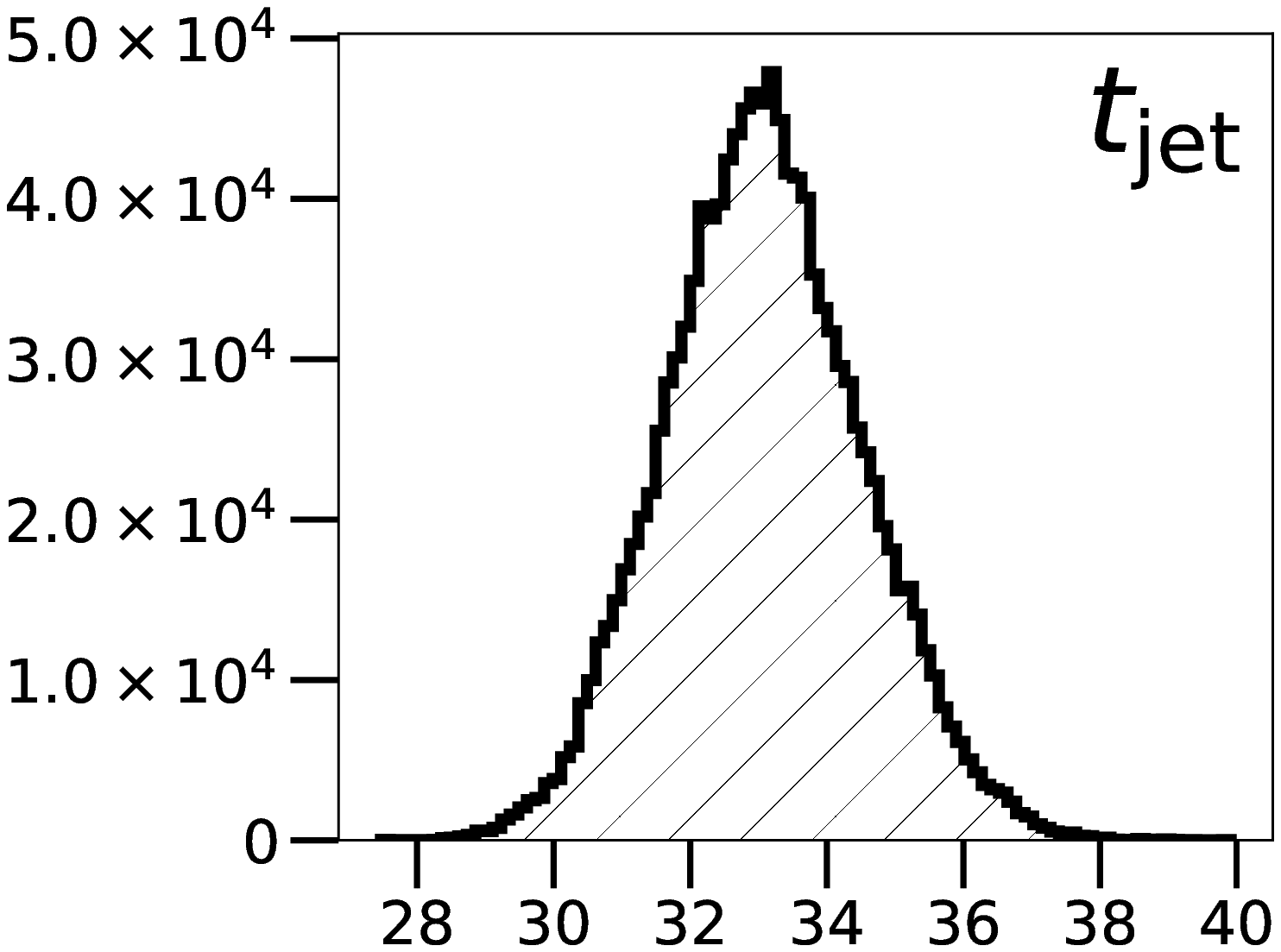} &  
 \includegraphics[width=0.23\textwidth]{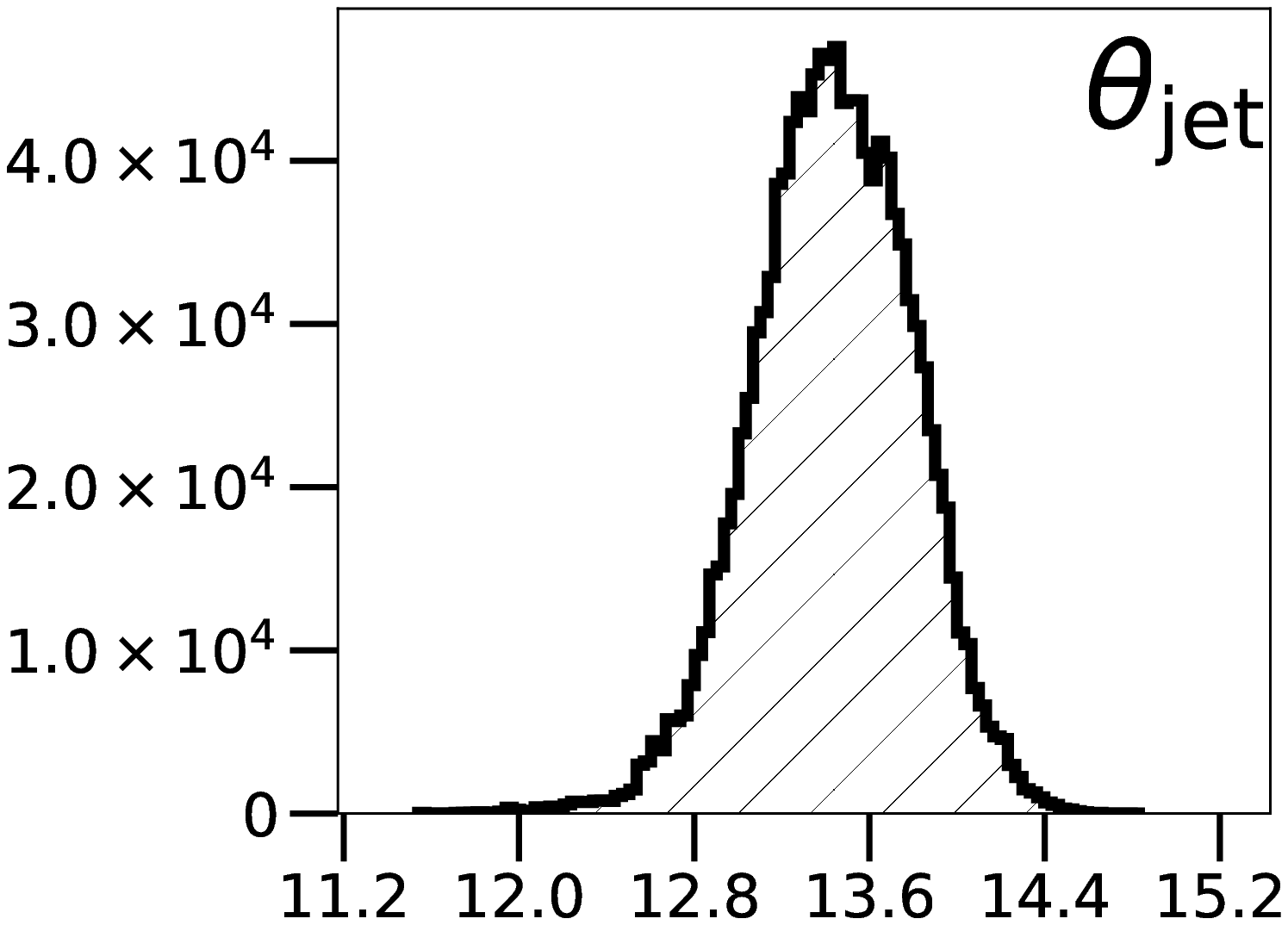} \\
 \includegraphics[width=0.23\textwidth]{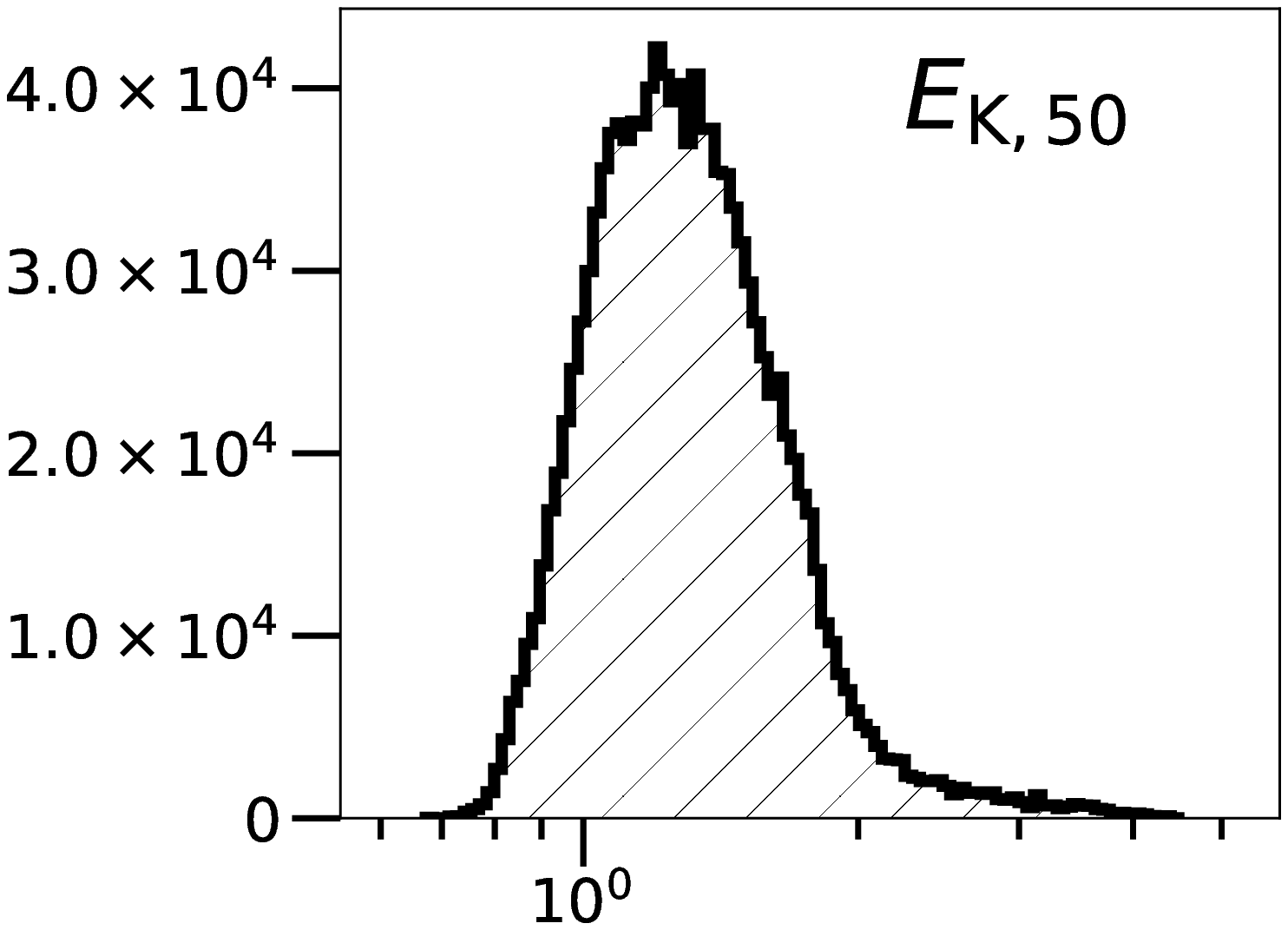} & 
 \includegraphics[width=0.23\textwidth]{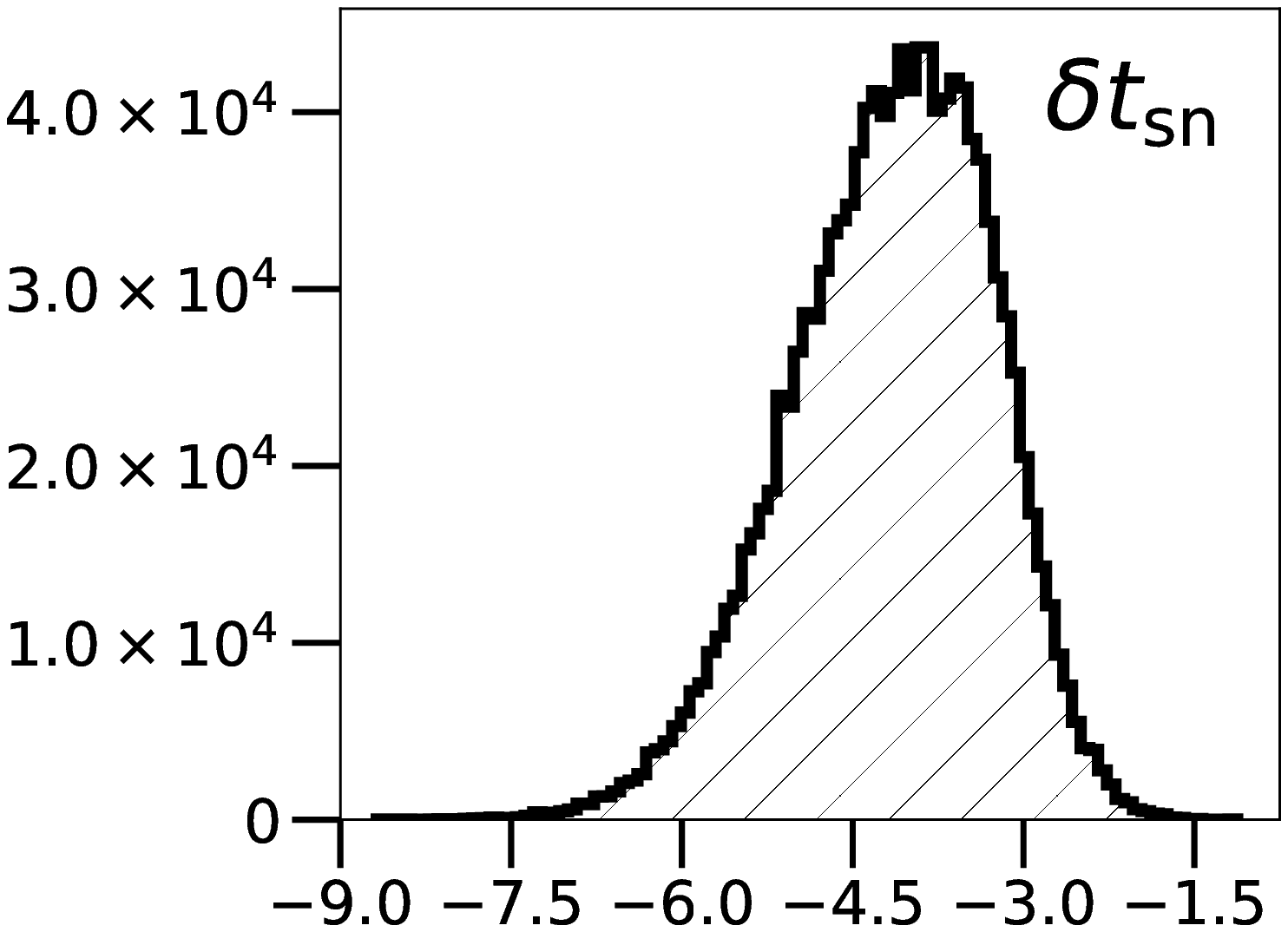} \\
 \includegraphics[width=0.23\textwidth]{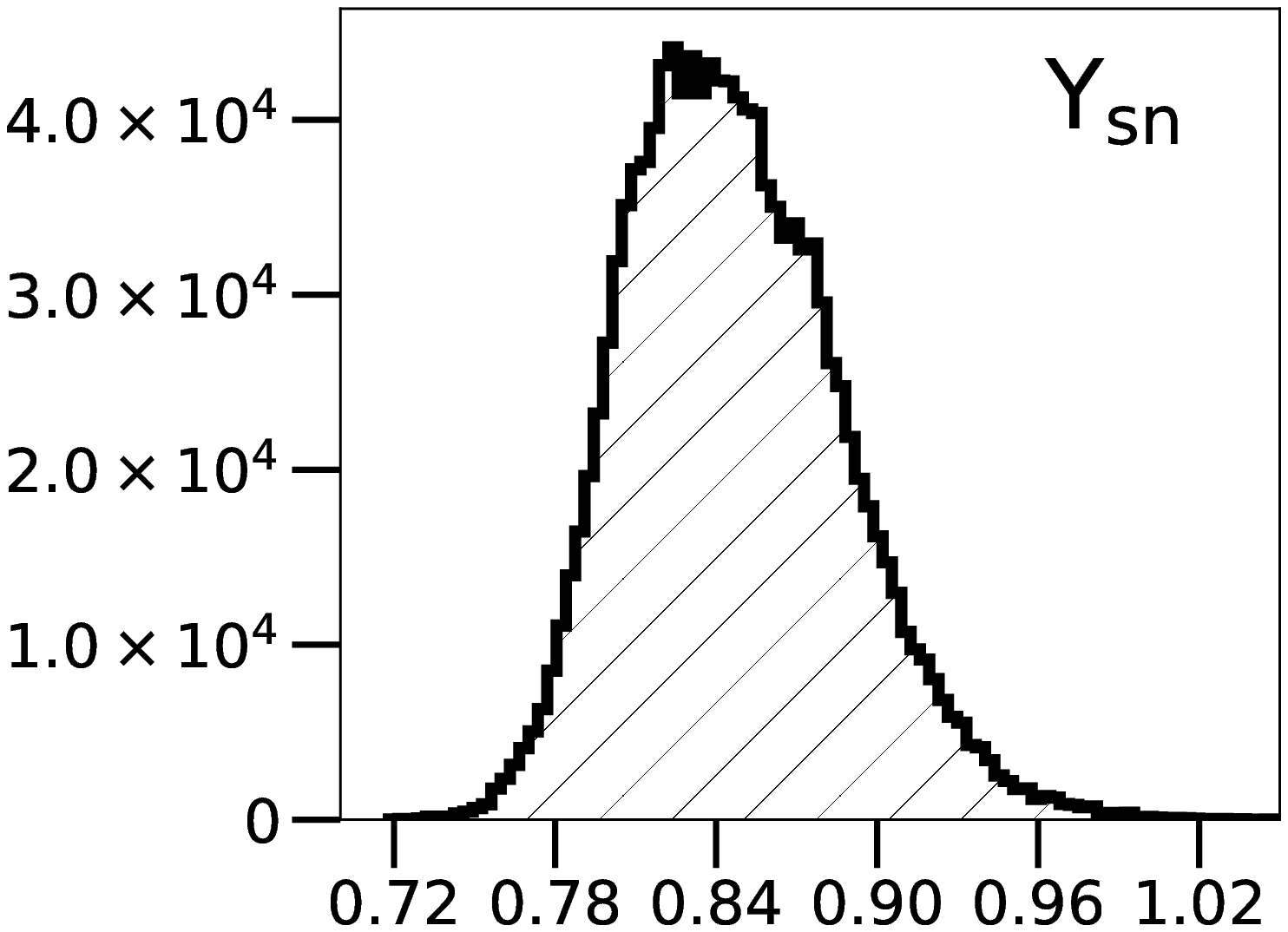} &
 \includegraphics[width=0.23\textwidth]{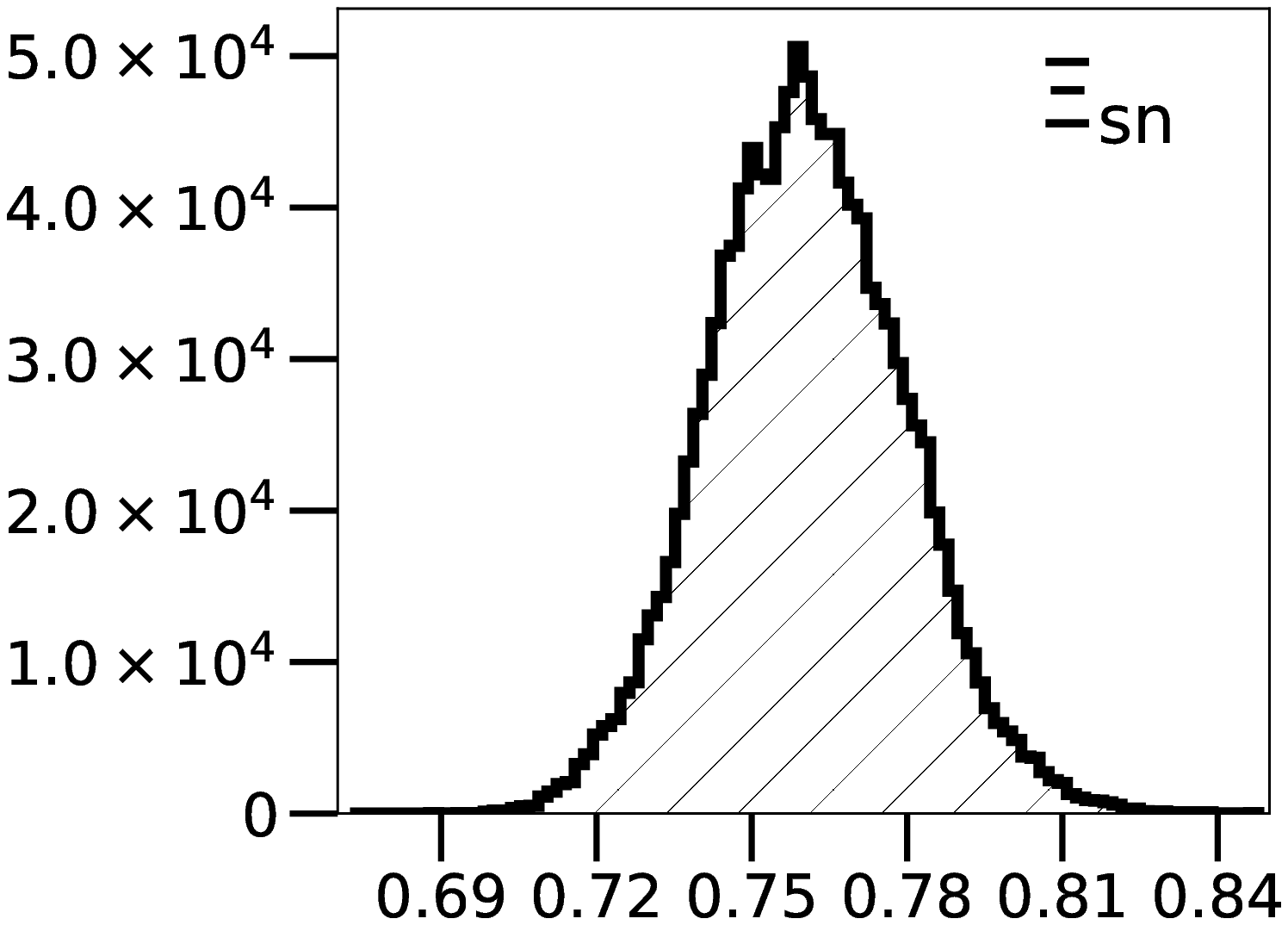} 
\end{tabular}
\caption{Posterior probability density functions for the physical parameters for GRB\,161219B and 
the light curve of SN\,2016jca.\label{fig:hists}}
\end{figure}

\begin{figure}
\begin{tabular}{cc}
\centering
 \includegraphics[width=0.48\columnwidth]{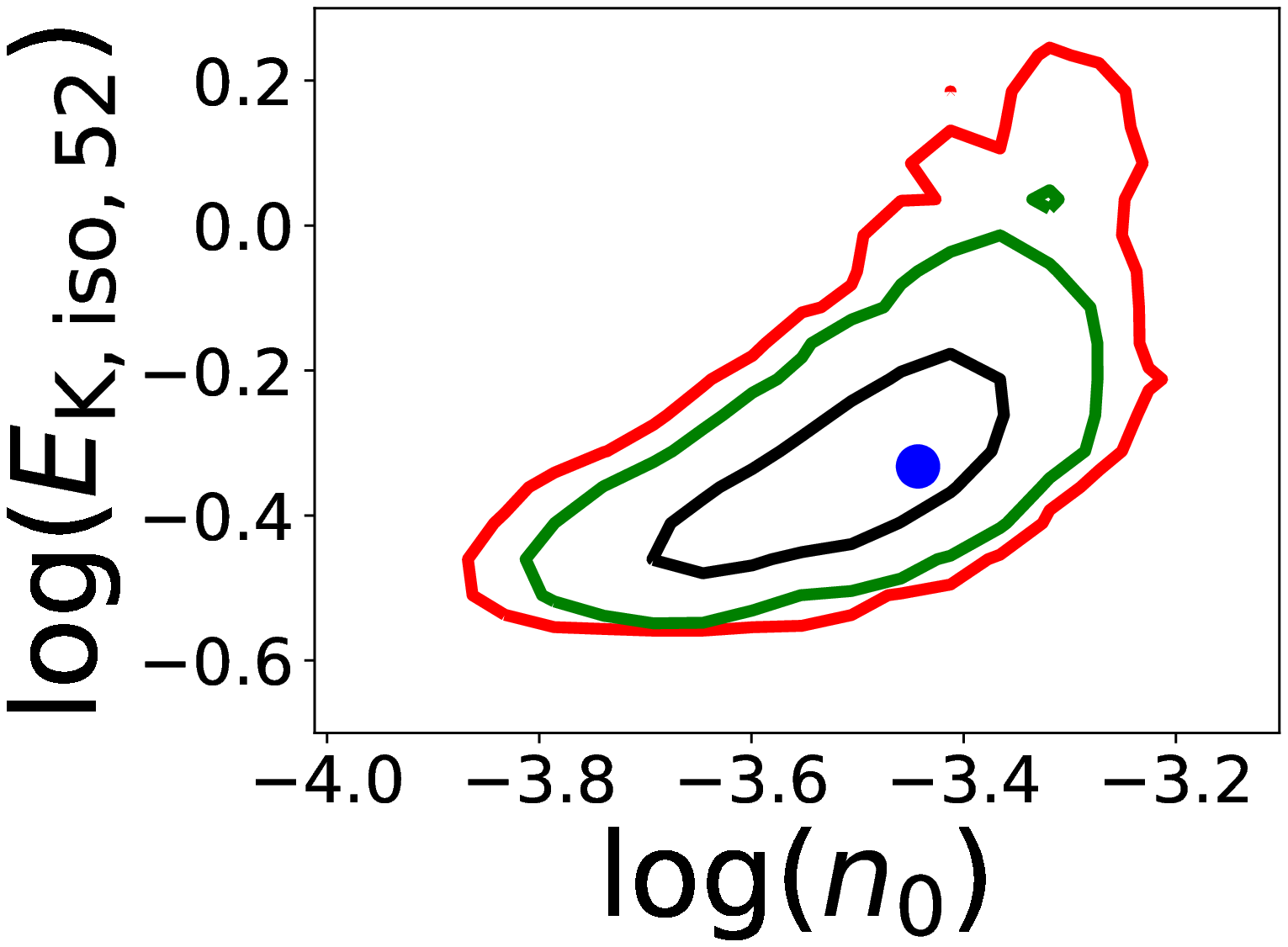} &
 \includegraphics[width=0.48\columnwidth]{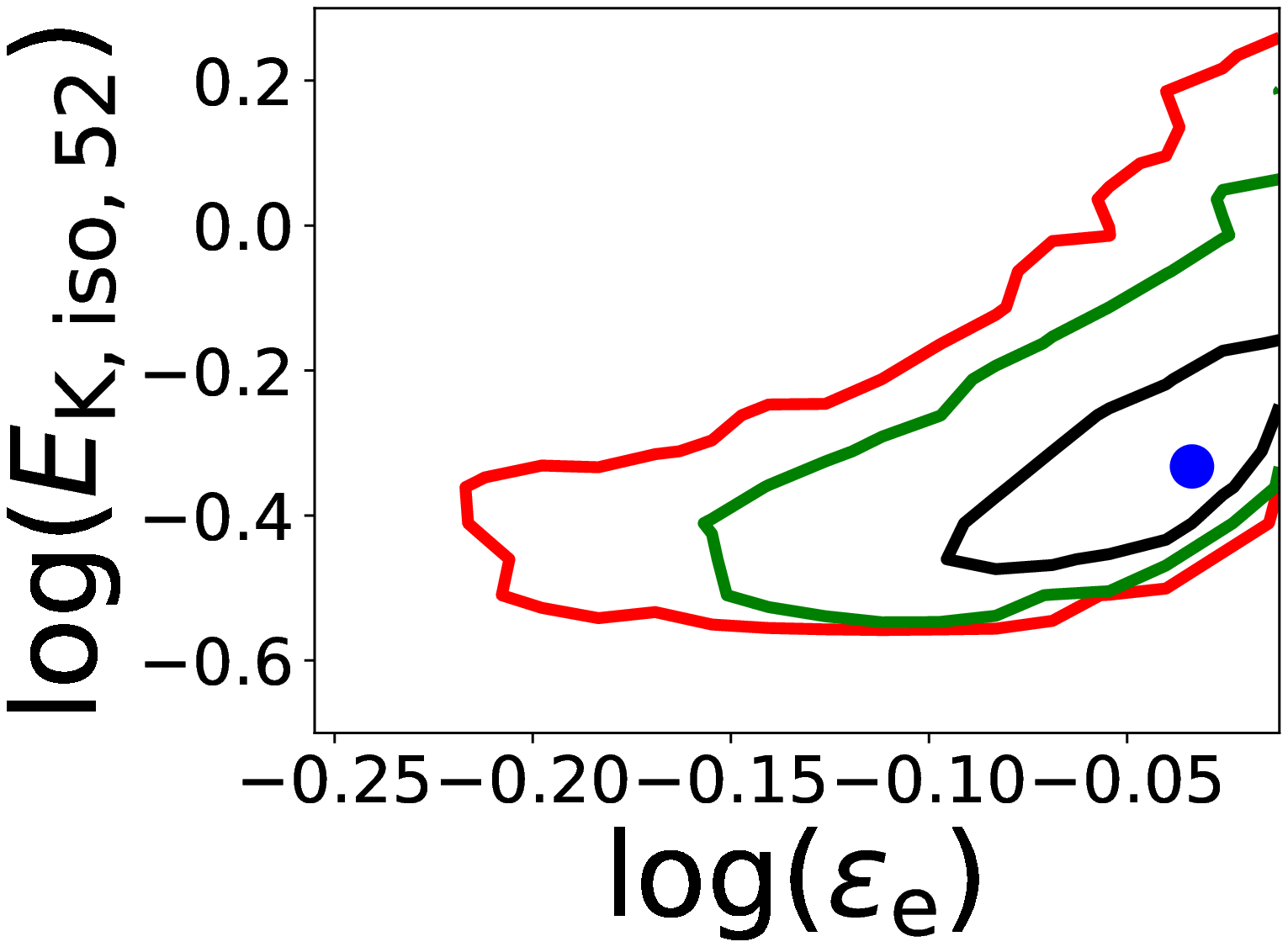} \\
 \includegraphics[width=0.48\columnwidth]{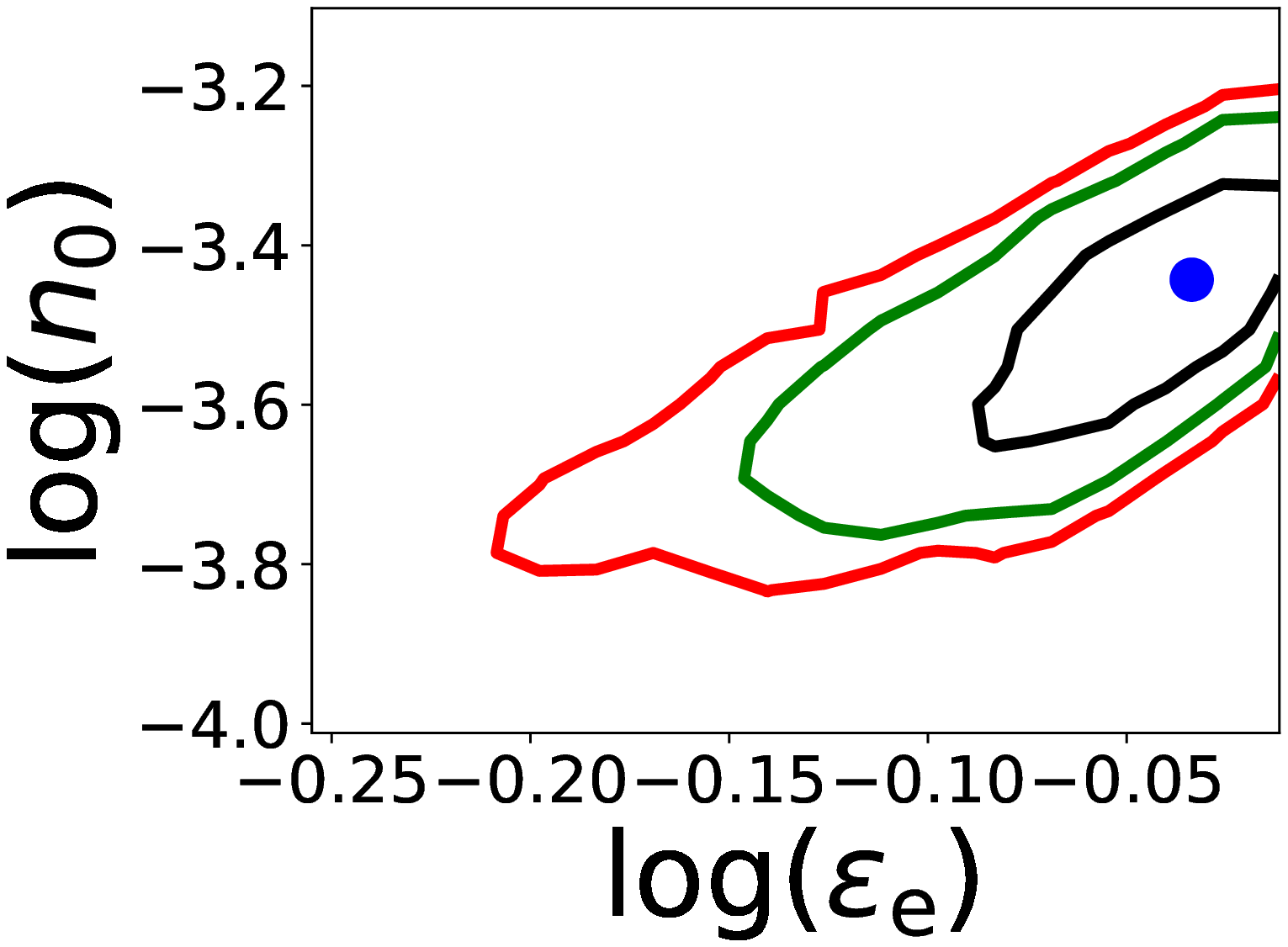} &
 \includegraphics[width=0.48\columnwidth]{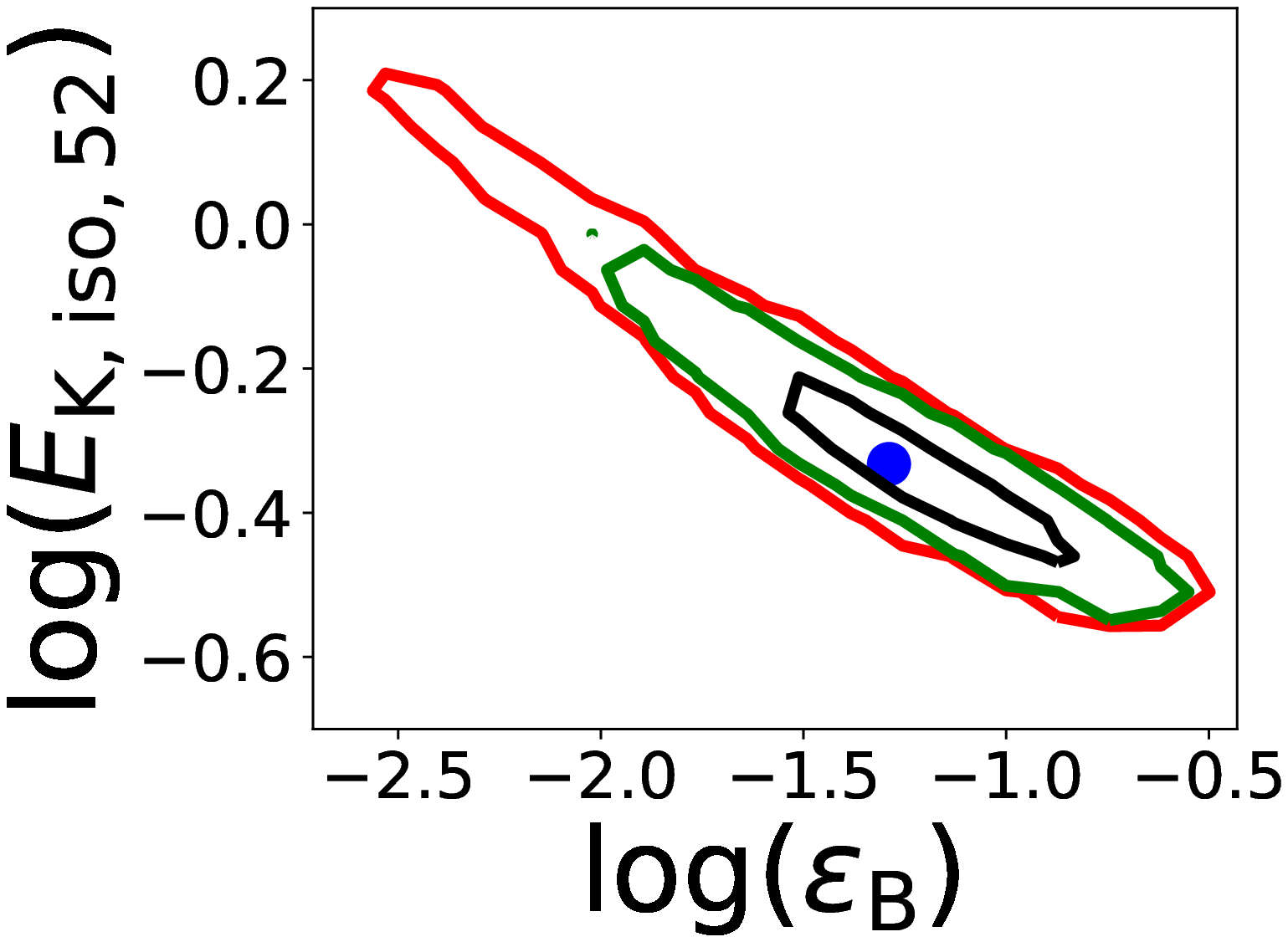} \\
 \includegraphics[width=0.48\columnwidth]{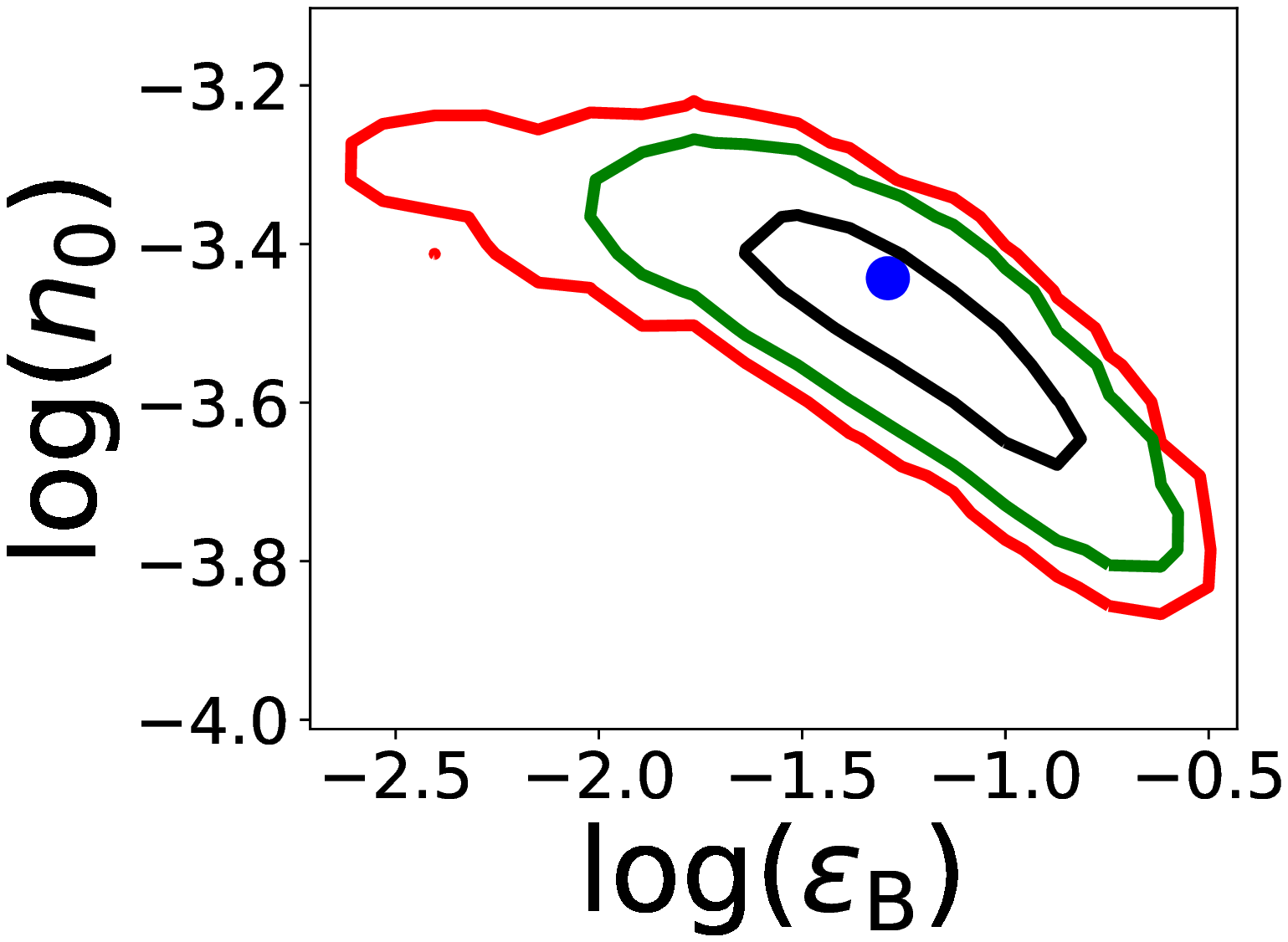} &
 \includegraphics[width=0.48\columnwidth]{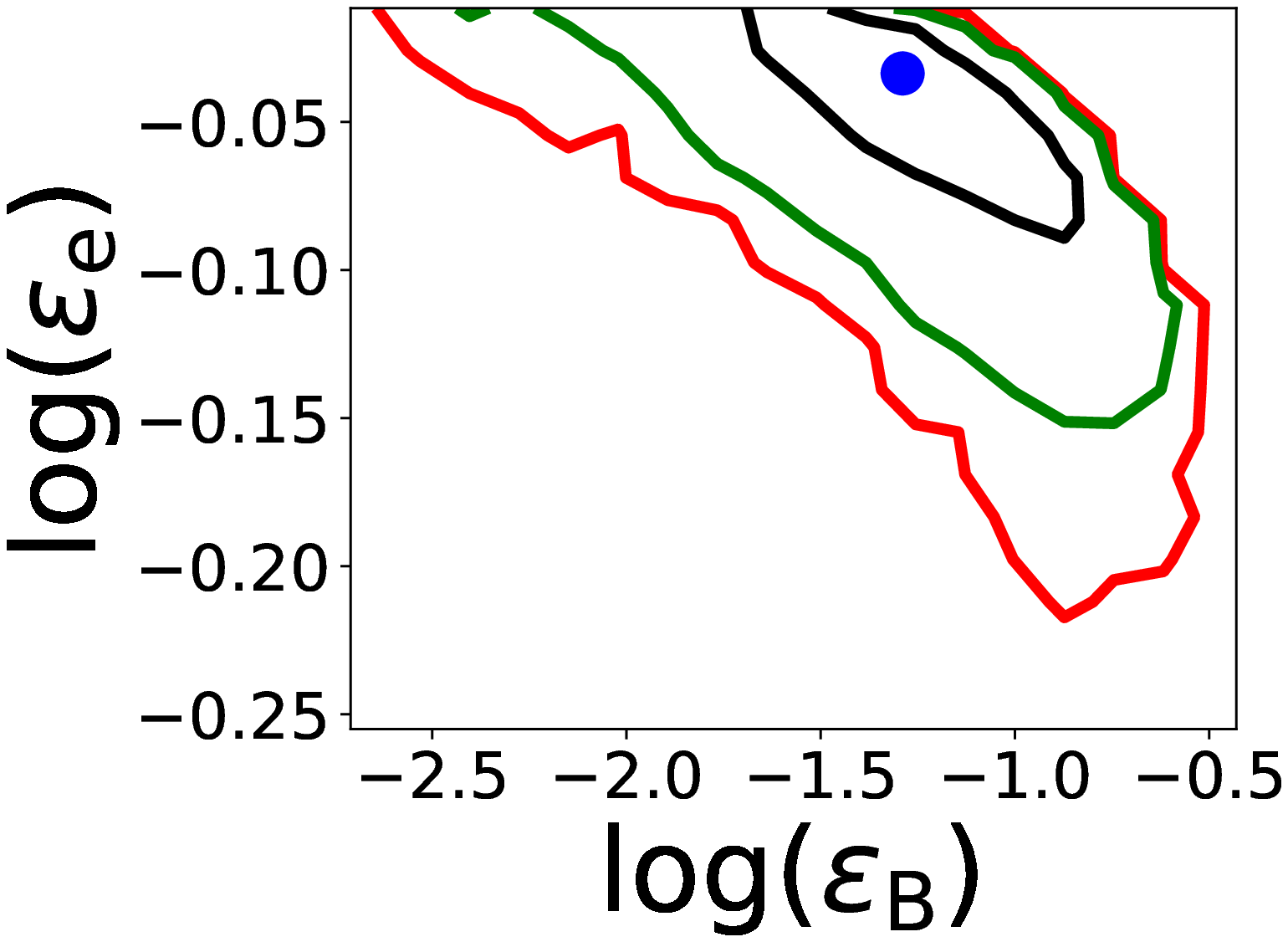} \\
\end{tabular}
\caption{1$\sigma$ (red), 2$\sigma$ (green), and 3$\sigma$ (black) contours for correlations
between the physical parameters, \EKiso, \dens, \epse, and \epsb\ for GRB\,161219B from Monte Carlo 
simulations. We have restricted $\epse+\epsb < 1$. See the on line version of this 
Figure for additional correlation plots. \label{fig:corrplots}}
\end{figure}

This model also requires a jet break at $\tjet\approx32$\,d, corresponding to an opening angle of 
$\approx13^\circ$ \citep{sph99}. The resulting beaming-corrected kinetic and $\gamma$-ray energies 
are $\EK\approx1.3\times10^{50}$\,erg and $\Egamma\approx4.9\times10^{48}$\,erg, respectively. The 
corresponding radiative efficiency is extremely low, $\eta\approx4\%$ (independent of the beaming 
angle). We discuss this further in Section \ref{text:RS}. This break time is later than derived 
from fitting the X-ray light curve alone (Section \ref{text:basic_x}), owing to the steeper 
post-break decline rate in the physical model compared to the simple power law fits of the X-ray 
light curve. The resulting model matches the X-ray data at $\gtrsim0.1$\,d fairly well (Figure 
\ref{fig:modellc_FS_splits}). We note 
the time of the jet break is partly driven by the ALMA light curve, which declines as 
$\alpha=-1.5\pm0.3$ between 24.5 and 78.2\,d, steeper than the expected value of $\approx-0.8$ for 
the ordering $\numax<\nu_{\rm ALMA}<\nuc$ and a spherical, adiabatic shock, as also discussed in 
Section \ref{text:basic_radio}. The resulting model light curve matches the ALMA flux density in the 
final 3 epochs, but underpredicts the mm- and cm-band observations before $\approx8.5$\,d (Figure 
\ref{fig:modelsed_FS}).
The parameters for the best fit model, together with the median and $68\%$ credible intervals from 
the MCMC analysis, are provided in Table \ref{tab:params} and histograms of the marginalized 
posterior density are presented Figure \ref{fig:hists}. We note that $\nua$ is not constrained by 
the data, resulting in some degeneracies between the physical parameters (Figure 
\ref{fig:corrplots}). 

\begin{deluxetable}{lcc}
 \tabletypesize{\footnotesize}
 \tablecolumns{3}
 \tablecaption{Results of multi-wavelength modeling}
 \tablehead{   
           \colhead{Parameter} &
           \colhead{Best-fit}  &
           \colhead{MCMC}
   }
 \startdata   
   \multicolumn{3}{c}{\textit{Forward Shock}} \\   
   \cmidrule{1-3} \\ [-3 pt]
   $p$                  & 2.08 & $2.079^{+0.009}_{-0.006}$ \\ [3 pt]
   \epse                & 0.93 & $0.89^{+0.05}_{-0.07}$\\ [3 pt]
   \epsb                & $5.1\times 10^{-2}$ & $(5.8^{+5.4}_{-3.0})\times10^{-2}$\\ [3 pt]
   \dens (\pcc)         & $3.6\times 10^{-4}$ & $(3.2^{+1.4}_{-1.2})\times10^{-4}$\\ [3 pt]  
 
   $E_{\rm K, iso, 52}$ (erg)
                        & 0.47 & $0.46^{+0.14}_{-0.09}$\\ [3 pt] 
   \tjet\ (d)           & 31.5 & $33.0^{+1.5}_{-1.4}$\\ [3 pt]
   \thetajet\ (deg)     & 13.5 & $13.44\pm0.35$ \\ [3 pt]
   \AV\ (mag)           & $3.0\times 10^{-2}$ & $(2.1^{+2.0}_{-2.1})\times10^{-2}$\\ [3 pt]
   $\EK$ (erg)    & $1.3\times 10^{50}$ & $(1.27^{+0.36}_{-0.25})\times10^{50}$ \\ [3 pt] 
   \cmidrule{1-3} \\ [-3 pt]
   \multicolumn{3}{c}{\textit{Prompt Emission}} \\   
   \cmidrule{1-3} \\ [-3 pt]   
   $E_{\gamma,\rm iso}$ & $(1.8\pm0.4)\times10^{50}$ & \ldots \\ [3 pt]
   $E_{\gamma}$ (erg)   & $4.9\times 10^{48}$ & $(4.9\pm1.9)\times10^{48}$ \\ [3 pt]
   $\etarad$     & $3.7\%$ & \ldots \\ [3 pt]   
   \cmidrule{1-3} \\ [-3 pt]
   \multicolumn{3}{c}{\textit{SN 2016jca}} \\   
   \cmidrule{1-3} \\ [-3 pt]
   $\delta t_{\rm sn, peak}$ (d) & $-3.7$ & $-4.10^{+0.80}_{-0.96}$\\ [3 pt]
   $\Upsilon$           &  0.83 & $0.84\pm0.04$\\ [3 pt]   
   $\Xi_{\rm f}$        & 0.73 & $0.76\pm0.02$\\
   \cmidrule{1-3} \\ [-3 pt]
   \multicolumn{3}{c}{\textit{Mean host contribution ($\mu$Jy)}} \\  
   \cmidrule{1-3} \\ [-3 pt]   
  $uw2$ & 1.64 & $1.57\pm0.34$\\ [3 pt]
  $um2$ & 1.48 & $1.12^{+0.30}_{-0.51}$ \\ [3 pt]
  $uw1$ & 2.35 & $2.39\pm0.27$ \\ [3 pt]
  $uwh$ & 5.03 & $4.90\pm0.35$ \\ [3 pt]
  $uvu$ & 3.61 & $3.26\pm0.39$ \\ [3 pt]
  $u\p$ & 2.92 & $2.77^{+0.59}_{-0.81}$ \\ [3 pt]
  $uvb$ & 8.35 & $8.03^{+0.38}_{-0.69}$ \\ [3 pt]
  $B$   & 3.89 & $3.64\pm0.25$ \\ [3 pt]
  $g\p$ & 9.58 & $9.28\pm0.27$ \\ [3 pt]
  $uvv$ & 6.61 & $5.32\pm0.88$ \\ [3 pt]
  $V$   & 15.3 & $14.0\pm0.8$\\ [3 pt]
  $r\p$ & 13.2 & $12.8\pm0.6$ \\ [3 pt] 
  $R$   & 11.9 & $10.6\pm1.0$ \\ [3 pt]
  $i\p$ & 17.1 & $16.4\pm0.6$\\ [3 pt]
  $I$   & 14.1 & $13.4\pm1.0$ \\ [3 pt]
  $z\p$ & 20.2 & $19.6\pm0.65$ \\ [3 pt]
  \it{UKIRT-J} & 30.6 & $31.5\pm2.1$ \\ [3 pt]
  $J$   & 19.2 & $17.3\pm1.2$\\ [3 pt]
  \it{UKIRT-H} & 28.7 & $29.3\pm2.5$\\ [3 pt]
  $H$   & 19.1 & $18.5\pm1.2$\\ [3 pt]
  $K$   & 17.3 & $18.4\pm2.0$\\ [3 pt]
  \it{UKIRT-K} & 34.5 & $37.5\pm2.5$\\ [3 pt]
  5\,GHz   & 60.2 & $52.0\pm16.4$ \\ [3 pt]
  7.4\,GHz & 42.0 & $41.2\pm13.2$
 \enddata
\label{tab:params}
\end{deluxetable}

\subsection{Supernova}
The supernova (SN 2016jca) associated with this burst has previously been studied by \citet{apm+17} 
and \citet{cidup+17}, who consider both magnetar and radioactive decay models for powering
the SN light curve. \citet{apm+17} argue for the magnetar model with an ejecta mass of 
$M_{\rm sn,ej} \approx 8 M_{\odot}$, despite the high isotropic-equivalent 
ejecta kinetic energy required, $E_{\rm sn,K,iso}\approx5.4\times10^{54}$\,erg.
Their afterglow model used to derive the SN light curve requires $p<2$, while the large
jet opening angle they infer, $\thetajet\approx40^{\circ}$, is based on an assumed 
circumburst density of $\dens\approx1$\pcc, over 3 orders of magnitude larger than the value
obtained here from multi-wavelength modeling. \citet{cidup+17} derive a lower ejecta mass,
$M_{\rm sn,ej}=5.8\pm0.3 M_{\odot}$, and a similar ejecta kinetic energy, 
$E_{\rm sn,K,iso}=(5.1\pm0.8)\times10^{54}$\,erg. Under the assumption that the 
SN light curve is powered by radioactive decay of ${}^{51}$Ni, they find a Nickel mass of
$M_{\rm Ni}=0.22\pm0.08 M_{\odot}$, and $\gamma$-ray opacity,
$\kappa_{\gamma}\approx0.034$\,cm$^2$\,g$^{-1}$. 
Our method, which assumes the same color evolution as the template, yields a 
stretch factor of $\Upsilon_{\rm sn}\approx0.8$ and a flux scale 
factor\footnote{These correspond to the parameters $k$ and $s$ of \citet{can14},
respectively. We use different symbols in this work to avoid confusion with 
the the ejecta Lorentz factor distribution (equation \ref{eq:sofm}) and
the circumburst medium density profile index.} 
of $\Xi_{\rm sn}\approx 0.8$, within $\approx 1\sigma$ of the correlation between
these parameters derived by \citet{can14}. Whereas our method does not allow us to derive specific 
physical parameters of SN2016jca, our results are broadly consistent with those of 
\citet{cidup+17}, who find (frequency-dependent) stretch factors of $\Upsilon_{\rm 
sn}\approx$0.6--0.9 and $\Xi_{\rm sn}\approx$0.7--0.8.

\subsection{Host galaxy}
\begin{figure}
 \includegraphics[width=\columnwidth]{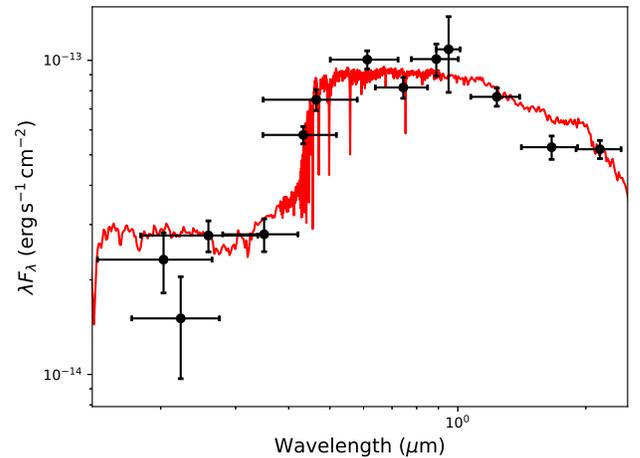}
 \caption{SED of the host of GRB~161219B derived from multi-wavelength modeling (Section 
\ref{text:model}), together with a best-fit model from \cite{bc03}.}
\label{fig:hostsed}
\end{figure}

We derive an SED for the host galaxy using five of the six narrow-band UVOT 
filters\footnote{Photometry in the UVOT $v$-band is most significantly affected by diffracted light 
from the nearby star, and is less reliable than in the other bands at late times. We therefore 
exclude this band from the SED fit.}, together with the pre-explosion PS1 $grizy$ host photometry 
\citep{cidup+17} 
and our $JHK$ data (Figure \ref{fig:hostsed}).  We fit a set of galaxy templates from \cite{bc03} 
using \texttt{FAST} \citep{kvdl+09}, assuming an exponentially declining star-formation history 
($\tau$-model), a \cite{cha03} IMF, and a stellar metallicity of $Z=0.008$ \citep[0.4 solar, 
corresponding to the value for the host obtained from H$\alpha$ and emission line 
diagnostics;][]{cidup+17}. 
Whereas the  extinction and $\tau$ are particularly susceptible to systematic photometric 
uncertainties in the \Swift\ photometry and are poorly constrained by the weak UV detections of the 
host, the stellar mass is well determined, $\log(M_*/M_{\odot})=8.92^{+0.04}_{-0.02}$. 
We derive a stellar population age, 
$\log{t_0}=9.0^{+0.2}_{-0.1}$, $\tau\approx0.3$\,Gyr, rest-frame extinction, 
$\AV=0.6^{+0.2}_{-0.6}$\,mag, and current star-formation rate 
${\rm SFR} = 0.19^{+0.02}_{-0.16}\,M_{\odot}\,{\rm yr}^{-1}$. These values are similar to 
those derived by \cite{cidup+17} using the PS1 photometry alone. The derived stellar mass is 
comparable to the mean stellar mass of GRB hosts at $z\lesssim1$ 
\citep[$\log(M_*/M_{\odot})=9.25^{+0.19}_{-0.23}$; ][]{lkbjz10}. On the other hand, the specific 
SFR, $\log{[\rm sSFR/Gyr^{-1}]}\approx-0.65$ appears an order of magnitude lower than the median 
sSFR of GRB hosts at $z\lesssim1$ \citep[$\log{[\rm 
sSFR/Gyr^{-1}]}\approx0.3$;][]{lkbjz10}. The possibility that the GRB occurred in an extreme 
star-forming region within an otherwise low sSFR host is disfavored by HST spectroscopy of the 
supernova site \citep{cidup+17}. Dust extinction may impact the derived SFR by extinguishing the 
light from young stars, especially in an edge-on system like the host of GRB~161219B; however, we 
derive a low extinction from afterglow modeling, consistent with the host SED fits. Since 
long-duration GRBs are typically associated with regions of the most intense 
star formation in their hosts \citep{bkd02,fls+06,slt+10,bbf16}, the lack of evidence for strong 
star-formation activity at the GRB site is puzzling.

\section{Energy injection}
\label{text:enj}
The optical and X-ray light curves exhibit an unusual achromatic break at $\approx0.1$\,d, which 
cannot be explained in the standard synchrotron framework. 
Furthermore, the model described in Section \ref{text:model} over-predicts the X-ray light curve 
before $\approx 0.1$\,d (Figure \ref{fig:modellc_FS_splits}). One of the simplest means to obtain a 
flatter light curve at earlier times 
is through the injection of energy into the forward shock due to extended activity of the central 
engine, the deceleration of a Poynting flux dominated outflow, or stratified ejecta with additional 
energy available at lower Lorentz factors 
\citep{dl98,rm98,kp00,sm00,zm01,zm02,gk06,zfd+06,dsg+11,uzh+12}. The effect on the FS is a gradual 
increase in the effective shock energy with time, $E\propto t^m\propto t^{1-q}$, where 
$-q\equiv m-1$ is the power law index of the injection luminosity, $L\propto t^{-q}$ 
\citep{zfd+06}. For energy injection due to accumulation from a distribution of ejecta Lorentz 
factors, the corresponding ejecta energy distribution is given by $E(>\Gamma)\propto \Gamma^{1-s}$, 
with 
\begin{equation}
\label{eq:sofm}
s = \frac{(7m+3)-k(2m+1)}{(3-k)-m}, 
\end{equation}
where the external density profile as a function of radius, $R$, is assumed to follow the 
general\footnote{We keep the discussion here general for completeness, and specialize to the ISM 
($k=0$) case later.} power law form, $\rho=AR^{-k}$. During this process, the FS Lorentz factor, 
shock radius, and post-shock 
magnetic field all evolve more slowly than the standard relativistic solution 
\citep{bm76,sm00,zfd+06},
\begin{align}
 \frac{\partial\ln\Gamma}{\partial\ln t}&=-\frac{q+2-k}{2(4-k)} = -\frac{3-k}{7+s-k}, \nonumber \\
 \frac{\partial\ln R}{\partial\ln t}&=  \frac{2-q}{4-k}         = \frac{1+s}{7+s-2k}, \nonumber \\
 \frac{\partial\ln B}{\partial\ln t}&= -\frac{q+k+2-kq}{2(4-k)} = -\frac{6+ks-k}{2(7+s-2k)},
\end{align}
with $s$ and $q$ related by
\begin{align}
 s &= \frac{10-3k-7q+2kq}{2+q-k},\nonumber \\
 q &= \frac{10-2s-3k+ks}{7+s-2k}.
\end{align}
The standard hydrodynamic evolution in the absence of energy injection can be recovered by setting 
$m=0$, $s=1$ or $q=1$ in the above expressions \citep[e.g.,][]{glz+13}. 

In our best-fit model, 
$\nux<\nuc$ at $\lesssim 0.1$\,d, whereupon $m=(4\alpha_{\rm X}+3p-3)/(p+3) = 0.35\pm0.09$ using 
$\alpha_{\rm X} = -0.37\pm0.09$ (Section \ref{text:basic_x}), which implies $s \approx 2$ for $k=0$ 
(equation \ref{eq:sofm}) in the massive ejecta model. 
No theoretical models yet exist of the expected distribution of ejecta Lorentz factors, and in fact 
the distribution need not follow a power law. However, our observations of energy injection in this 
event add to the growing collection of a measurement of $s$ in GRB jets \citep{lbm+15}.

We note that the forward shock cooling frequency, $\nu_{\rm c,f}\propto 
E^{-1/2}t^{-1/2}\propto t^{-(m+1)/2}\sim t^{-0.65}$. Thus, in our model $\nu_{\rm c,f}$ evolves from 
$\approx 3\times10^{18}$\,Hz to $\approx7\times10^{17}$\,Hz between the end of the flare at 
$\approx0.01$\,d and the end of energy injection at $\approx0.1$\,d. The presence of $\nu_{\rm 
c,f}$ within the \Swift\ X-ray band explains the observed X-ray spectral index, $\beta_{\rm 
X}\approx-0.86$, which is intermediate between $(1-p)/2\approx-0.54$ and $-p/2\approx-1.04$.

Since the peak flux density is $\approx0.5$\,mJy and $f_{\nu,\rm m}$ is constant in an ISM 
environment, a measured flux density greater than this value at any frequency and time cannot be 
explained by FS emission. Thus, as we previously argued, the optical light curve before 
$\approx3\times10^{-2}$\,d and the radio SEDs before $\approx8.5$\,d must be dominated by a 
distinct emission component. Whereas energy injection can explain the relatively flat X-ray light 
curve before $\approx 0.1$\,d, adding this to our model further worsens the fit to the optical 
light curves at that time. We address both concerns in the next section.

\section{Reverse shock}
\label{text:RS}
During the process of energy injection, a reverse shock mediates the transfer of energy from the 
ejecta into the FS. This RS, which is Newtonian or mildly relativistic, propagates for the period 
of the injection and (by definition) crosses the ejecta at the time ($\tE$) when energy injection 
terminates \citep{rm98,kp00a,zkm03}. Such a ``long-lasting'' RS 
propagating into the ejecta released during the GRB may produce detectable 
synchrotron radiation \citep{sm00,uhm11}. We now show that such an RS can reproduce the observed 
excess in both the optical light curves at $\lesssim3\times10^{-2}$\,d and the radio SEDs 
at $\lesssim8.5$\,d, beginning first with the theoretical model (Section \ref{text:RS_theory}), 
followed by the results from our data (Section \ref{text:RS_obs}), and consistency checks 
between theory and observations (Section \ref{text:RS_selfconsistency}).

\subsection{Energy injection RS -- theoretical prescription}
\label{text:RS_theory}
A detailed calculation of the hydrodynamics of the double shock system requires numerical 
simulations or semi-analytic modeling. Here, we follow previous analytic work \cite{sm00,uhm11} and 
make the simplifying assumption that the pressure behind the RS is equal to that at the FS, 
$P\propto \Gamma^2\rho$ (however, see 
\citealt{uzh+12} for a discussion of situations where this assumption is relaxed). The 
characteristic frequency, cooling frequency, and peak flux density of the radiation 
from the RS and FS are then related during the shock crossing ($t<\tE$) by
\begin{align}
 \frac{\numr}{\numf} &\sim \Gamma^{-2}\RB\RE^2, \nonumber \\
 \frac\nucr\nucf &\sim \RB^{-3} \left(\frac{1+Y_{\rm f}}{1+Y_{\rm r}}\right)^2, \nonumber \\
 \frac\fnumr\fnumf &\sim \Gamma\RB,
\end{align}
where $\Gamma$ is the Lorentz factor of the FS, $\RB\equiv(\epsilon_{\rm B,r}/\epsilon_{\rm 
B,f})^{1/2}$ is the ejecta magnetization, $Y_{\rm r}$ and $Y_{\rm f}$ are the Compton 
$Y$-parameters for the RS and FS, respectively, and $\RE\equiv\overline{\epsilon}_{\rm 
e,r}/\overline{\epsilon}_{\rm e,f}$, with $\overline{\epsilon}_{\rm e} \equiv (p-2)\epsilon_{\rm 
e}/(p-1)$ \citep{zfd+06}. We assume the same value of $p$ for both the RS and FS, so that 
$\RE=\epsilon_{\rm e,r}/\epsilon_{\rm e,f}$. As for the FS, we assume that the microphysical 
parameters of the RS (and hence $\RE$ and $\RB$) remain constant with time. Thus, the RS spectral 
parameters are directly proportional to those of the FS during shock crossing:
\begin{align}
\label{eq:fsrs_prop}
 \numr &\propto \Gamma^{-2}\numf, \nonumber \\
 \nucr &\propto \nucf, \nonumber \\
 \fnumr &\propto \Gamma\fnumf.
\end{align}

The number of electrons swept up by the FS (prior to the jet break) is given by $N_{\rm e,f} 
\propto R^3\rho\propto R^{3-k}$. Since $\numf\propto \Gamma\gammae^2 B$, 
$\nucf\propto\Gamma^{-1}B^{-3}t^{-2}$, and $\fnumf\propto N_{\rm e,f}B\Gamma$, while the minimum 
Lorentz factor of accelerated electrons, $\gammae\propto\Gamma$, the spectral 
parameters of the FS at $t<\tE$ are \citep{zfd+06},
\begin{align}
\label{eq:fs_synch_enj}
 \frac{\partial \ln\numf}{\partial \ln t}  &= -\frac{q+2}{2}, \nonumber \\
 \frac{\partial \ln\nucf}{\partial \ln t}  &=  \frac{(3k-4)(2-q)}{2(4-k)}, \nonumber \\
 \frac{\partial \ln\fnumf}{\partial \ln t} &=  \frac{3kq-4k-8q+8}{2(4-k)}.
\end{align}
These equations reduce to the standard results in the absence of energy injection ($q=1$), and can 
also be recovered by setting $E\propto t^{m}$ in the expressions given by \cite{gs02}.
Combining equations \ref{eq:fsrs_prop} and \ref{eq:fs_synch_enj}, the spectral parameters of the RS 
at $t<\tE$ are,
\begin{align}
\label{eqn:rsfsrelation}
 \frac{\partial \ln\numr}{\partial \ln t}  &= -\frac{2q-kq+4}{2(4-k)}, \nonumber \\
 \frac{\partial \ln\nucr}{\partial \ln t}  &=  \frac{(3k-4)(2-q)}{2(4-k)}, \nonumber \\
 \frac{\partial \ln\fnumr}{\partial \ln t} &=  \frac{3(kq-k-3q+2)}{2(4-k)}, 
\end{align}
which yield the expressions of \cite{sm00} for $k=0$.

The evolution of the RS self-absorption frequency during energy injection is more complex, and 
depends on the relative ordering of \nuar, \numr, and \nucr. When both the RS and FS are in the 
slow cooling regime ($\numr<\nucr$ and $\numf<\nucf$), we expect $\nuar\propto\Gamma^{8/5}\nuaf$ 
at $t<\tE$ \citep{sm00}, so that 
\begin{align}
 \frac{\partial \ln\numr}{\partial \ln t} &= -\frac85\left[\frac{q+2-k}{2(4-k)}\right]
 + \frac{\partial \ln\numf}{\partial \ln t}
\end{align}
which equals $-\frac{q+2}{5}$ (slower than the evolution of $\numr$) for the ISM case. We later 
show (Section \ref{text:RS_obs}) that $\numr\approx\nuar$ at $\approx1.4$\,d, so that $\nuar$ does 
not affect the light curve at any observed frequency prior to the end of energy injection at \tE. 
We therefore ignore self-absorption in the RS prior to \tE.

After injection ends, the residual RS spectrum fades according to the standard RS prescription 
\citep{kob00,zwd05}. The evolution of \nuar, \numr, \nucr, and \fnumr\ at $t>\tE$ depend on whether 
the RS was Newtonian or relativistic. For a relativistic RS, no additional parameters are 
necessary, while for a Newtonian RS, we follow \cite{ks00} in parameterizing the evolution of the 
ejecta Lorentz factor as $\Gamma\propto R^{-g}$. Since the shocked ejecta lag the FS and the FS 
Lorentz factor evolves with radius as $t^{(3-k)/2}$, we expect $g > (3-k)/2$. On the other hand, 
the fluid Lorentz factor in the adiabatic \cite{bm76} solution evolves as $\gamma_{\rm 
f}\propto t^{(2k-7)/2}$ \citep{wdhl03}; since a Newtonian RS does not decelerate the ejecta 
effectively, its Lorentz factor is expected to evolve with radius slower than the Blandford-McKee 
solution. Thus $(3-k)/2 \le g \le (2k-7)/2$, or $3/2 \le g \le -7/2$ for the ISM environment and 
$1/2 \le g \le 3/2$ in the wind case.

Using numerical simulations, \cite{ks00} found $g\approx2$ for a standard Newtonian RS not 
associated with energy injection in the ISM environment, and $g\approx3$ for a relativistic RS. 
Recent observations of radio afterglows have constrained $g\approx5$ for GRB\,130427A 
\citep{lbz+13,pcc+14} in the wind environment (outside the canonical range) and $g\approx2$ for 
GRB\,160509A in an ISM environment \citep{lab+16}. 
We consider both the relativistic and Newtonian prescriptions for evolution at $t>\tE$ in our 
analysis, discussing self-consistency in Section \ref{text:RS_selfconsistency}. Following 
the jet break, the evolution of \fnumr\ steepens further by a factor of $\Gamma^2$ due 
to geometric effects \citep{rho99,dcrgl12,gp12,lab+16} and we include this in our modeling.

\subsection{Energy injection RS -- observational constraints}
\label{text:RS_obs}
A detailed study of spectro-temporal variability in the radio afterglow in our companion 
paper, ALB18, indicates the variability peaks at 10--30\,GHz at $t\lesssim8.5$\,d, but is minimal 
at lower frequencies and in the ALMA bands (Figure \ref{fig:var}). We therefore anchor our RS 
model to the LSC bands at cm wavelengths ($1\sim10$\,GHz) and to the ALMA bands at $\approx 
100$\,GHz. From the observed cm- to mm-band SEDs (Figure \ref{fig:modelsed_FS}), we require the RS 
spectral peak, $\numr\approx 10$\,GHz at 1.4\,d, with $\fnumr\approx4$\,mJy. Since the flux density 
in the first epoch at $\approx0.5$\,d is $\lesssim 1$\,mJy at all bands and the RS light curves 
rise below $\nuar$ and fade below $\numr$, we expect $\nuar$ to be located at $1$--$10$\,GHz at 0.5 
and 1.4\,d to explain the observed brightening from between the first two epochs. Since the mm-band 
data are brighter than the prediction from the FS at both 1.4\,d and 3.4\,d, the RS must contribute 
some flux at those frequencies, and hence $\nucr \gtrsim 100$\,GHz at 3.4\,d. On the other hand, we 
require $\nucr\lesssim\nu_{\rm opt}$ at $\approx 10^{-2}$\,d so as to not over-predict the 
UV/optical light curves before $\approx0.1$\,d, implying $\partial\ln\nucr/\partial\ln t\gtrsim-2$. 
Thus the RS break frequencies should be ordered as $\nuar\lesssim\numr<\nucr$ at $\approx1.4$\,d.
This is challenging to achieve with highly relativistic RS models, for which $\nucr\propto 
t^{-15/8}$ only marginally satisfies the above condition. Upon detailed consideration, no 
relativistic RS models are able to reproduce the observations, and we focus in the rest of this 
section on models involving Newtonian or mildly relativistic shocks. 

From the energy injection model in Section \ref{text:enj}, 
$m\approx0.35$, implying $q\approx0.65$ in an ISM environment. 
For this value of $q$, the RS spectrum evolves as 
$\frac{\partial\ln\numr}{\partial\ln t} \approx -0.66$, 
$\frac{\partial\ln\nucr}{\partial\ln t} \approx -0.68$, and
$\frac{\partial\ln\fnumr}{\partial\ln t} \approx 0.02$ at $t<\tE$. Thus the RS peak flux is 
approximately constant during shock crossing. Evolution after shock crossing depends on the 
value of $g$.

\begin{figure*} 
 \begin{tabular}{cc}
  \includegraphics[width=0.47\textwidth]{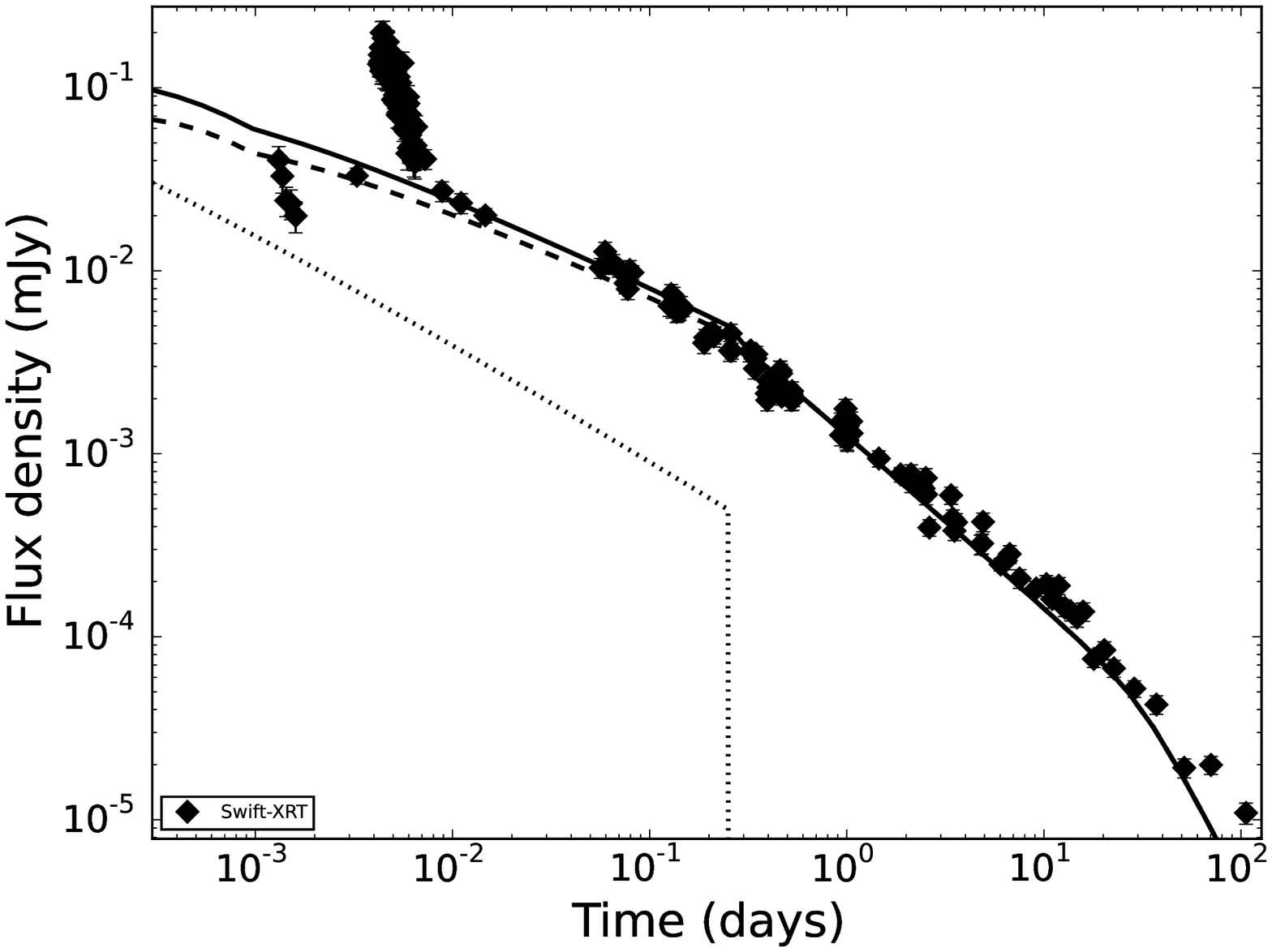} &
  \includegraphics[width=0.47\textwidth]{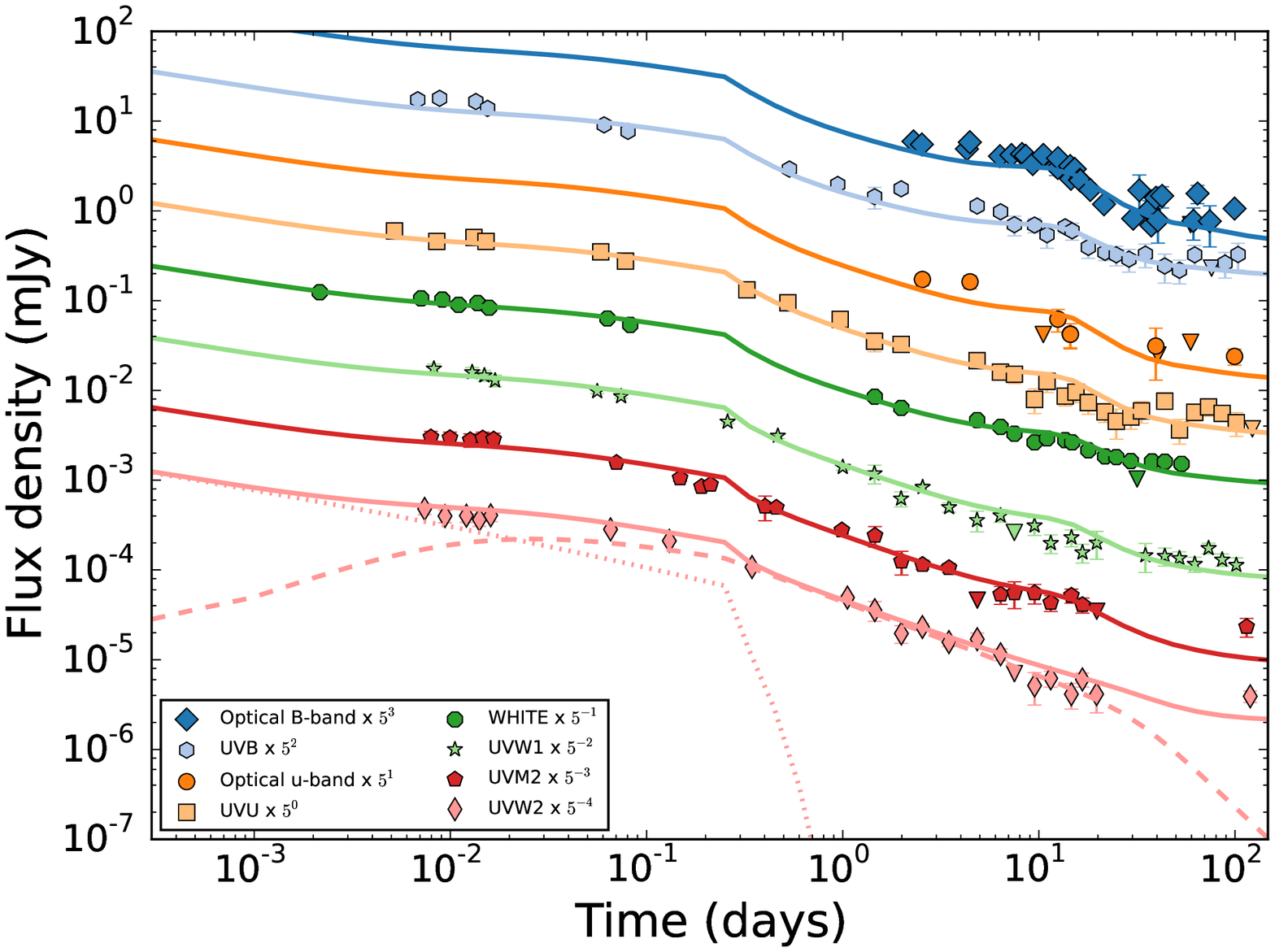} \\
  \includegraphics[width=0.47\textwidth]{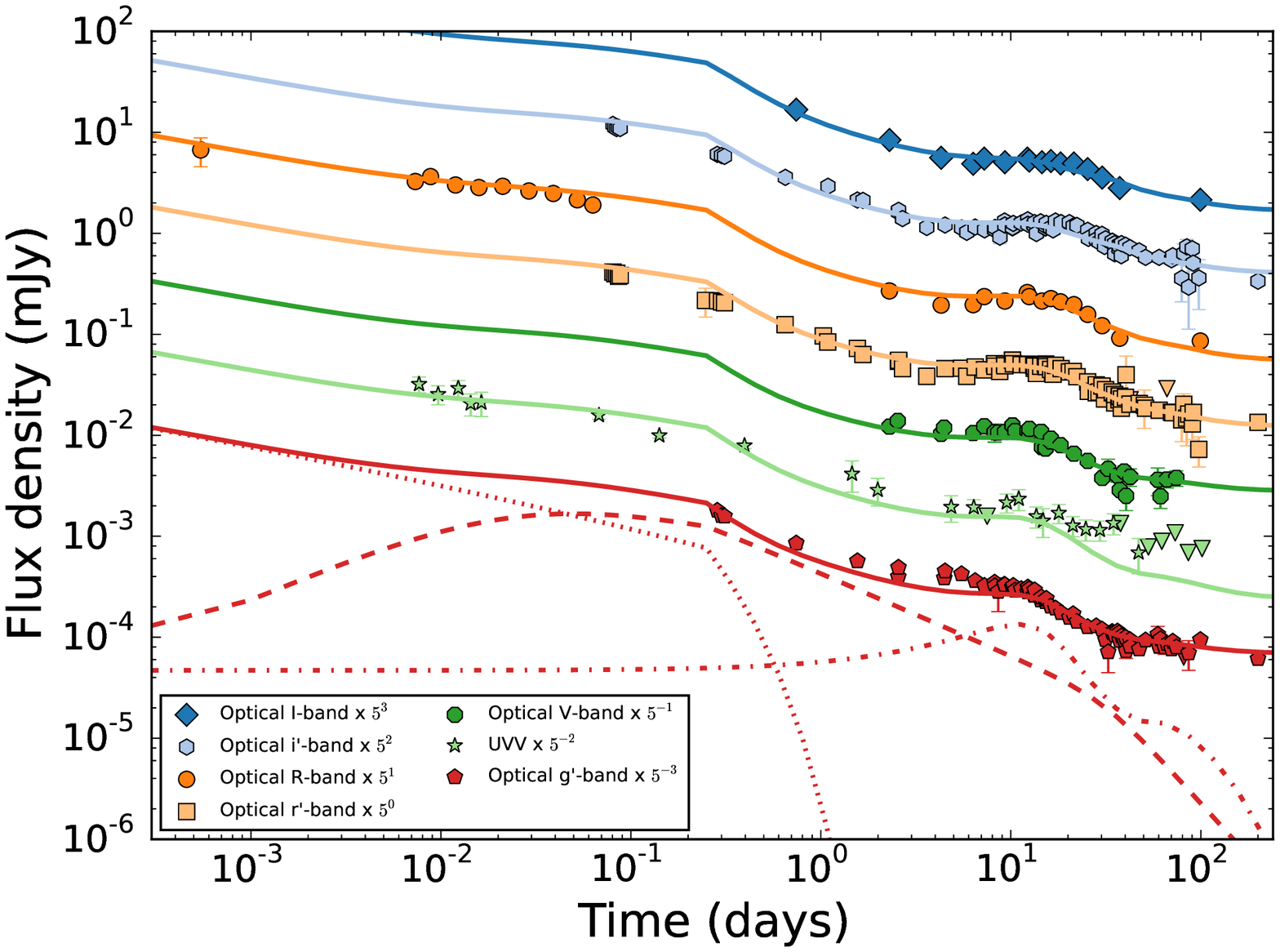} &
  \includegraphics[width=0.47\textwidth]{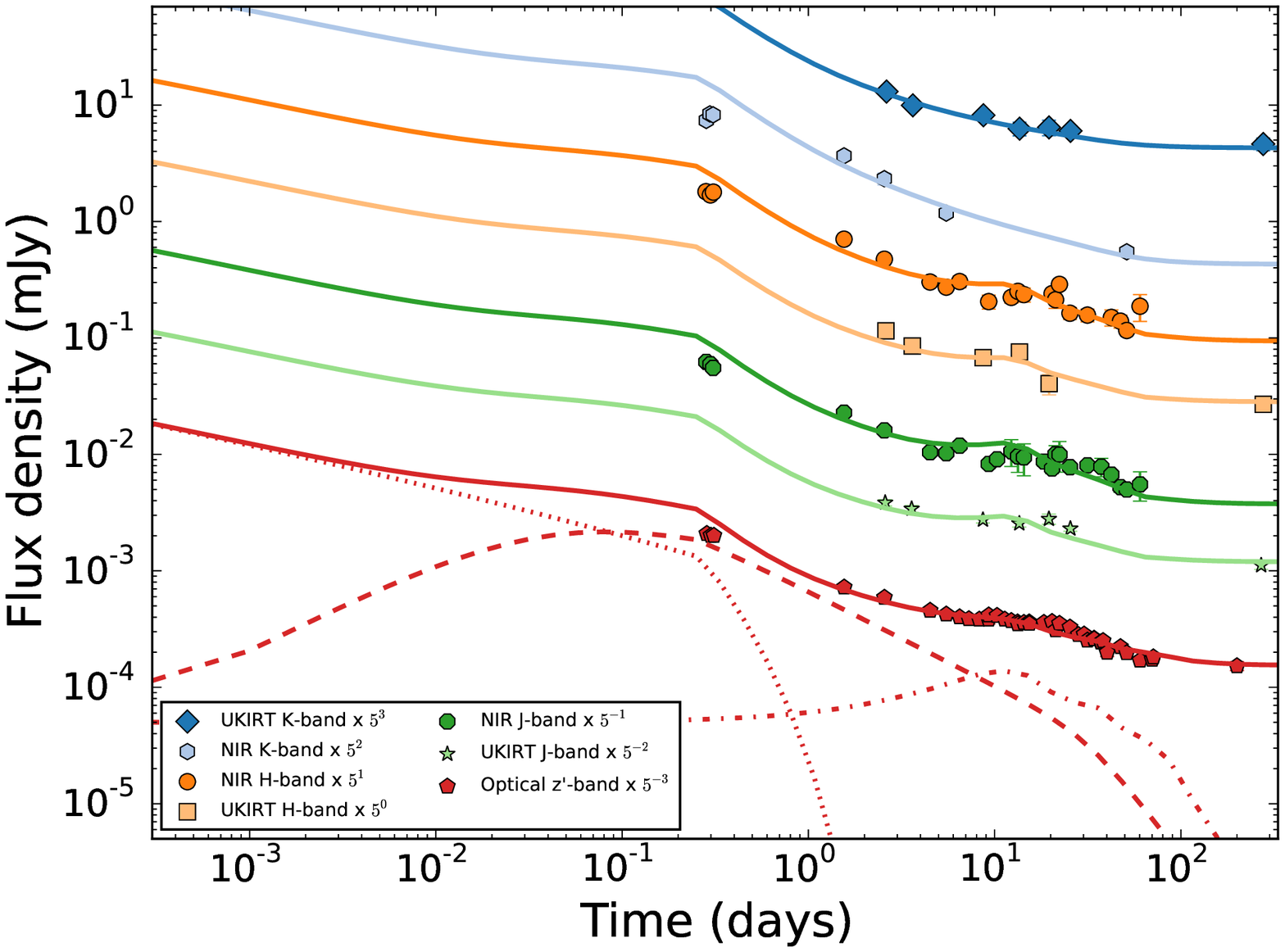} \\
  \includegraphics[width=0.47\textwidth]{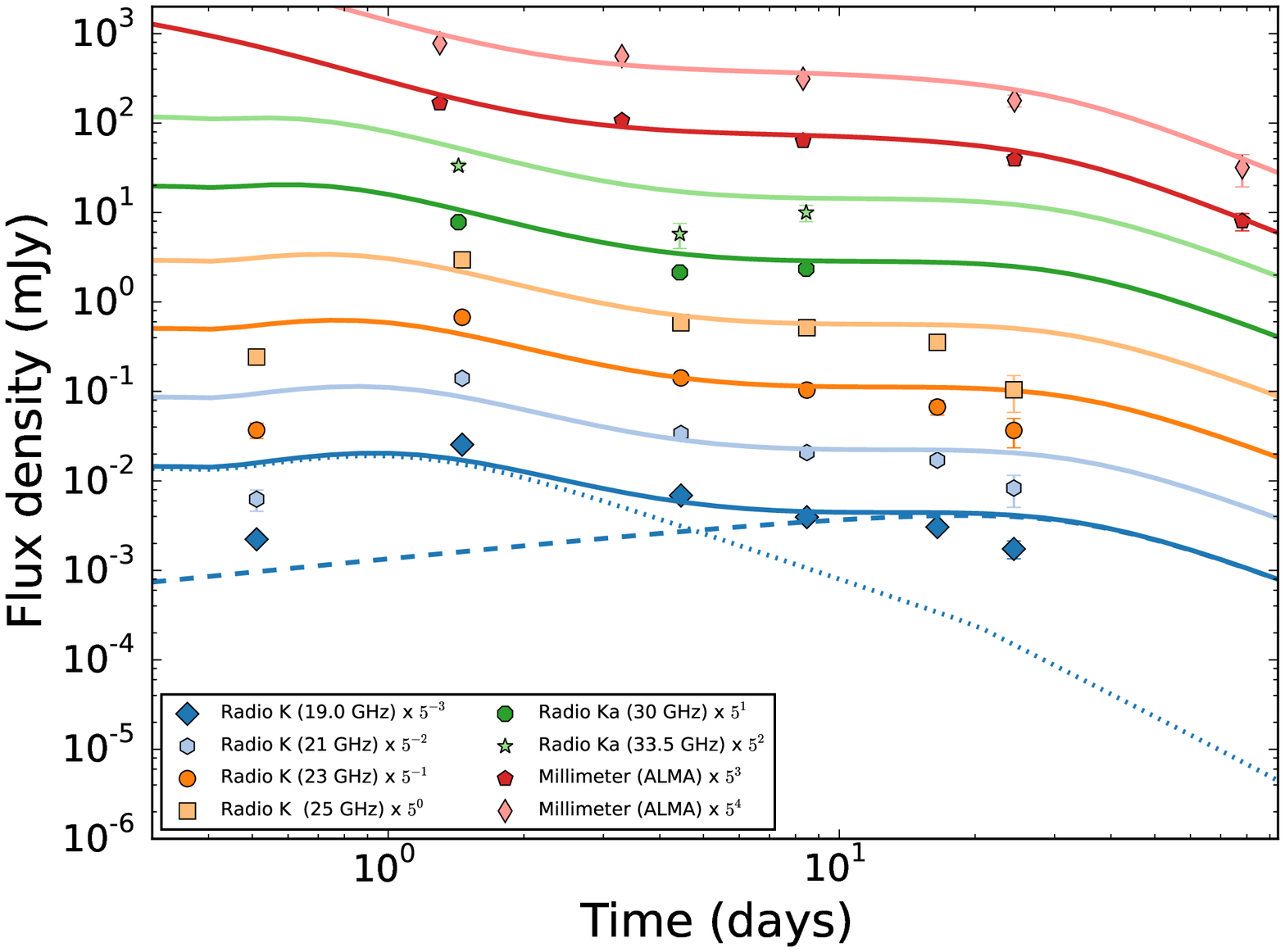} &
  \includegraphics[width=0.47\textwidth]{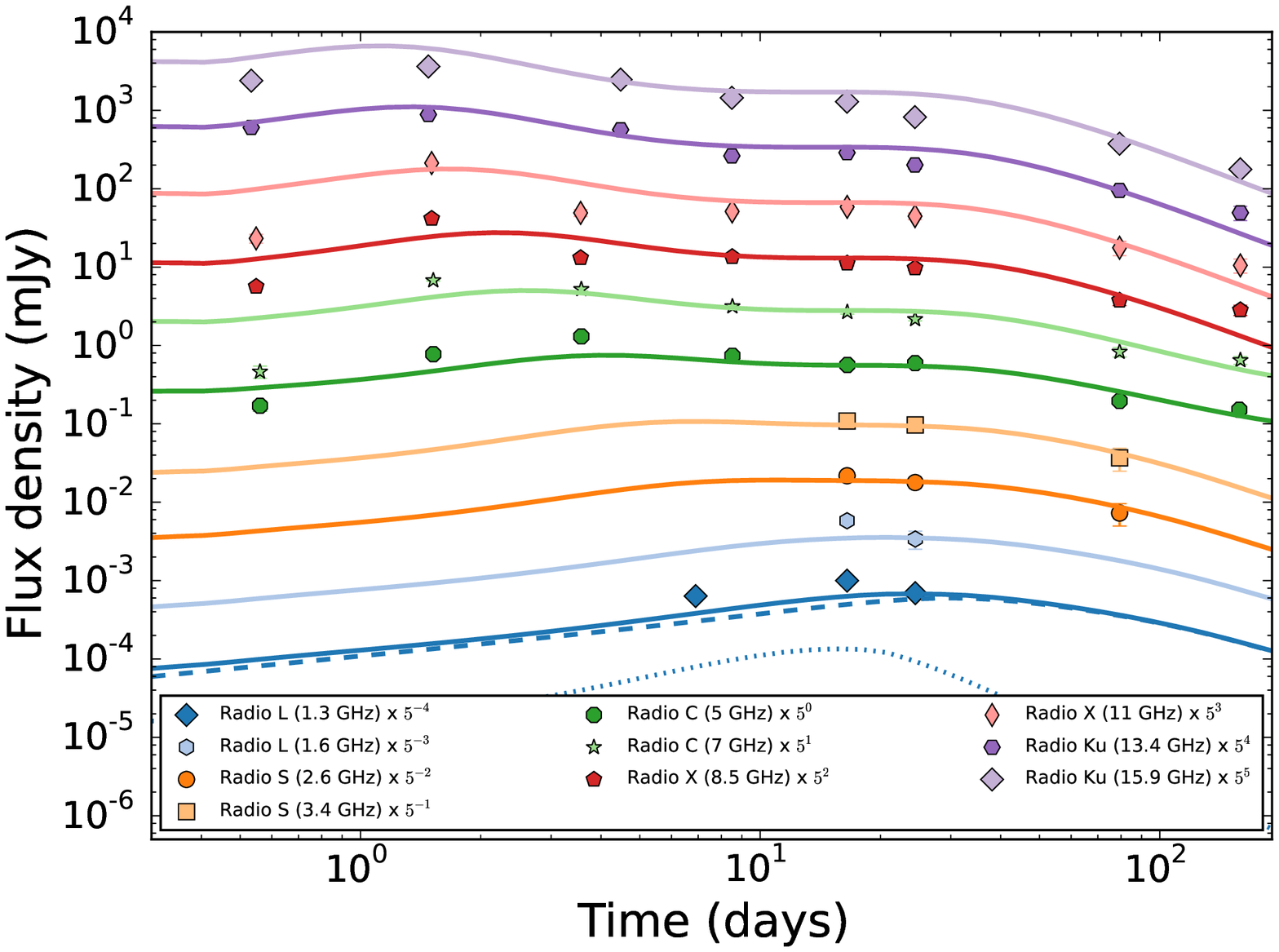} \\
 \end{tabular}
 \caption{X-ray (top left), UV (top right), optical (center left), NIR (center right), and radio 
(bottom) light curves of GRB\,161219B/SN2016jca, together with a full FS+RS model with 
energy injection (solid lines). We show a decomposition of the X-ray, \Swift/{\textit w2}-band, 
optical $g^{\prime}$-band, optical $z^{\prime}$-band, 19\,GHz and 1.3\,GHz light curves into FS 
(with energy injection; dashed), refreshed RS (dotted) and supernova (dash-dotted) components.
The combined model overcomes the deficiencies of the FS-only model (without energy 
injection; Section \ref{text:model}; Figures \ref{fig:modellc_FS_splits} and \ref{fig:modelsed_FS}), 
and explains the overall behavior of the light curves at all 41 observing frequencies over 5 orders 
of magnitude in time. Residual differences in the $10$--$30$\,GHz VLA light curves are likely 
related to the rapid cm-band variability observed for this event (Section \ref{text:basic_radio}). 
See Figure~\ref{fig:modellc_RS} in the appendix for a combined plot showing all 41 observing 
frequencies.}
\label{fig:modellc_RS_splits}
\end{figure*}

\begin{figure}
 \includegraphics[width=\columnwidth]{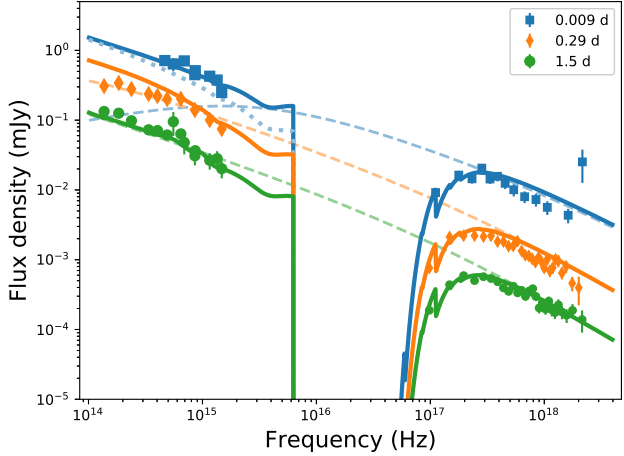}
 \caption{NIR to X-ray spectral energy distributions of the afterglow of GRB\,161219B at 
$9\times10^{-3}$\,d (blue), 0.29\,d (orange) and 1.5\,d (green), together with the best-fit 
synchrotron model to the entire multi-band data set (Section \ref{text:model}; solid lines). The 
dip in the UV is the combined effect of extinction in the Galaxy and in the host. The optical data 
have been interpolated using the average UV light curve (Table \ref{tab:uvlcfit}); the contribution 
of the host has been removed. The dashed lines represent the FS model without photoelectric 
absorption or optical extinction; the peak at $\approx 3\times10^{15}$\,Hz in the first epoch is 
$\numax$, and the break at $\approx 2\times10^{17}$\,Hz in the later epochs is $\nuc$. The SED at 
$9\times10^{-3}$\,d is dominated by the RS in the optical (dotted) and the FS in the X-rays 
(Section 
\ref{text:RS}); RS 
contribution at later times is negligible in the optical and X-rays. The slight discrepancy in the 
X-ray SED in the first two epochs may arise from Klein-Nishina corrections to the light curve above 
$\nuc$ (Section \ref{text:KN}).
}
\label{fig:optxrtsed}
\end{figure}

\begin{figure*}
\begin{tabular}{ccc}
 \centering
 \includegraphics[width=0.31\textwidth]{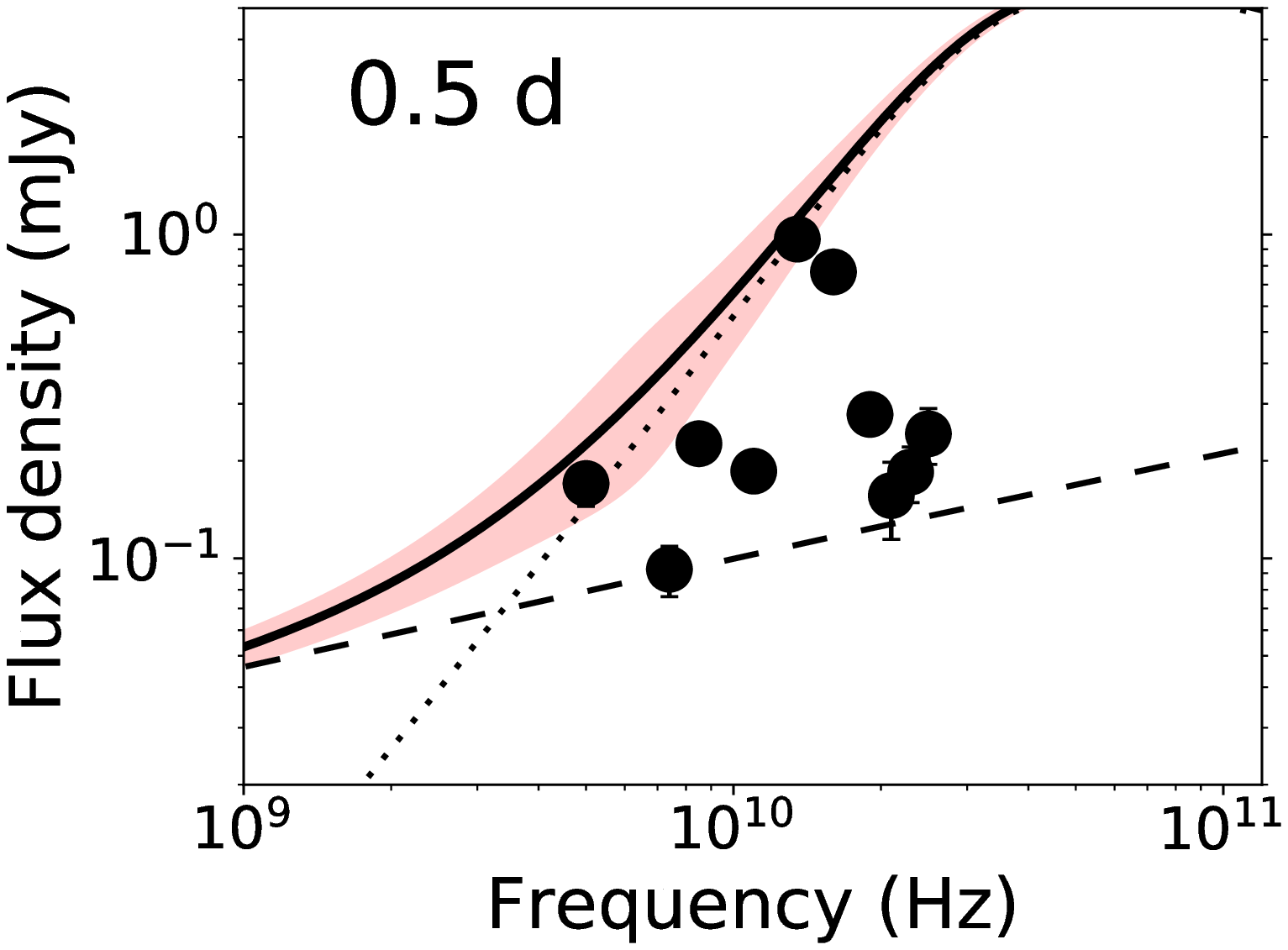} &
 \includegraphics[width=0.31\textwidth]{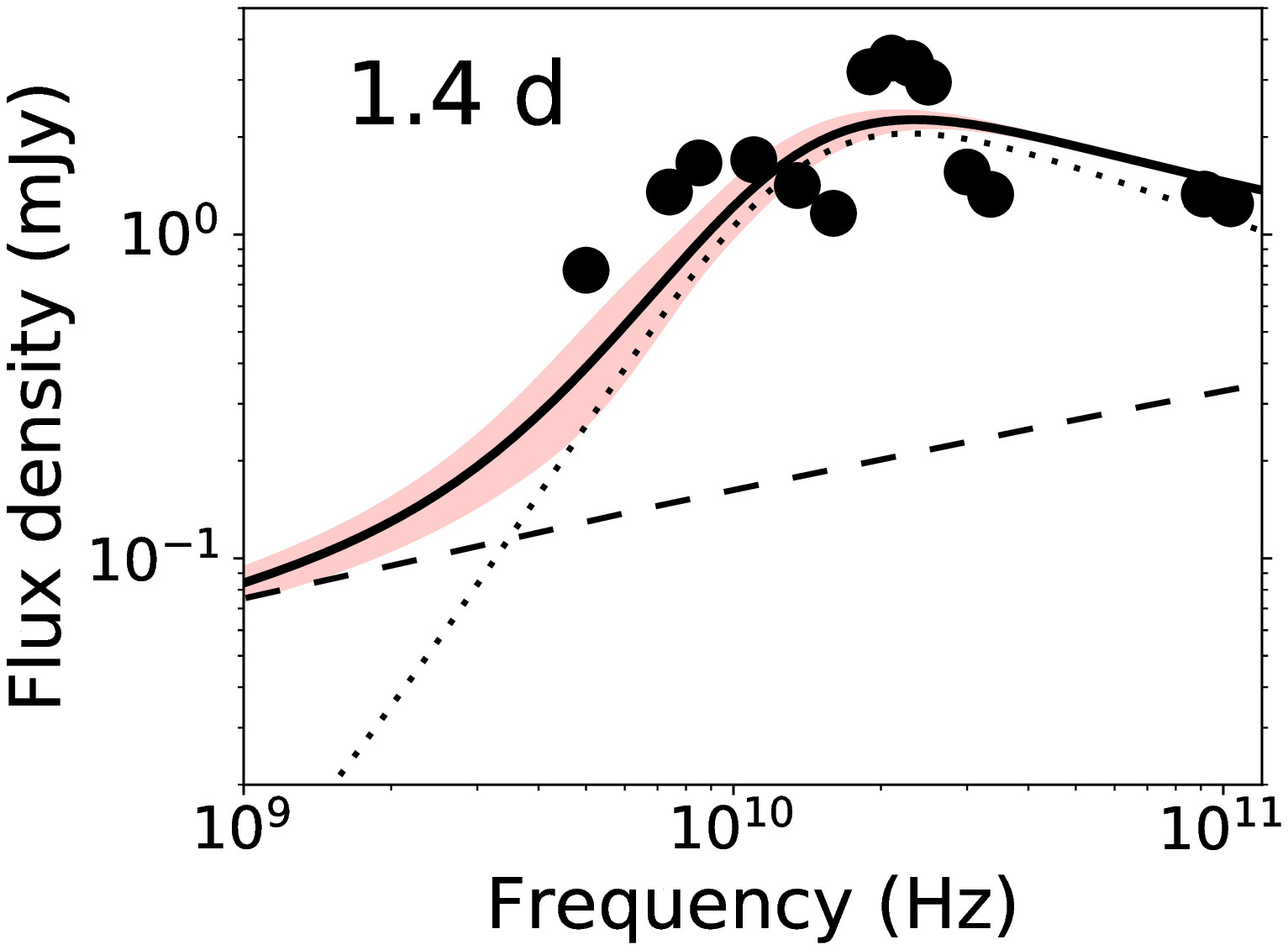} &
 \includegraphics[width=0.31\textwidth]{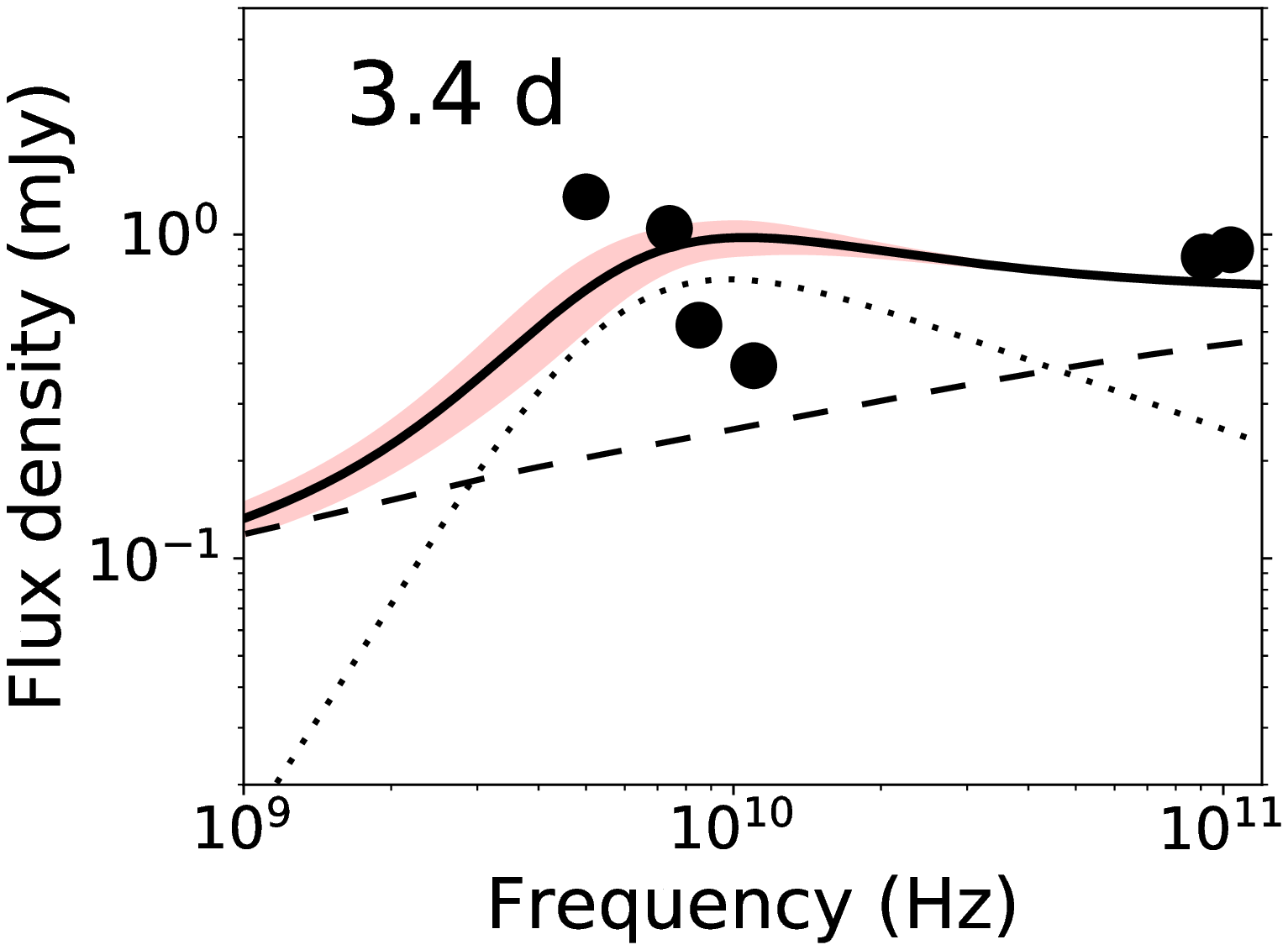} \\
 \includegraphics[width=0.31\textwidth]{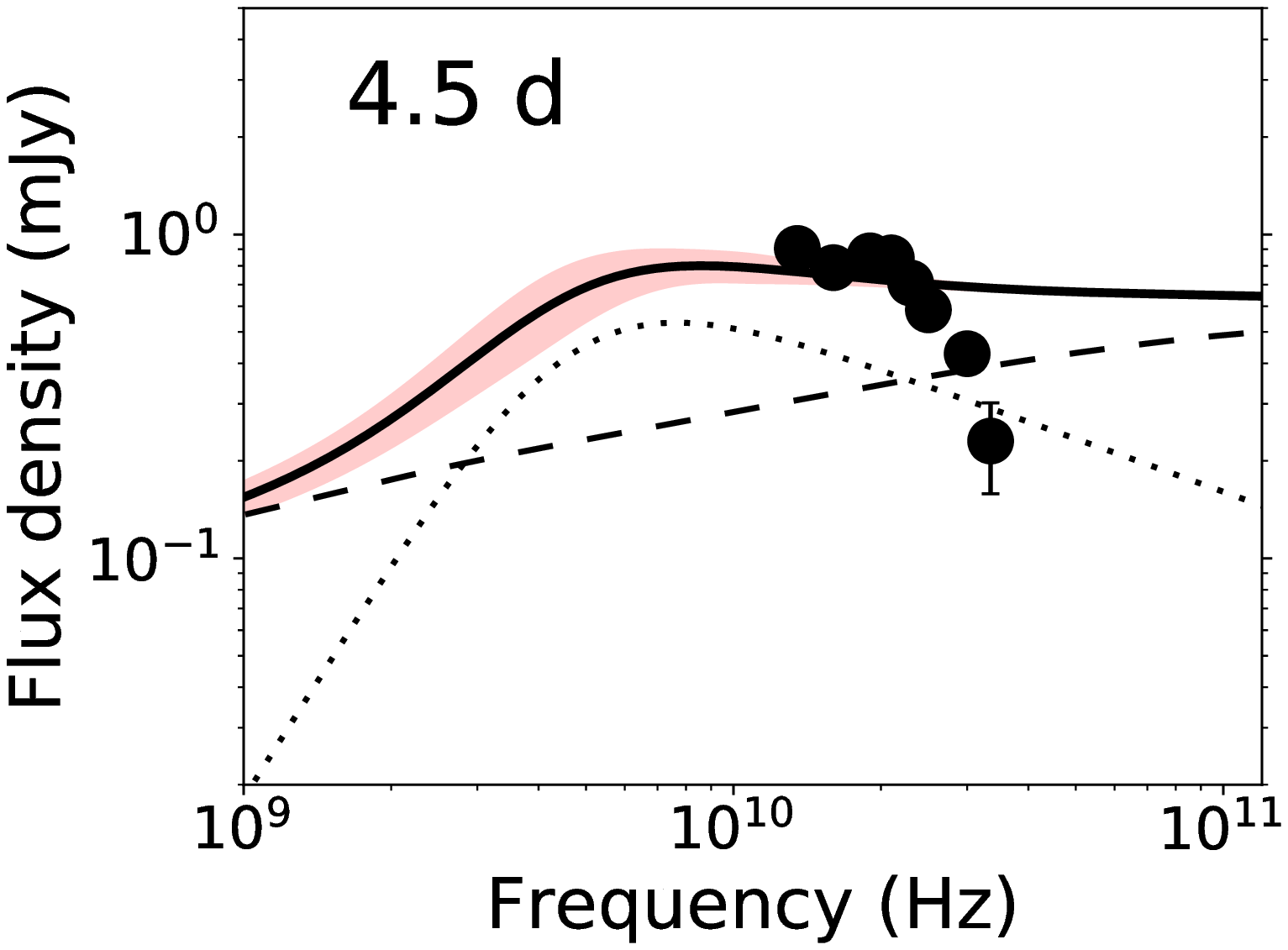} &
 \includegraphics[width=0.31\textwidth]{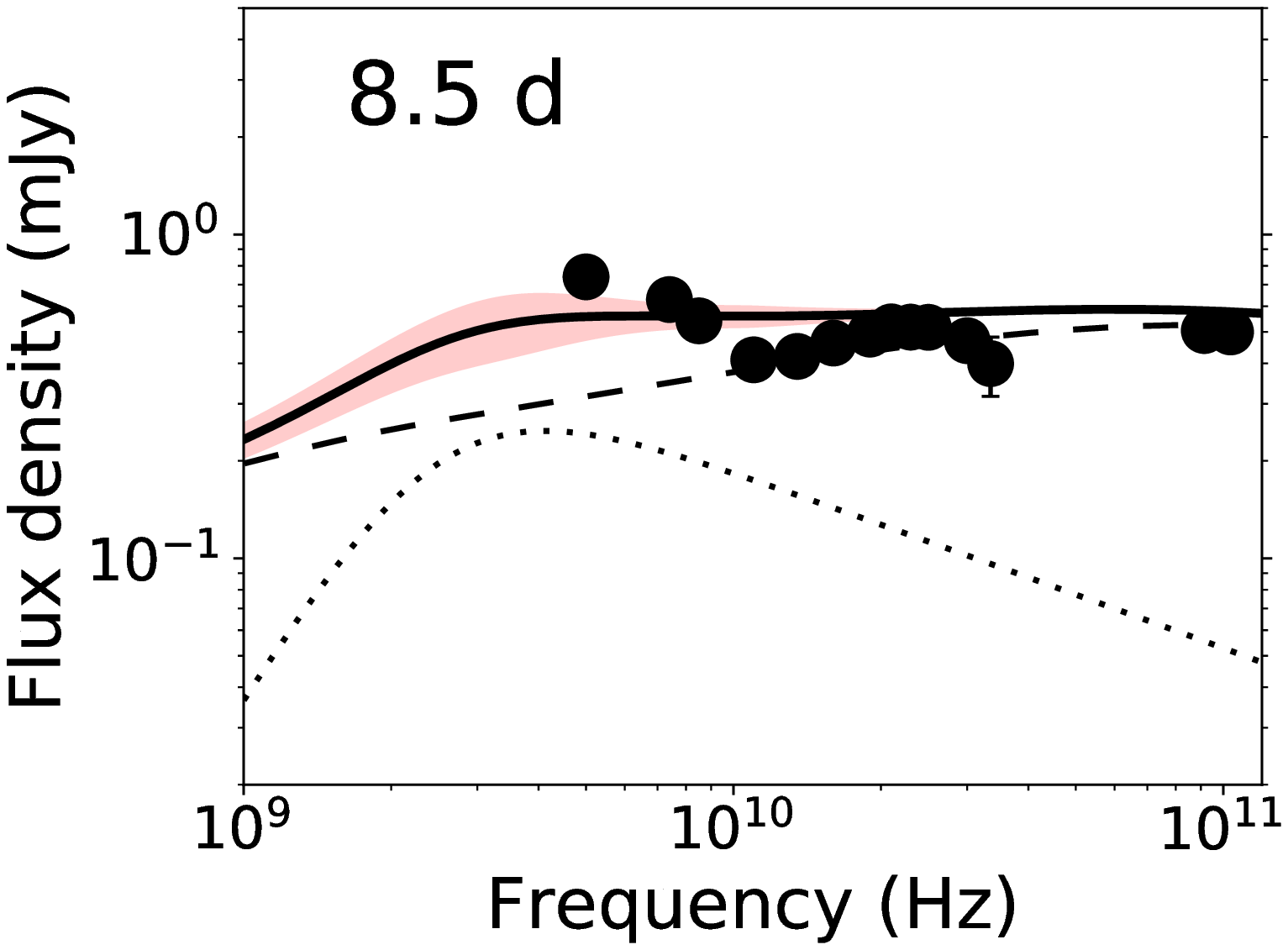} &
 \includegraphics[width=0.31\textwidth]{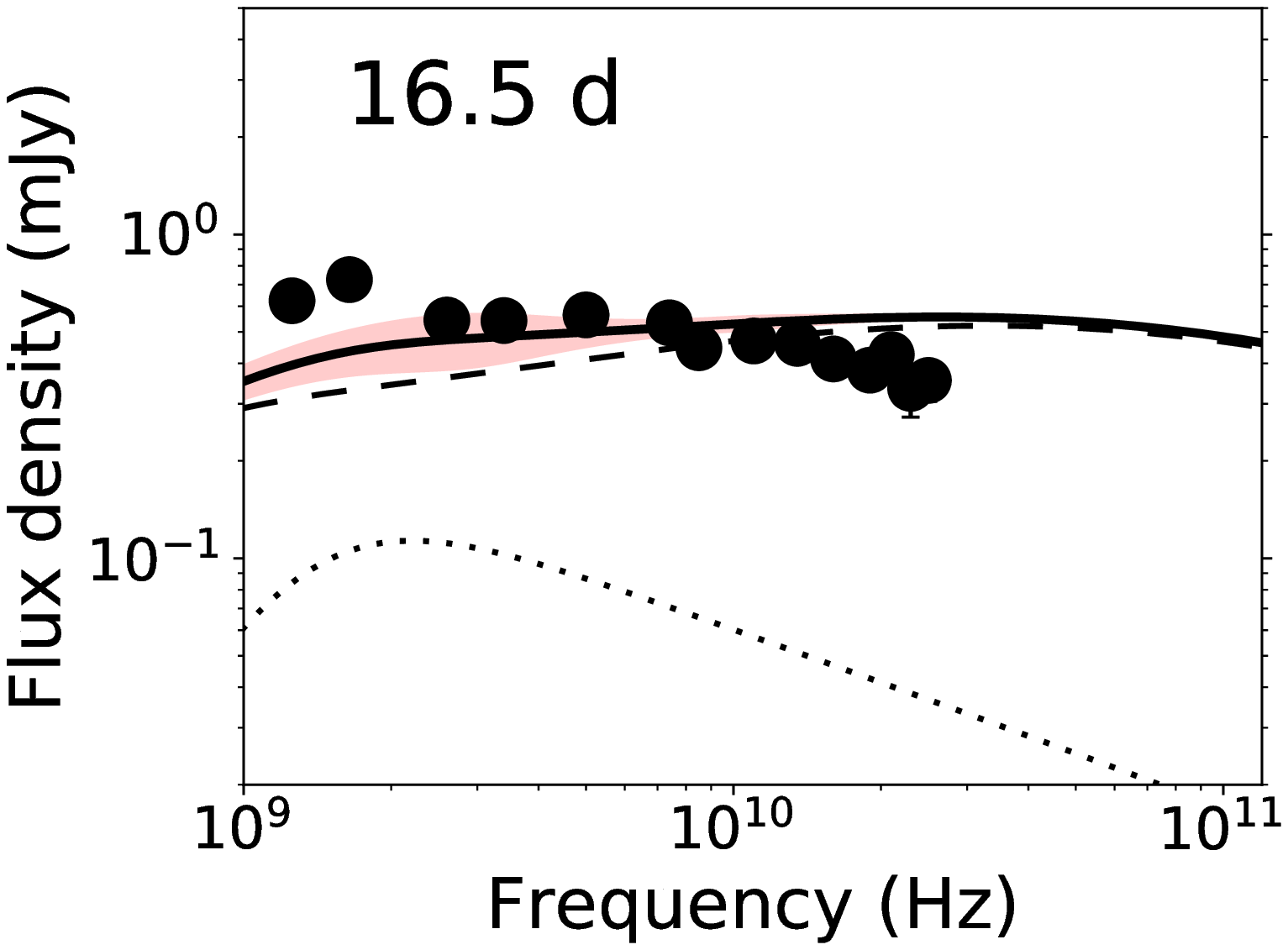} \\
 \includegraphics[width=0.31\textwidth]{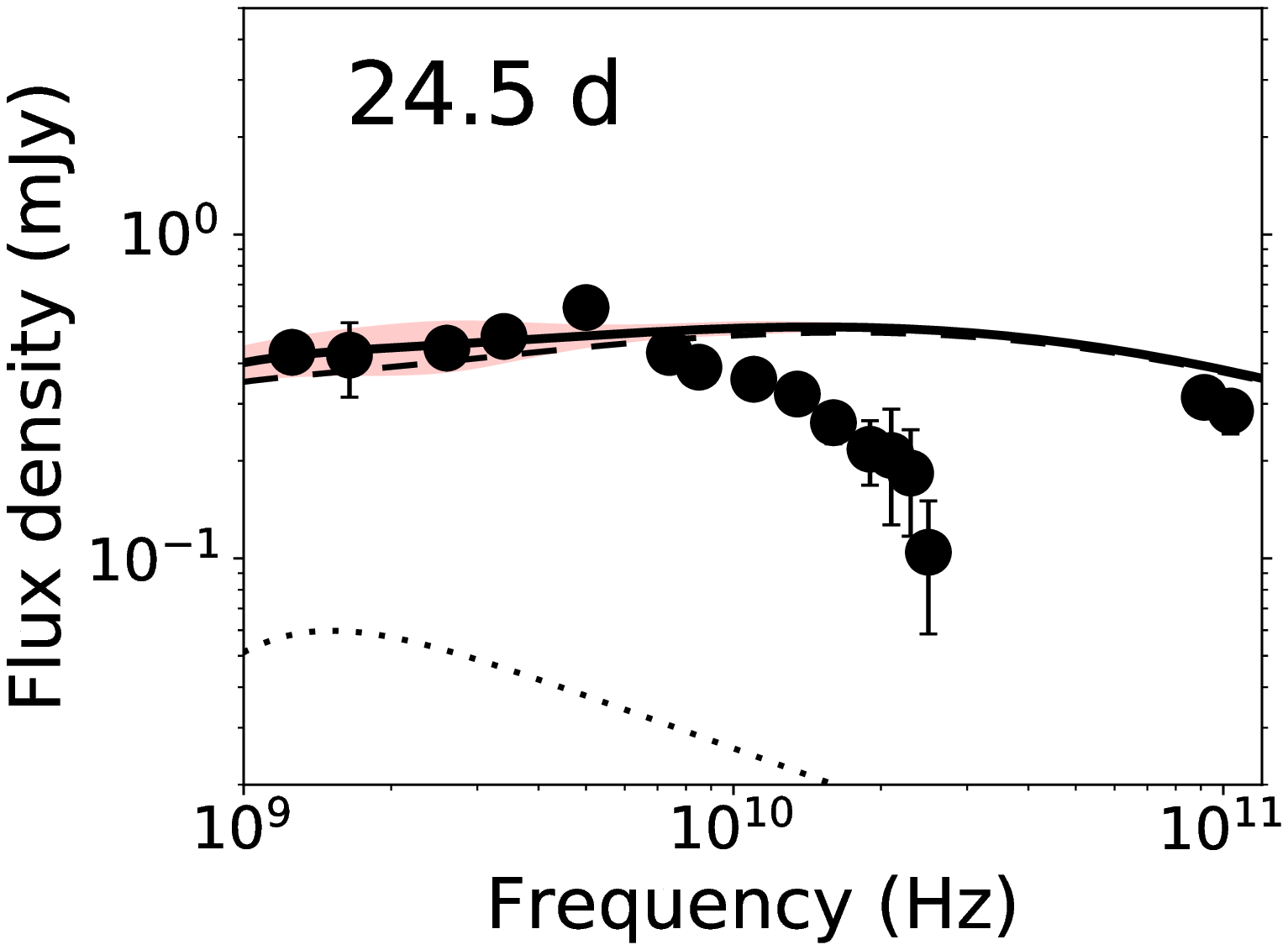} &
 \includegraphics[width=0.31\textwidth]{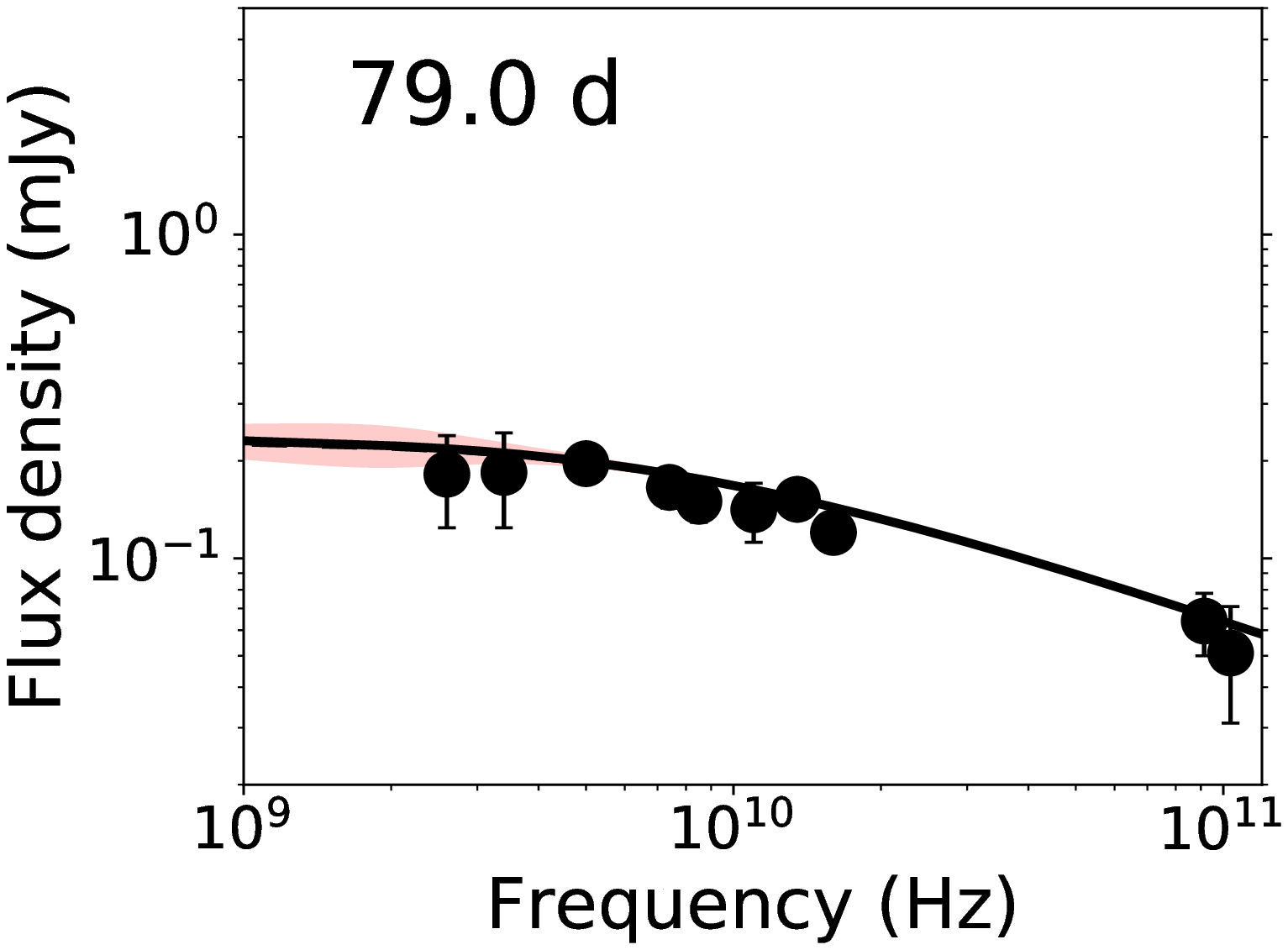} &
 \includegraphics[width=0.31\textwidth]{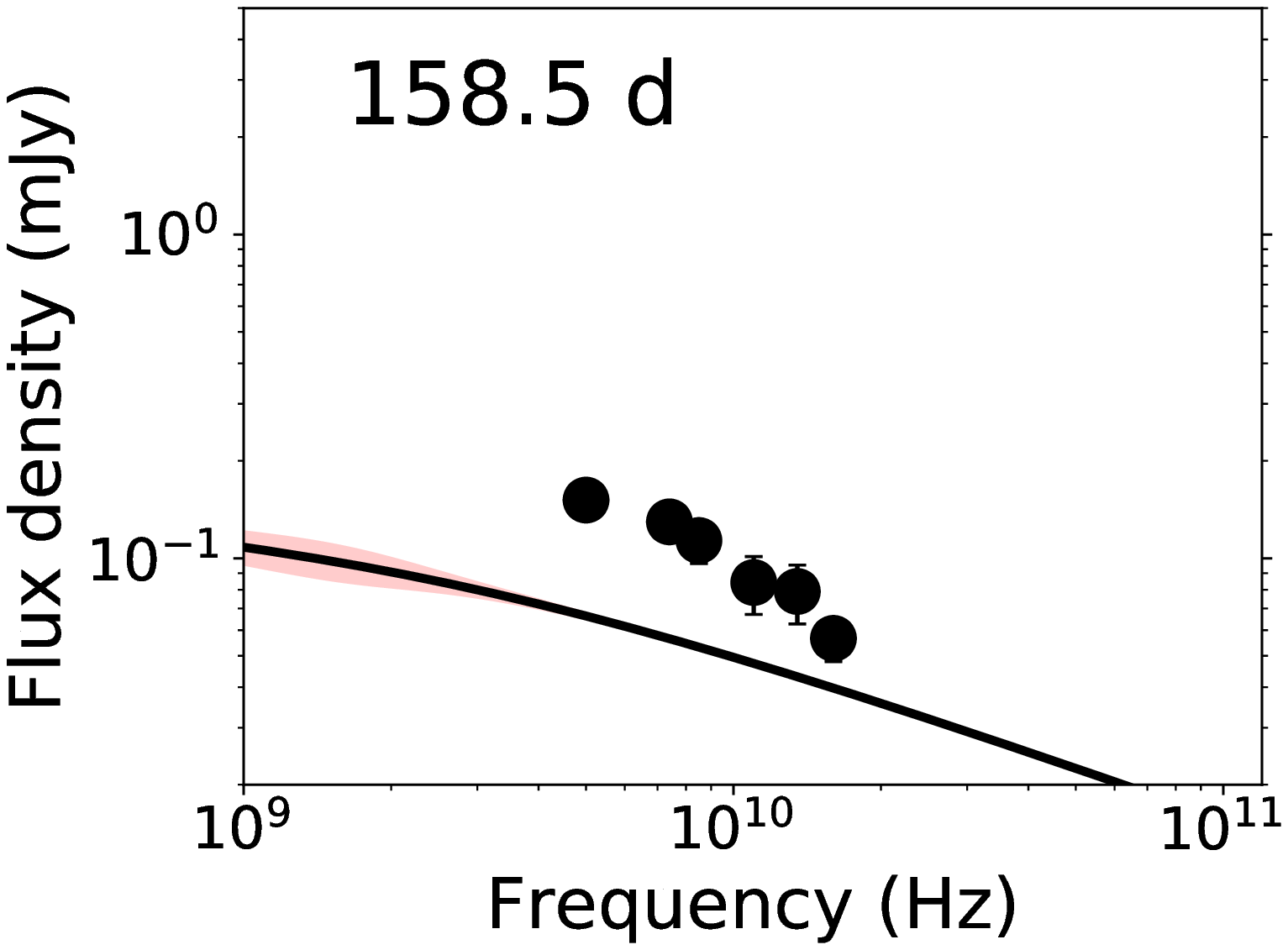} 
\end{tabular}
\caption{
VLA cm-band and ALMA mm-band spectral energy distributions of the afterglow of 161219B at 
multiple epochs starting at 0.5~d, together with the same FS+RS ISM model as in Figure 
\ref{fig:modellc_RS_splits} (solid), decomposed into FS (dashed) and RS (dotted) contributions. The 
red shaded regions represent the expected variability due to scintillation.
}
\label{fig:modelsed_RS}
\end{figure*}

\begin{figure} 
  \includegraphics[width=\columnwidth]{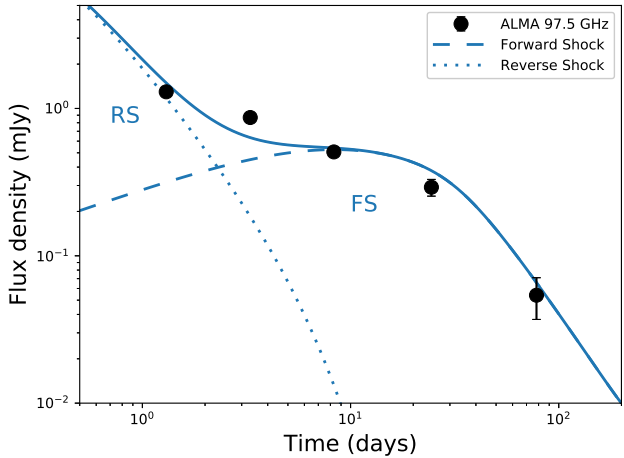}
 \caption{Side-band averaged ALMA 97.5\,GHz light curve of GRB\,161219B/SN2016jca, together 
with the final FS+RS ISM model (solid), decomposed into FS (dashed) and RS (dotted) components. The 
RS (Section \ref{text:RS}) explains the excess in the mm-band before $3.4$\,d. The ALMA 
observations also provide critical constraints on the the FS peak frequency, the peak flux, and the 
jet break time, as the cm-band data exhibit significant variability likely due to extreme 
scintillation.}
\label{fig:ALMA_lc}
\end{figure}

Under this spectral evolution and the observational constraints described above, we find that an RS 
model with $g\approx2.8$, $\tE\approx0.25$\,d, $\nucr(\tE)\approx1.2\times10^{15}$\,Hz, 
$\numr(\tE)\approx9.4\times10^{10}$\,Hz, $\nuar(\tE)\approx5.9\times10^{10}$\,Hz, and 
$\fnumr(\tE)\approx22$\,mJy fits the early optical and X-ray data well (Figure 
\ref{fig:modellc_RS_splits}). This value of $g$ is intermediate between the values expected for a 
Newtonian ($g\approx2.2$) and relativistic RS ($g\approx3$) for the case of no energy injection.
In this model, the X-ray light curve is dominated by the FS at all times, with the suppression 
prior to $0.25$\,d arising from energy injection with $m\approx0.35$. The UV/optical/NIR 
light curves are dominated by the FS after the end of energy injection at $\approx0.25$\,d and 
exhibit significant contribution from the RS arising from injection process prior to this time 
(Figure \ref{fig:optxrtsed}). The radio SEDs at $\approx 1.4$, 3.4, 4.5, and 8.5\,d are well 
matched 
by the same RS, propagated to the times of the radio observations (Figure \ref{fig:modelsed_RS}). 
Whereas the model does over-predict the 18--26\,GHz observations at 0.5\,d, we caution that 
these frequencies also exhibit the greatest cm-band variability before $\approx 8.5$\,d, 
possibly due to extreme interstellar scintillation (ALB18). The large scatter in flux 
density observed between individual frequencies in the SED further complicates the comparison 
against the model prediction at this time. 
Finally, the model also explains the excess in the ALMA light curve\footnote{See 
Figure~\ref{fig:modellc_RS} in the appendix for a combined plot showing all 41 observing 
frequencies.} at $\lesssim3.4$\,days (Figure \ref{fig:ALMA_lc}).

\subsection{Energy injection RS -- self-consistency with FS}
\label{text:RS_selfconsistency}
If the excess flux density in the early optical and radio observations of GRB\,161219B arises from 
an RS mediating energy injection, we expect the parameters of the RS and FS to be related at the 
time of cessation of the injection ($\tE\approx0.25$\,d). From the FS parameters derived 
from multi-wavelength modeling (Table \ref{tab:params}), the Lorentz factor of the FS is 
$\Gamma(\tE)\equiv\Gamma_{\rm E}\approx22.7$ \citep{bm77}. We present a comparison between the 
observed RS parameters and the values expected by scaling the FS parameters by $\Gamma_{\rm E}$ in 
Table \ref{tab:rsparams}. We find the peak flux density to match within 5\%(!), suggesting that the 
RS is not magnetized relative to the FS. The characteristic frequencies also agree upon scaling to 
within 50\%, a stronger match than previously obtained for the Newtonian RS detected in 
GRB\,130427A \citep{lbz+13,pcc+14}. Whereas the scaled ratio of the self-absorption frequencies 
$\nuar/\Gamma_{\rm E}^{8/5}\nuaf$ is too large by a factor of $\approx20$, we note that $\nuaf$ is 
not well constrained, being below the radio band at all times. 

The ratio of the cooling frequencies $\nucf/\nucr\equiv \nucratio \approx 3.6\times10^2$ is 
larger than 
expected, which is difficult to explain if $\RB\approx1$. Since $\nucr$ does not have a strong 
observational signature being hidden inside the FS spectrum, it could be higher by a factor of 
several; however, increasing $\nucr(\tE)$ beyond $\approx10^{16}$\,Hz does begin to affect the 
X-ray light curve at $10^{-2}$ to $2\times10^{-1}$\,d. Another possibility is that the Compton 
$Y$-parameter of the RS is higher than the FS. In the regime $\epse \gg \epsb$, we expect 
$Y\approx\sqrt{\eta_{\rm Y}\epse/\epsb}$, where $\eta_{\rm Y}=(\nuc/\numax)^{-(p-2)/2}$. 
Since this ratio is lower than the corresponding ratio for the FS, we expect $Y_{\rm r}\lesssim 
Y_{\rm f}$, and thus inverse Compton cooling cannot explain the observed high ratio of $\nucf$ to 
$\nucr$. The third option is that $\nucf$ is lower than predicted from the model. A reduced value 
of $\nucf$ would require the spectral index above $\nucf$ to be shallower than $-p/2$ in order to 
continue to match the X-ray light curve. Such a flatter spectrum is indeed afforded by the 
Klein-Nishina correction to the synchrotron spectrum. We discuss this further in Section 
\ref{text:KN} and Appendix \ref{appendix:KN}. We note that recent numerical work suggests the 
analytical relations over-predict the RS flux by factors of a few to $\approx10$ \citep{np04,hk13}; 
thus, it is possible that the equivalence between these quantities derived above may arise from a 
coincidence, and that the ejecta are magnetized at the level $\RB\approx5$--10.
Since $\nucf/\nucr\propto\RB^3$ is strongly dependent on $\RB$, even a slight RS magnetization (say 
$\RB\approx2$, which may be feasible given the uncertainties in the RS parameters), could alleviate 
the problem. Thus, a combination of a higher value of $\nucr$, a lower value of $\nucf$, or slight 
ejecta magnetization may explain the apparent discrepancy.

The observed strong RS signature in GRB\,161219B can be used to place constraints on the 
circumburst density. The UV/optical light curves prior to the end of energy injection at 
$\approx0.25$ suggest $\nucr\gtrsim10^{15}$\,Hz at this time. Combining equation 
\ref{eqn:rsfsrelation} with the expression for FS cooling frequency from \cite{gs02},
\begin{align}
 \nucf &= 6.37\times10^{13}(p-0.46)e^{-1.16p}(1+z)^{-1/2} \nonumber\\
       &\times  \epsb^{-3/2}\dens^{-1}E_{\rm K,iso,52}^{-1/2}(1+Y_{\rm f})^{-2}\td^{-1/2}
         {\rm Hz} \nonumber \\
       &\approx 5\times10^{13} \epsilon_{\rm B,-1}^{-3/2}\dens^{-1} {\rm Hz} \nonumber \\
       &\sim \RB^{3}\nucr 
\end{align}
at the redshift of GRB\,161219B for $p\approx2.1$, $\EKiso\approx5\times10^{51}$\,erg, 
$t\approx0.25$\,d and $Y_{\rm f} \approx3$ yields,
\begin{align}
 \dens &\approx \frac{5\times10^{13}\,{\rm Hz}}{\nucr}\epsilon_{\rm B,-1}^{-3/2}\RB^{-3}\pcc 
\nonumber\\
 &\lesssim 5\times10^{-2}\epsilon_{\rm B,-1}^{-3/2}\RB^{-3}\pcc.
\end{align}
For values of $\RB\gtrsim1$, this constraint becomes stronger. The measured density for 
GRB\,161219B is $\dens\approx4\times10^{-4}\pcc$, which satisfies this constraint, providing 
further evidence that bright reverse shock emission is more likely to be detectable in GRBs that 
occur in low density environments.

\begin{deluxetable}{cc}
 \tabletypesize{\footnotesize}
 \tablecolumns{2}
 \tablecaption{RS parameters at $\tE$}
 \tablehead{   
           \colhead{FS (scaled to $\tE$)} &
           \colhead{RS}  
   }
 \startdata      
 $\Gamma_{\rm E}^{-2} \numf  \approx 6.2\times10^{10}$\,Hz&
 $\numr \approx 9.4\times10^{10}$\,Hz\\[3pt]
 $\nucf\approx4.3\times10^{17}$\,Hz & $\nucr\approx1.2\times10^{15}$\,Hz\\[3pt]
 $\Gamma_{\rm E}\fnumf \approx 23$\,mJy & $\fnumr\approx22$\,mJy\\[3pt]
 $\Gamma_{\rm E}^{8/3}\nuaf\approx2.6\times10^{9}$\,Hz & $\nuar\approx5.9\times10^{10}$\,Hz
 \enddata
\label{tab:rsparams}
\end{deluxetable}


\section{Discussion}
\label{text:discussion}
We have presented multi-wavelength observations of GRB\,161219B and SN2016jca. The X-ray to 
radio afterglow can be modeled well as a combination of a forward shock with energy injection prior 
to $\approx0.25$\,d, and a Newtonian reverse shock arising from the injection process. 
The peak frequency and peak flux of the two shocks are fully self-consistent at the deceleration 
time, indicating low ejecta magnetization. This is the first direct detection of an energy 
injection RS in a GRB afterglow. 

\subsection{Radiative efficiency}
Comparing the radiated $\gamma$-ray energy with the isotropic-equivalent kinetic energy derived 
from afterglow modeling results in an extremely low radiative efficiency, 
$\etarad\approx4\%$. In our previous work on energy injection in GRB afterglows, we found 
that all events exhibiting a late-time achromatic re-brightening (`extreme re-brightening 
events', or EREs) in the optical and X-rays also exhibit low efficiencies, ranging from $43\%$ to 
as low as $3\%$ \citep{lbm+15}. Those events were interpreted in the context of energy injection, 
with fast-moving ejecta responsible for the $\gamma$-ray emission and slow-moving ejecta carrying 
the bulk of the kinetic energy.

Extrapolating the energy of the FS in GRB\,161219B from $\approx 5\times10^{51}$\,erg at 
0.25\,d to the time of the first optical detection at $\approx5\times10^{-4}$\,d, we expect 
$\EKiso(t_{\rm X})\approx 6\times10^{50}$\,erg. Thus, if the energy injection has been carrying 
on since that time and only the highest Lorentz factor material is responsible for producing the 
observed $\gamma$-rays, then the required efficiency is higher, $\etarad\approx20\%$, similar 
to values obtained in other events \citep{zlp+07,bndp15}. The FS Lorentz factor decreases by a 
factor of $\approx10$ from $t_{\rm X}$ to $\tE$. In the framework of the internal shock model, if 
we associate this with the range of Lorentz factors ejected by the central engine, we expect a 
theoretical efficiency of $\approx15\%$ from internal shocks, comparable to the extrapolated value 
\citep{kps97}. With the caveat that the precise value of the computed efficiency depends on the time 
at which the injection starts, our observations may provide an independent validation of the 
internal shock model.

\subsection{Energy injection and RS}
Our observations indicate a relatively slow injection rate, $E\propto t^{0.35}$. If this 
arises from a distribution of ejecta Lorentz factors, $m\approx0.35$ corresponds to a shallow 
ejecta profile, $s\approx2$. This is similar to the value obtained in GRB\,010222 from the X-ray 
and optical light curves \citep{bhpf02}, but lower than the values obtained from multi-wavelength 
modeling of the EREs \citep{lbm+15}, suggesting that GRB ejecta span a range of Lorentz factor 
distributions.

We note that the EREs of \citet{lbm+15} did not exhibit reverse shock signatures, a fact that we 
suggested may have been due to a gentle injection process. Another explanation could be 
differences in the ejecta Lorentz factor at the time of the interaction -- in the case of 
GRB\,161219B, the RS appears to have been observable from the earliest times in the optical (at 
$\lesssim 10^{-3}$\,d) when the ejecta Lorentz factor is high ($\Gamma\gtrsim 100$). On the other 
hand, the onset of injection in the EREs occurred at $\Gamma\sim20$, which may have been responsible 
for yielding a fainter RS, or RS emission peaking at frequencies too high to be observable (e.g., in 
the sub-millimeter). Since the RS is long-lasting and mildly relativistic, it continuously 
decelerates the ejecta; therefore it is only possible to determine a lower bound on the initial 
Lorentz factor of the outflow from these observations, $\Gamma_0\gtrsim100$.

\begin{figure}
   \centering
   \includegraphics[width=\columnwidth]{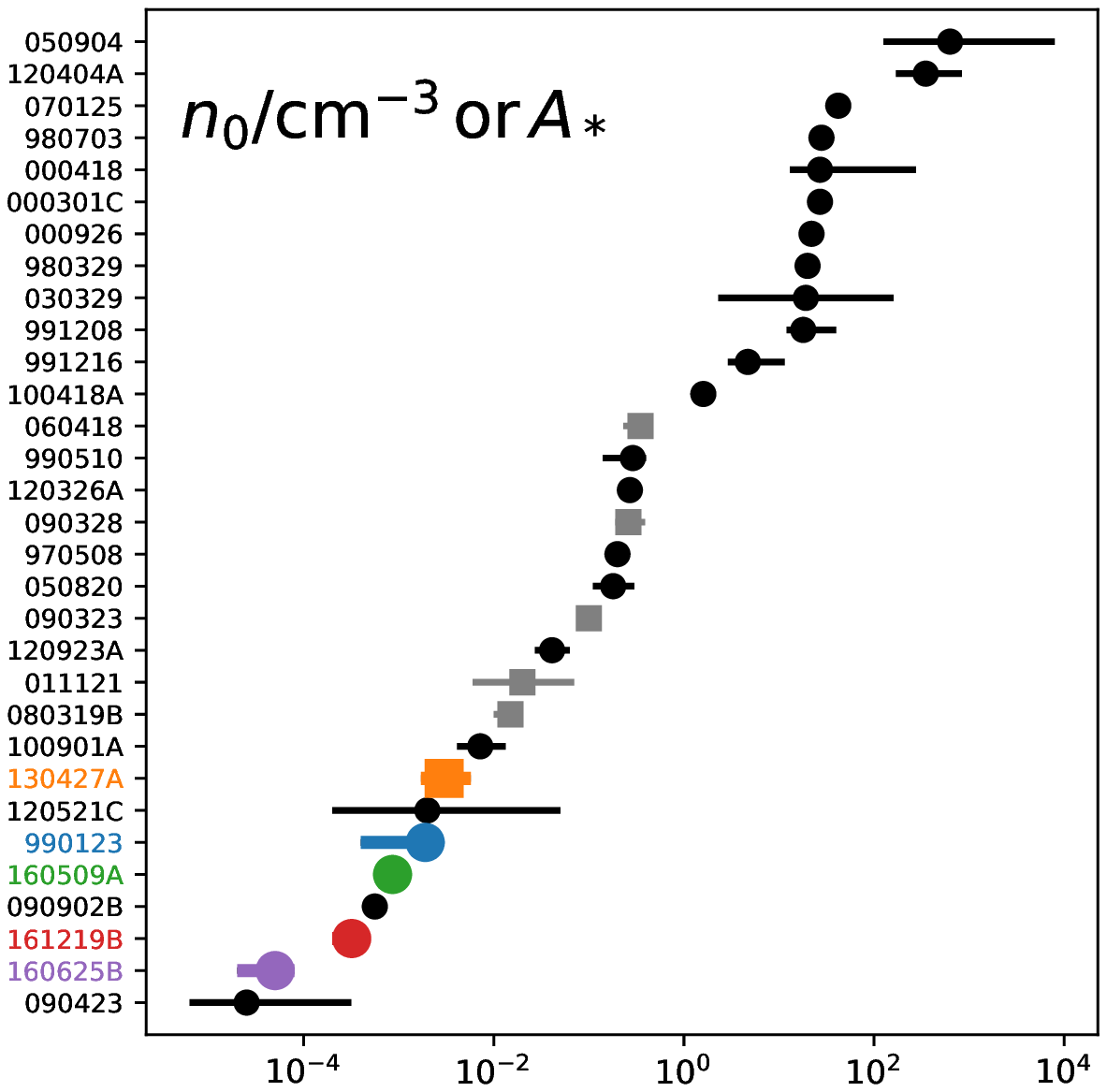}
 \caption{Circumburst density for both ISM (black circles) and wind-like environments (grey 
squares) for GRBs with multi-wavelength observations and modeling \citep{pk02, yhsf03, ccf+08, 
cfh+10, cfh+11,lbz+13,lbt+14,lab+16,alb+17,tll+17}. The GRBs with strong reverse shocks (highlighted 
as colored points) also exhibit some of the lowest circumburst densities of the sample. We note 
that the remaining 3 GRBs with the lowest densities (090423, 090902B, and 120521C) have also been 
suggested to exhibit RS signatures \citep{cff+10,cfh+11,lbt+14}.}
 \label{fig:dens}
\end{figure}

The multi-wavelength modeling of the EREs was consistent with a constant density circumburst 
environment in each case, and the events considered exhibited densities ranging from $10^{-2}$ to 
$10^3$\,\pcc. In contrast, the inferred circumburst density of GRB\,161219B is 
extremely low, $\dens\approx3\times10^{-4}$\,\pcc. 
One possible mechanism for evacuating the environment around massive stars prior to core 
collapse may be late shell ejections due to super-Eddington winds in the Wolf-Rayet phase or 
LBV-like eruptions that sweep up the ambient medium; however, the precise degree to which this 
mechanism is operational and effective for GRB progenitors remains an open question 
\citep{wmc+77,mar97,mhs00,rdmt01,clf04,kfd+13,smi14,mkm+17}.
We note that all observed instances of 
detectable RS emission, without exception, have been in low-density environments ranging from 
$n_0\approx5\times10^{-5}$ to $\approx 10^{-2}$\,\pcc\ (Figure \ref{fig:dens}), and an overwhelming 
majority of the cases exhibiting strong RS signatures have been detected in constant density 
environments \citep[5 out of 6: GRBs 990123, 021211, 041219A, 
160509A, and 160625B;][]{wdl00,pk02,kp03,wei03,np04,fzw05,lbz+13,lab+16,alb+17}. We have previously 
speculated that the low density medium may be responsible for a slow cooling reverse shock, allowing 
the RS emission to be detectable for longer \citep{lbz+13,lab+16,lab+17,kmk+15,alb+17}. Indeed, we 
find a slow cooling RS in a low density 
medium in the case of GRB\,161219B also, lending credence to this hypothesis.

\begin{figure}
   \centering
   \includegraphics[width=\columnwidth]{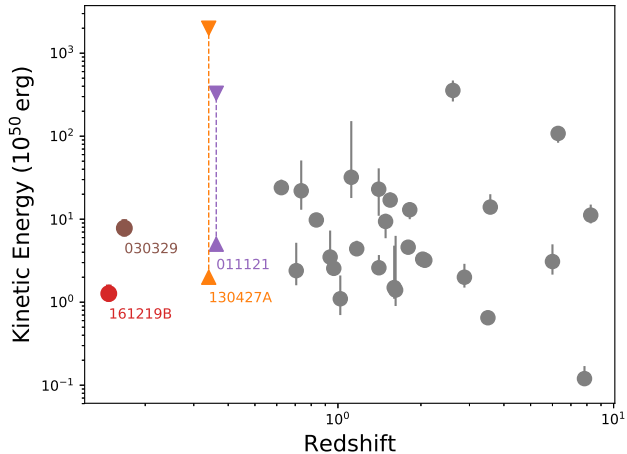}
 \caption{Beaming-corrected kinetic energy of GRB jets as a function of redshift for events with 
multi-wavelength modeling and measured opening angles. We include the 
supernova-associated GRBs\,011121 and 130427A with published lower limits on $\EK$; the 
corresponding upper limits correspond to $\EKiso$.}
 \label{fig:EKz}
\end{figure}

\subsection{Comparison with nearby GRB-SNe and low-luminosity GRBs}
Owing to the relative faintness of GRB-supernovae compared to the afterglow light, only a small 
fraction (30 out of $\gtrsim1000$ bursts, or $\lesssim5\%$) of GRBs have detected supernovae. A 
still smaller number (18) of these have been spectroscopically confirmed to date \citep{cwdw17}. At 
the same time, a large fraction of these spectroscopically confirmed GRB-SNe (6/18) appear to have 
low peak $\gamma$-ray luminosities ($L_{\gamma,iso}\lesssim10^{48.5}$\,erg\,s$^{-1}$), raising the 
question of whether these are representative of the cosmological ($z\gtrsim1$) population 
\citep{cow05,gdv07,vlz09,bnp11}.
In this context, GRB\,161219B can be classified as an intermediate luminosity event, and is an 
outlier in the $E_{\gamma,\rm peak}$--$E_{\gamma,iso}$ relation \citep{ama06}, together with 
several low- and intermediate-luminosity GRBs \citep{cidup+17}. 

Of the 12 discovered GRB-SNe at $z<0.5$, only one has a measured jet opening 
angle \citep[GRB~030329;][]{fsk+05}, while two others have lower limits, yielding lower limits on 
$\EK$ \citep[GRBs~011121 and 130427A;][]{pbr+02,lbz+13,pcc+14}. Thus, GRB\,161219B is the second 
supernova-associated GRB with a well-determined kinetic energy, and it has the lowest kinetic 
energy 
of these four events (Figure \ref{fig:EKz}). 
Future observations of GRB-supernovae at higher redshifts and at later 
times to measure their degree of collimation are thus essential for understanding the population of 
these low-energy events in the context of their cosmological counterparts.

\subsection{High value of \texorpdfstring{\epse}{ee}}
The derived value of $\epse\approx0.9$ is quite close to 1, significantly higher than the values 
of $\epse\approx0.1$ derived from simulations of particle acceleration in relativistic shocks 
\citep{ss09,ss11} and also larger than the equipartition value of $\epse\approx1/3$. Since we 
expect a non-zero fraction of the shock energy to be transferred to ions, large values of the shock 
microphysical parameters are problematic. We note here that while the values for $\epse$ and 
$\epsb$ given in Table \ref{tab:params} are the best fit parameters, lower values of $\epse$ are 
feasible, and are correlated with a lower energy, lower density, and higher value of $\epsb$ (Figure 
\ref{fig:corrplots}). On the other hand, values of $\epse$ near or above equipartition have been 
found previously in other works, suggesting that the discrepancy could also arise due to missing 
physics in the modeling process \citep{yhsf03,cfh+10,cfh+11,lbm+15,lab+16,lbc+18}. For instance, 
accounting for a fraction $f\approx 1$\% of the electrons accelerated by the shock would reduce 
$\epse$ by a corresponding amount \citep{ew05,rl17}. Unfortunately, the degeneracy in the physical 
parameters introduced by $f$ precludes a unique determination of this quantity. Relaxing the 
assumption that all electrons fall into a power law distribution would also change the definition of 
$\epse$ by the factor $(p-2)/(p-1)$, and thus alleviate this discrepancy. The observational 
signatures of such modifications to the electron energy distribution are under investigation 
\citep{webn17,rl17}.

\subsection{Unusual X-ray properties}
\label{text:KN}
\begin{figure}
 \includegraphics[width=\columnwidth]{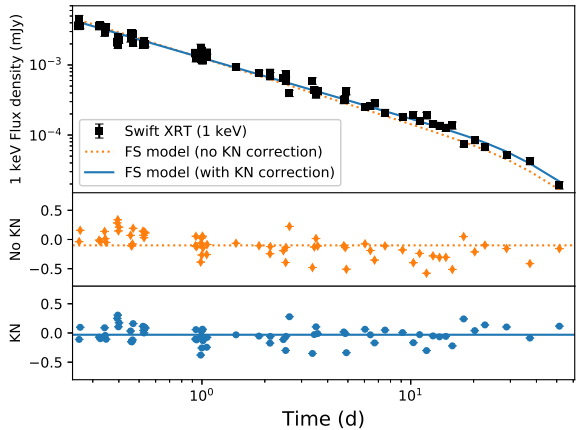}
 \caption{Swift/XRT light curve at 1\,keV (top panel) between 0.25\,and 51.5\,d, together with the 
FS model described in Section \ref{text:model} (orange, dotted) and an FS model including KN 
corrections to the spectrum above $\nuc$ as well as to the evolution of $\nuc$ itself (blue, 
solid). The residuals from the original FS model (center) exhibit systematic trends with time. 
Including the effects of the KN correction (lower panel) reduces the trends in the X-ray residuals 
and yields a better fit to the data. The two models use the same value of $p$.}
\label{fig:kncor}
\end{figure}

The observed X-ray spectral index of $\beta_{\rm X}=-0.86\pm0.03$ falls between the values $\beta 
= (1-p)/2\approx -0.5$ and $\beta=-p/2\approx-1$ for $p\approx2$, while the X-ray light curve 
decline rate also lies between the values expected on either side of the cooling break. Our 
best-fit model described above requires $\nuc \approx \nu_{\rm X}$ for the majority of the X-ray 
light curve, with $\nuc\approx10$\,keV at $4\times10^{-3}$\,d and $\nuc\approx1$\,keV at 0.6\,d. 
Since the cooling break is a gentle transition \citep{gs02}, this may explain the 
intermediate spectral index and decline rates measured. We note that the hard spectrum above $\nuc$ 
may also be the result of Klein-Nishina corrections to the synchrotron spectrum, where we expect a 
spectral index $\beta_{\rm X}\approx-3(p-1)/4\approx-0.8$ and $\alpha_{\rm 
X}\approx-7(p-1)/8\approx-0.9$, closer to the observed values (Appendix 
\ref{appendix:KN}). With these modifications, the X-ray light curve is modeled much better (Figure 
\ref{fig:kncor}). Similar effects may explain the slight discrepancy noted between the expected and 
measured values of $\beta_{\rm X}$ in the case of GRB\,160625B, for which \cite{alb+17} also find 
$\nuc\approx\nux$ in an ISM-like environment. A detailed analysis incorporating this effect requires 
a modified synchrotron spectrum including Klein-Nishina corrections \citep[e.g.,][]{nas09}, and is 
beyond the scope of this work. 

\subsection{Radio excess at 158.5 d}
\label{text:radioexcess}
Whereas the model fits the radio SED at 79\,d extremely well, it under-predicts the final radio SED 
at 158.5\,d. The radio light curve decline rate between these final two epochs steepens from 
$\alpha=-0.4\pm0.2$ at 5\,GHz to $-1.1\pm0.3$ at 16\,GHz, but is shallower than the expected value 
of $\alpha\approx-2$ for $\nu\gtrsim\numax$ at $t>\tjet$. The FS model also under-predicts the 
last two \Swift/XRT observations at $\approx70.4$--106\,d, suggesting the effect may be 
pan-chromatic (Figure \ref{fig:modellc_RS}). Since the Lorentz factor of the FS at these late times 
is low ($\Gamma\approx1.2$), it is possible that electrons at $\gamma\lesssim\gamma_{\rm min}$ 
are contributing significantly to the observed radiation, thus invalidating the premise of the 
radiation model. Another way to achieve a shallower light curve is through a transition to 
non-relativistic expansion \citep{fwk00,lw00,sg13}; however, this is not expected to occur until 
$\approx240$\,d \citep{wkf98}, while the transition to the deep Newtonian phase takes place even 
later, near the Sedov time, $t_{\rm ST}\sim(E/\rho c^5)^{1/3}\sim7$\,yr. \cite{lbc+18} found a 
similar late-time flattening in the cm-band light curves of GRB\,140311A, and considered an early 
transition to non-relativistic expansion (such as the FS encountering a density enhancement) as a 
possible solution. Our model assumes a rapid spreading of the outflow following \tjet, whereas 
recent numerical work suggests the decollimation process may be more gradual 
\citep{zm09,vezm10a,vem12a,dl17}. A detailed study of this effect requires numerically calibrated 
models of the evolution of the synchrotron spectrum during the spreading phase. Analytical 
calculations in this regime, combined with future late-time X-ray and radio observations of GRB 
afterglows, will be crucial for clarifying the observed discrepancy. Here, we consider the 
possibility that the flattening is due to emerging contribution from the underlying host galaxy, and 
include an additive constant at these frequencies in the multi-band modeling. Further observations 
of this source at cm-band frequencies several years hence would allow distinguishing between these 
possibilities.

\section{Conclusions}
We have presented detailed multi-wavelength observations of GRB\,161219B, SN\,2016jca, and their 
host galaxy, including the first ALMA light curve of a GRB afterglow, and the first direct 
detection of an energy injection RS. Through simultaneous multi-frequency modeling, we constrain 
the properties of the afterglow, supernova, and host, and determine that the GRB occurred in an 
extremely low density environment, $\dens\approx3\times10^{-4}$\,\pcc. The data constrain the 
beaming angle of the relativistic outflow, allowing us to derive the degree of ejecta collimation 
($\thetajet\approx13^{\circ}$) and to correct the $\gamma$-ray and kinetic energy for beaming, 
$E_{\gamma}\approx4.9\times10^{48}$\,erg and $\EK\approx1.3\times10^{50}$\,erg. The prompt 
efficiency is low, $\etarad\approx4\%$. The early radio and optical data require an additional 
emission component, which we interpret as synchrotron radiation arising from a refreshed reverse 
shock, powered by injection of energy into the forward shock through slow-moving ejecta. The 
combined model explains the X-ray to radio light curves over 8 orders of magnitude in frequency and 
5 orders of magnitude in time. We measure a low ejecta magnetization, and our observations provide 
another confirmation for the internal shock model of GRB prompt emission. The supernova component is 
fainter and evolves faster than SN\,1998bw, while the stellar mass of the host galaxy is comparable 
to that of GRB hosts at $z\lesssim1$. We conclude that detailed multi-frequency radio observations 
and early optical detections are key to constraining refreshed reverse shocks in GRBs, and may 
yield crucial insight into the production and nature of GRB jets.

\acknowledgements
We thank the anonymous referee for their helpful comments on improvements of the manuscript.
TL is a Jansky Fellow of the National Radio Astronomy Observatory. 
The Berger Time-Domain Group at Harvard is supported in part by the NSF under grant 
AST-1411763 and by NASA under grant NNX15AE50G.
CGM acknowledges support from the Science and Technology Facilities Council.
We thank Peter Nugent for generating the supernova model.
This paper makes use of the following ALMA data: ADS/JAO.ALMA\#2016.1.00819.T and 
ADS/JAO.ALMA\#2016.A.00015.S. ALMA is a partnership of ESO (representing its member states), NSF 
(USA) and NINS (Japan), together with NRC (Canada), NSC and ASIAA (Taiwan), and KASI (Republic of 
Korea), in cooperation with the Republic of Chile. The Joint ALMA Observatory is operated by
ESO, AUI/NRAO and NAOJ. 
VLA observations for this study were obtained via project 15A-235. The National Radio Astronomy 
Observatory is a facility of the National Science Foundation operated under cooperative agreement by 
Associated Universities, Inc.
This work makes use of data supplied by the UK Swift Science Data Centre at the 
University of Leicester and of data obtained through the High Energy Astrophysics Science Archive 
Research Center On-line Service, provided by the NASA/Goddard Space Flight Center.
This work includes data obtained with the Swope Telescope at Las Campanas Observatory, Chile, as 
part of the Swope Time Domain Key Project (PI Piro; co-PIs Shappee, Drout, Madore, Phillips, Foley, 
and Hsiao). 
Some of the data presented herein were obtained at the W. M. Keck Observatory, which 
is operated as a scientific partnership among the California Institute of Technology, the University 
of California and the National Aeronautics and Space Administration. The Observatory was made 
possible by the generous financial support of the W. M. Keck Foundation.
This research has made use of the SVO Filter Profile 
Service\footnote{\url{http://svo2.cab.inta-csic.es/theory/fps/}}, supported from the Spanish MINECO 
through grant AyA2014-55216.

\appendix
\section{The Klein-Nishina correction}
\label{appendix:KN}
The critical energy at which electrons effectively Compton scatter off their own synchrotron 
photons is given by,
\begin{equation}
 \gamma_{\rm self} = \frac{B_{\rm QED}}{B},
\end{equation}
where $B_{\rm QED} = 4.4\times10^{13}$\,G is the quantum critical field and $B$ is the post-shock 
magnetic field \citep{nas09}. Writing $B = \left(16\pi\epsb m_{\rm p} n_0 c^2 
\Gamma^2\right)^{1/2}$ for the ISM environment (cgs units) and substituting the relativistic 
hydrodynamic solution for the Lorentz factor of the FS ($\Gamma$) as a function of observer time 
\citep{bm76}, we have
\begin{equation}
 \gamma_{\rm self} = 3.5\times10^4 E_{52}^{-1/24} n^{-1/8} \epsb^{-1/6} 
 \left(\frac{t_{\rm d}}{1+z}\right)^{1/8}, 
\end{equation}
where $t_{\rm d}$ is the observer time in days.
For the FS parameters in Table \ref{tab:params}, the ordering of the critical Lorentz factors at 
$\approx 1$\,d is $\gamma_{\rm m} \approx \widehat{\gamma}_{\rm c}<\gamma_{\rm self}<\gamma_{\rm 
c}<\widehat{\gamma}_{\rm m}$, where $\gamma^2\widehat{\gamma}=\gamma_{\rm self}^3$ \citep{nas09}. 
Thus, the spectral slope above $\nuc$ is expected to be 
$\beta = -3(p-1)/4\approx -0.8$ (rather than $\beta =-p/2\approx-1.0$), agreeing better 
with the measured X-ray spectral index of $\beta_{\rm X} = -0.86\pm0.03$ at this time. In this 
regime, we expect $\nuc\propto t^{-(8-3p)/(8-2p)}$ \citep{nas09}. The decline rate of the 
resulting light curve is, therefore, expected to be marginally shallower; $\alpha \sim 7(1-p)/8 
\approx -0.94$ rather than $\alpha \sim (2-3p)/4\approx -1.05$, in slightly better agreement with 
the observed decline rate of $\alpha_{\rm X} \approx-0.82$ during this time. The residual 
differences may be related to variations in $Y(\gamma_c)$ as the KN-corrected SED transitions
between spectral regimes.


\begin{figure*} 
  \includegraphics[width=\textwidth]{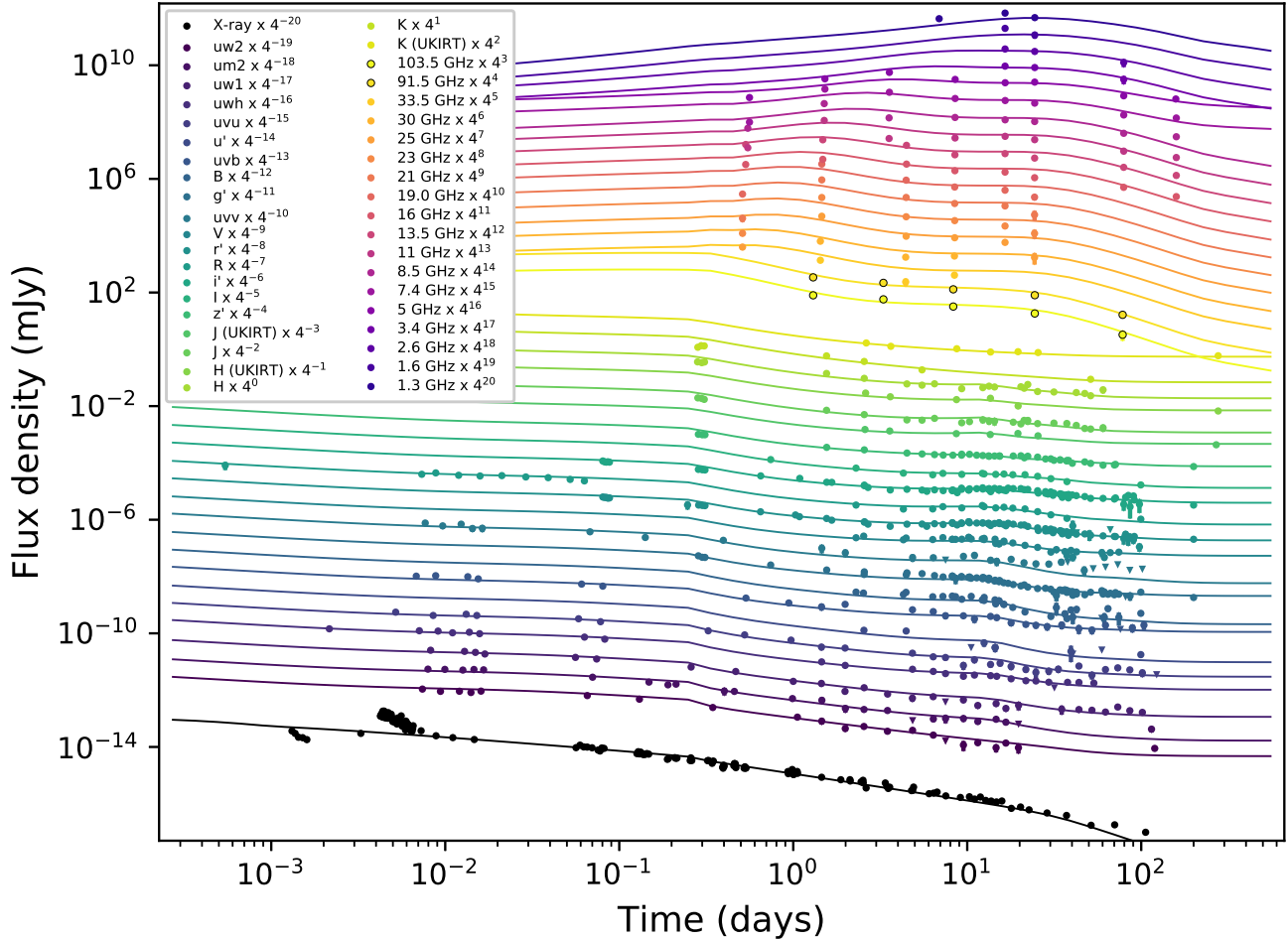}
 \caption{X-ray to radio light curves of GRB\,161219B/SN2016jca, together with the FS+RS model 
presented in Section \ref{text:RS}. The combined model overcomes the deficiencies of the FS-only 
model (Figures 
\ref{fig:modellc_FS_splits} and \ref{fig:modelsed_FS}), and explains the 
overall behavior of the light curves at all 41 observing frequencies over 5 orders of magnitude in 
time.}
\label{fig:modellc_RS}
\end{figure*}

\bibliographystyle{apj}
\bibliography{grb_alpha,gcn}

\end{document}